\documentclass[11pt]{article}
\usepackage{epic}
\usepackage{eepic}
\usepackage{amsmath,makeidx,epsfig}
\usepackage{amsfonts}
\usepackage[amsmath]{ntheorem}
\makeindex
\newtheorem{theorem}{Theorem}[subsection]
\newtheorem{corollary}[theorem]{Corollary}
\newtheorem{lemma}[theorem]{Lemma}

\newtheorem{proposition}[theorem]{Proposition}
\newtheorem{definition}[theorem]{Definition}

\newtheorem{warning}[theorem]{Warning}

\newcommand{\qed}{\rule{7pt}{7pt}}
\newenvironment{proof}{\noindent{\bf Proof}\hspace*{1em}}{\qed\bigskip}
\newenvironment{proof-sketch}{\noindent{\bf Sketch of Proof}\hspace*{1em}}{\qed\bigskip}
\newenvironment{proof-idea}{\noindent{\bf Proof Idea}\hspace*{1em}}{\qed\bigskip}
\newenvironment{proof-of-lemma}[1]{\noindent{\bf Proof of Lemma #1}\hspace*{1em}}{\qed\bigskip}
\newenvironment{proof-of-theorem}[1]{\noindent{\bf Proof of Theorem #1}\hspace*{1em}}{\qed\bigskip}
\newenvironment{proof-of-proposition}[1]{\noindent{\bf Proof of Proposition #1}\hspace*{1em}}{\qed\bigskip}

\makeatother

\def\orig#1{\bar{#1}}

\def\aa{\pmb{\mathit{a}}}
\newcommand\bb{\boldsymbol{\mathit{b}}}
\newcommand\cc{\boldsymbol{\mathit{c}}}
\newcommand\dd{\boldsymbol{d}}
\newcommand\ddelta{\boldsymbol{d}}
\def\gg{\boldsymbol{\mathit{g}}}
\newcommand\hh{\boldsymbol{\mathit{h}}}
\newcommand\pp{\boldsymbol{\mathit{p}}}

\newcommand\qq{\boldsymbol{\mathit{q}}}
\newcommand\rr{\boldsymbol{\mathit{r}}}
\def\tt{\boldsymbol{\mathit{t}}}
\newcommand\uu{\boldsymbol{\mathit{u}}}
\newcommand\vv{\boldsymbol{\mathit{v}}}
\newcommand\xx{\boldsymbol{\mathit{x}}}
\newcommand\yy{\boldsymbol{\mathit{y}}}
\newcommand\zz{\boldsymbol{\mathit{z}}}
\newcommand\aalpha{\boldsymbol{\alpha}}

\newcommand\ttau{\boldsymbol{\tau}}
\newcommand\oomega{\boldsymbol{\omega}}
\newcommand\ppsi{\boldsymbol{\psi}}

\newcommand\zzero{\mathbf{0}}
\newcommand\oone{\mathbf{1}}

\newcommand\aap{\boldsymbol{\mathit{a}}^{+}}

\newcommand\yyp{\boldsymbol{\mathit{y}}^{+}}
\newcommand\zzp{\boldsymbol{\mathit{z}}^{+}}

\def\AA{\boldsymbol{\mathit{A}}}
\def\GG{\boldsymbol{\mathit{G}}}
\newcommand\BB{\boldsymbol{\mathit{B}}}

\newcommand\MM{\boldsymbol{\mathit{M}}}

\newcommand\RR{\boldsymbol{\mathit{R}}}

\newcommand\AAo{\orig{\boldsymbol{\mathit{A}}}}
\newcommand\AAt{\boldsymbol{\tilde{\mathit{A}}}}
\newcommand\GGt{\boldsymbol{\tilde{\mathit{G}}}}

\newcommand{\htil}{\tilde{h}}
\newcommand{\yt}{\tilde{y}}
\newcommand{\yp}{y^{+}}

\newcommand\aao{\orig{\boldsymbol{\mathit{a}}}}
\newcommand\aat{\tilde{\boldsymbol{\mathit{a}}}}
\newcommand\ggt{\tilde{\boldsymbol{\mathit{g}}}}

\newcommand\ggh{\hat{\boldsymbol{\mathit{g}}}}
\newcommand\pph{\hat{\boldsymbol{\mathit{p}}}}
\newcommand\ph{\hat{p}}
\newcommand\hhat{\hat{h}}
\newcommand\nuh{\hat{\nu}}

\newcommand\yyt{\tilde{\boldsymbol{\mathit{y}}}}
\newcommand\hht{\tilde{\boldsymbol{\mathit{h}}}}

\newcommand\aalphao{\boldsymbol{\orig{\alpha}}}
\newcommand\aalphat{\boldsymbol{\tilde{\alpha}}}

\newcommand\tto{\boldsymbol{\mathit{t}}'}
\newcommand\xxo{\boldsymbol{\mathit{x}}'}
\newcommand\yyo{\boldsymbol{\mathit{y}}'}

\newcommand\xxs{\boldsymbol{\mathit{x}}^{*}}

\newdimen\pIR
\newcommand\StevesR{{\rm I\kern\pIR R}}

\newcommand{\calC}{\mathcal{C}}
\newcommand{\calD}{\mathcal{D}}
\newcommand{\calF}{\mathcal{F}}
\newcommand{\calG}{\mathcal{G}}
\newcommand{\calI}{\mathcal{I}}
\newcommand{\calK}{\mathcal{K}}
\newcommand{\calM}{\mathcal{M}}
\newcommand{\calP}{\mathcal{P}}
\newcommand{\calS}{\mathcal{S}}
\newcommand{\calT}{\mathcal{T}}

\newcommand{\calTp}{\mathcal{T}^{+}}
\newcommand{\calSp}{\mathcal{S}^{+}}

\def\vs#1#2#3{#1_{#2},\ldots , #1_{#3}}

\def\vsb#1#2#3#4{#1_{#2_{#3}},\ldots , #1_{#2_{#4}}}

\def\vsDiv#1#2#3#4{#1_{#3} / #2_{#3},\ldots , #1_{#4} / #2_{#4}}

\def\defeq{\stackrel{\mathrm{def}}{=}}

\def\simp#1{\bigtriangleup \left(#1  \right)}

\def\ind#1{\left[#1 \right]}

\def\convhull#1{\mbox{\textbf{ConvHull}} \left(#1 \right)}

\def\Form#1#2{\left<#1 \Big| #2 \right>}
\def\form#1#2{\left<#1 | #2 \right>}
\def\vol#1{\mbox{{\bf Vol}}\left(#1  \right)}

\def\det#1{\mbox{{\bf det}}\left(#1  \right)}

\def\union{\bigcup }
\def\iff{\Leftrightarrow}

\def\setof#1{\left\{#1  \right\}}

\def\Reals#1{\StevesR^{#1}}
\def\sphere#1{S^{#1}}

\def\aff#1{\mbox{{\bf Aff}}\left(#1  \right)}

\def\Span#1{\textbf{Span}\left(#1  \right)}

\def\cone#1{\mbox{{\bf Cone}}\left(#1  \right)}

\def\shadow#1#2{\mbox{{\bf Shadow}}_{#1}\left(#2 \right)}

\def\abs#1{\left|#1  \right|}
\def\sizeof#1{\left|#1  \right|}

\def\norm#1{\left\| #1 \right\|}

\newcommand\symProb{\operatorname{\textbf{Pr}}\displaylimits}
\newcommand\symExpec{\operatorname{\textbf{E}}\displaylimits}
\def\prob#1#2{\symProb_{#1}\left[ #2 \right]}
\def\expec#1#2{\symExpec_{#1}\left[ #2 \right]}

\def\dist#1#2{\mbox{{\bf dist}}\left(#1, #2 \right)}

\def\from#1#2{\stackrel{\, #1}{\leftarrow} #2}

\def\diff#1{\, d #1 \,}

\def\optVert#1#2{\mbox{\bf optVert}_{#1 }{\left(#2 \right)}}
\def\optvert#1{\mbox{\bf optVert}_{#1}}

\def\optsimp#1{\mbox{\bf optSimp}_{#1}}
\def\optSimp#1#2{\mbox{\bf optSimp}_{#1}{\left(#2 \right)}}

\def\angle#1{\mbox{{\bf angle}}\left(#1  \right)}

\def\angTo#1#2{\mbox{{\bf ang}}\left(#1, #2 \right)}
\def\angZ#1#2{\mbox{{\bf ang}}_{#1}\left(#2 \right)}

\def\warning#1{}

\def\dastag#1{}

\reversemarginpar

\def\Jacob#1#2{\abs{\det{\frac{\partial (#1)}{\partial (#2)}}}}
\def\jacob#1#2{\abs{\det{\frac{\partial #1}{\partial #2}}}}
\def\jacobn#1#2{\abs{\frac{\partial #1}{\partial #2}}}

\def\smin#1{\mbox{\bf s}_{\textbf{min}}\left(#1  \right)}

\newcommand{\floor}[1]{\left\lfloor #1 \right\rfloor}
\newcommand{\ceiling}[1]{\left\lceil #1 \right\rceil}

\title{Smoothed Analysis of Algorithms:\\
Why the Simplex Algorithm Usually Takes
Polynomial Time \thanks{An extended abstract of this paper
appeared in the {\it Proceedings of the 33rd Annual ACM Symposium on
  Theory of Computing}, pp. 296-305, 2001.}}

\author{Daniel A. Spielman 
\thanks{Partially supported by an Alfred P. Sloan Foundation Fellowship,
NSF CAREER award CCR-9701304, NSF grant CCR-0112487,
  and a Junior Faculty Research Leave sponsored by the M.I.T.
School of Science}\\ 
Department of Mathematics \\ 
Massachusetts Institute of Technology\\
\and 
Shang-Hua Teng 
\thanks{
Partially suppoted by an Alfred P. Sloan Foundation Fellowship,
and NSF grant CCR: 99-72532. 
Part of this work was done while
at UIUC and visiting the department of 
mathematics at M.I.T.}\\
Department of Computer Science\\
Boston University, and\\
Akamai Technologies Inc.\\
}

\begin{document}

\maketitle

\begin{abstract}
{\bf
We introduce the \textit{smoothed analysis of algorithms}, which
  continuously interpolates between 
  the worst-case and average-case analyses of
  algorithms.
In smoothed analysis, we measure the maximum over inputs of
  the expected performance of an algorithm under small random
  perturbations of that input.
We measure this performance in terms of both the input
  size and the magnitude of the perturbations.
We show that the simplex algorithm has \emph{smoothed complexity}
  polynomial in the input size and the standard deviation
  of Gaussian perturbations.
}
\end{abstract}

\newpage

\tableofcontents
\newpage
\section*{List of Theorems, Lemmas, Corollaries and Propositions}

\newtheoremlisttype{mine}%
 {\begin{trivlist}\item}
 {\item[##1 ##2]\ \textbf{(##3)\dotfill ##4}}%
 {\end{trivlist}}
\theoremlisttype{mine}

\listtheorems{theorem}
\listtheorems{lemma}
\listtheorems{corollary}
\listtheorems{proposition}
%\linenumbers
\newpage

\section{Introduction}\label{sec:intro}
The Analysis of Algorithms community has been challenged by the
  existence of remarkable algorithms that are known by scientists and
  engineers to work well in practice, but whose theoretical analyses
  are negative or inconclusive.  
The root of this problem is that algorithms
  are usually analyzed in one of two ways: by worst-case or average-case
  analysis.  
Worst-case analysis can improperly suggest that an
  algorithm will perform poorly by examining its performance under
  the most contrived circumstances.
Average-case analysis was introduced to
  provide a less pessimistic measure of the performance of algorithms,
 and many practical algorithms perform well on the random
  inputs considered in average-case analysis.
However, average-case analysis may be unconvincing as
  the inputs encountered in many application domains
  may bear little resemblance to the random inputs
  that dominate the analysis. 

We propose an analysis that we call smoothed analysis which
  can help explain the
  success of algorithms that have poor worst-case complexity
  and whose inputs look sufficiently different from random that
  average-case analysis cannot be convincingly applied.
In smoothed analysis, we measure the
  performance of an algorithm under slight random perturbations of
  arbitrary inputs.  
In particular, we consider 
  Gaussian perturbations of inputs to algorithms that take real 
  inputs, and we measure the running times of algorithms in terms
  of their input size and the standard deviation of the Gaussian perturbations.

We show that the simplex method has polynomial smoothed
  complexity.  
The simplex method is the classic example of an
  algorithm that is known to perform well in practice but which takes
  exponential time in the worst case
\cite{KleeMinty,Murty,GoldfarbSit,Goldfarb,AvisChvatal,Jeroslow,AmentaZiegler}.
In the late 1970's and early 1980's the simplex method was shown
  to converge in expected polynomial time on various distributions of
  random inputs by researchers including Borgwardt, Smale, Haimovich, Adler,
  Karp, Shamir, Megiddo, and Todd
\cite{Borg82,Borg77,SmaleRand,Haimovich,AdlerKarpShamir,AdlerMegiddo,ToddRand}.
These works introduced novel probabilistic tools to the analysis
  of algorithms, and provided some intuition as to why the
  simplex method runs so quickly.
However, these analyses are dominated by
  ``random looking'' inputs: even if one were to prove
  very strong bounds on the higher moments of the distributions
  of running times on random inputs,
  one could not prove that an algorithm performs well
  in any particular small neighborhood of inputs.

To bound expected running times on small neighborhoods of inputs,
  we consider linear programming problems in the form
\begin{eqnarray}\label{prg:A}
 &  \mbox{maximize} & \zz ^{T} \xx  \nonumber \\
 & \mbox{subject to} & \AA  \xx  \leq \yy,
\end{eqnarray}
 and prove that for every vector $\zz$
  and every matrix $\AAo$ and vector $\orig{\yy}$,
  the expectation over standard deviation
  $\sigma \left(\max_{i}\norm{(\orig{y}_{i}, \aao_{i})} \right)$
  Gaussian perturbations $\AA$ and $\yy$ of
  $\AAo $ and $\orig{\yy}$
  of the time taken by a two-phase shadow-vertex simplex method
  to solve such a linear program
  is polynomial in $1/\sigma$ and the dimensions of $\AA$.

\subsection{Linear Programming and the Simplex Method}\label{ssec:lp}
It is difficult to overstate the importance of linear programming
  to optimization.
Linear programming problems arise in innumerable industrial contexts.
Moreover, linear programming is often used as a fundamental step
  in other optimization algorithms.
In a linear programming problem, one is asked to maximize or
  minimize a linear function over a polyhedral region.

Perhaps one reason we see so many linear programs is that we
  can solve them efficiently.
In 1947, Dantzig~\cite{Dantzig} introduced the simplex method,
  which was the first practical approach to solving linear programs
  and which remains widely used today.
To state it roughly, the simplex method proceeds by walking from
  one vertex to another of the polyhedron defined by the inequalities
  in \eqref{prg:A}.
At each step, it walks to a vertex that is better with respect to
  the objective function.
The algorithm will either determine that 
  the constraints are unsatisfiable, determine that the objective function is
  unbounded, or  reach a vertex from which it cannot make
  progress, which necessarily optimizes the objective function.

Because of its great importance, other algorithms for 
  linear programming have been invented.
In 1979, Khachiyan~\cite{Khachiyan} applied the
  ellipsoid algorithm to linear programming and proved that
  it always converged in time polynomial in
  $d$, $n$, and $L$---the number of
  bits needed to represent the linear program.
However, the ellipsoid algorithm has not been competitive  
  with the simplex method in practice.
In contrast, the interior-point method introduced in 1984
  by  Karmarkar~\cite{Karmarkar}, which also runs in time polynomial
  in $d$, $n$, and $L$, has performed very well:
 variations of the interior point method are competitive with
  and occasionally superior to the simplex method in practice.

In spite of half a century of attempts to unseat it,
  the simplex method remains the most popular method
  for solving linear programs.
However, there has been no satisfactory theoretical 
  explanation of its excellent performance.
A fascinating approach to understanding the performance of the
  simplex method has been the attempt to
  prove that there always exists a short
  walk from each vertex to the optimal vertex.
The Hirsch conjecture states that there should
  always be a walk of length at most $n - d$.
Significant progress on this conjecture was 
  made by Kalai and Kleitman~\cite{KalaiKleitman}, who proved that
  there always exists a walk of length
  at most $n ^{\log_{2}d + 2}$.
However, the existence of such a short walk does not imply
  that the simplex method will find it.

A simplex method is not completely defined until one
  specifies its \textit{pivot rule}---the method by which
  it decides which vertex to walk to 
  when it has many to choose from.  
There is no deterministic pivot rule under which the
  simplex method is known to take a sub-exponential
  number of steps.
In fact, for almost every deterministic
  pivot rule there is a family of polytopes 
  on which it is known to take an exponential number of 
  steps
\cite{KleeMinty,Murty,GoldfarbSit,Goldfarb,AvisChvatal,Jeroslow}.
  (See~\cite{AmentaZiegler} for a survey and a 
  unified construction of these polytopes).
The best present analysis of randomized pivot rules shows
  that they take expected time $n^{O (\sqrt{d})}$%
\cite{KalaiSubexp,Matousek},
  which is quite far from the polynomial complexity
  observed in practice.
This inconsistency between the exponential worst-case behavior of the
  simplex method and its everyday practicality leave us wanting
  a more reasonable theoretical analysis.

%% from STOC version

Various average-case analyses of the simplex method
  have been performed.
Most relevant to this paper is the analysis of
  Borgwardt~\cite{Borg77,Borg82}, who
  proved that the simplex method with the shadow
  vertex pivot rule runs in expected polynomial time
  for polytopes whose constraints are drawn independently from 
  spherically symmetric distributions 
  (\textit{e.g.} Gaussian distributions centered at the origin).
Independently, 
  Smale~\cite{SmaleRand,SmaleRand2} proved bounds on the 
  expected running time of Lemke's self-dual parametric simplex algorithm
  on linear programming problems 
  chosen from a spherically-symmetric distribution.
Smale's analysis was substantially improved by Megiddo~\cite{Megiddo}.

While these average-case analyses are significant
  accomplishments, it is not clear whether they
  actually provide intuition for what happens
  on typical inputs.
Edelman~\cite{EdelmanRoulette} writes on this point:
\begin{quotation}
What is a mistake is to psychologically link a random
  matrix with the intuitive notion of a ``typical'' matrix
  or the vague concept of ``any old matrix.''
\end{quotation}

Another model of random linear programs was studied in
  a line of research initiated independently
  by Haimovich~\cite{Haimovich} and Adler~\cite{Adler}.
Their works
  considered the maximum over matrices, $\AA$,
  of the expected time taken by parametric simplex
  methods to solve linear programs over these matrices
  in which the directions of the
  inequalities are chosen at random.
As this framework considers the maximum of an average,
  it may be viewed as a precursor to smoothed 
  analysis---the distinction being that 
  the random choice of
  inequalities cannot be viewed as a perturbation,
  as different choices yield radically different linear programs.
Haimovich and Adler both proved that 
  parametric simplex methods
  would take an expected linear number of steps
  to go from the vertex minimizing the objective function
  to the vertex maximizing the objective function,
  even conditioned on the program being feasible.
While their theorems confirmed the intuitions of many practitioners, 
  they were geometric rather than algorithmic%
\footnote{Our results in Section~\ref{sec:shadow} are analogous to
  these results.}
 as it 
  was not clear how an algorithm would locate either vertex.
Building on these analyses, Todd~\cite{ToddRand},
  Adler and Megiddo~\cite{AdlerMegiddo},
  and Adler, Karp and Shamir~\cite{AdlerKarpShamir}
  analyzed parametric algorithms for linear programming under this model
  and proved quadratic
  bounds on their expected running time.
While the random inputs considered in these analyses are
  not as special as the random inputs obtained from spherically
  symmetric distributions,
  the model of randomly flipped inequalities provokes some
  similar objections.

\subsection{Smoothed Analysis of Algorithms
 and Related Work}\label{ssec:smooth}
We introduce the \textit{smoothed analysis of algorithms} in the hope that
  it will help explain the good practical performance of many
  algorithms that worst-case does not and for which average-case analysis 
  is unconvincing.
Our first application of the smoothed analysis of algorithms will be to
  the simplex method.
We will consider the maximum over $\AAo$
 and $\orig{\yy}$ of the expected running time
  of the simplex method on inputs of the form
\begin{eqnarray}
 &  \mbox{maximize} & \zz ^{T} \xx \nonumber \\
 & \mbox{subject to} & (\AAo + \GG) \xx  \leq (\orig{\yy} + \hh),  \label{prg:AG}
\end{eqnarray}
where we let $\AAo$ and $\orig{\yy}$ be arbitrary 
  and $\GG$ and $\hh$ be a matrix and a vector of independently chosen
  Gaussian random variables of mean $0$ and 
  standard deviation $\sigma \left(\max_{i}\norm{(\orig{y}_{i}, \aao_{i})} \right)$.
If we let $\sigma $ go to $0$, then we obtain the worst-case
  complexity of the simplex method; whereas, if we let $\sigma $
  be so large that $\GG$ swamps out $\AA$, we obtain the
  average-case analyzed by Borgwardt.
By choosing polynomially small $\sigma $, this analysis combines
  advantages of worst-case and average-case analysis, and roughly
  corresponds to the notion of imprecision in low-order digits.

In a smoothed analysis of an algorithm, we assume that the inputs
  to the algorithm are subject to slight random perturbations,
  and we measure the complexity of the algorithm in terms of the input
  size and the standard deviation of the perturbations.
If an algorithm has low smoothed complexity, then one should expect it to
  work well in practice since most real-world problems are generated
  from data that is inherently noisy.
Another way of thinking about smoothed complexity is to observe that if an
  algorithm has low smoothed complexity, then one must be unlucky
  to choose an input instance on which it performs poorly.

We now provide some definitions for the smoothed analysis of algorithms
  that take real or complex inputs.
For an algorithm $A$ and input $\xx $, let 
\[
   \calC_{A} (\xx )
\]
be a complexity measure of $A$ on input $\xx$.
Let $X$ be the domain of inputs to $A$, and let
  $X_{n}$ be the set of inputs of size $n$.
The size of an input can be measured in various ways.
Standard measures are the number of real variables
  contained in the input and the sums of the bit-lengths  
  of the variables.
Using this notation, one can say that $A$ has worst-case
  $\calC$-complexity $f (n)$ if 
\[
  \max _{\xx \in X_{n}} (\calC_{A} (\xx )) = f (n).
\]
Given a family of distributions $\mu_{n} $ on $X_{n}$, we say that $A$
  has average-case $\calC$-complexity $f (n)$ under $\mu $ if
\[
  \expec{\xx  \from{\mu _{n}}{X_{n}}}{\calC_{A} (\xx )} = f (n).
\]
Similarly, we say that $A$ has \textit{smoothed $\calC$-complexity} 
  $f (n, \sigma )$ if
\begin{equation}\label{eqn:smoothedcomplexity}
 \max _{\xx  \in X_{n}} 
  \expec{\gg }{\calC_{A} (\xx + \left(\sigma \norm{\xx}_{?} \right) \gg  )} = f (n, \sigma ),
\end{equation}
\index{smoothed-complexity}%
 where $\left( \sigma \norm{\xx}_{?} \right) \gg$ is a vector of Gaussian random variables of mean $0$ and
  standard deviation $\sigma \norm{\xx}_{?}$ and $\norm{\xx}_{?}$ is a measure of the magnitude
  of $\xx$, such as the largest element or the norm.
We say that an algorithm has \textit{polynomial smoothed complexity}
  if its smoothed complexity is polynomial in $n$ and $1/\sigma $.
\index{polynomial smoothed complexity}
In Section~\ref{sec:conclusions}, we present some 
  generalizations of the definition of smoothed complexity that
  might prove useful.
To further contrast smoothed analysis with average-case analysis, 
  we note that the probability mass in \eqref{eqn:smoothedcomplexity} is
  concentrated in a region of radius $O (\sigma \sqrt{n})$ and
  volume at most $O (\sigma \sqrt{n})^{n}$,
  and so, when $\sigma$ is small, this region contains an exponentially small fraction
  of the probability mass in an average-case analysis.
Thus, even an extension of average-case analysis to higher moments
  will not imply meaningful bounds on smoothed complexity.

A discrete analog of smoothed analysis has been studied in a collection
  of works inspired by Santha and Vazirani's \textit{semi-random source}
  model~\cite{SanthaVazirani}.
In this model, an adversary generates an input, and each bit of this input
  has some probability of being flipped.
Blum and Spencer~\cite{BlumSpencer} design a polynomial-time 
  algorithm that $k$-colors
  $k$-colorable graphs generated by this model.
Feige and Krauthgamer~\cite{FeigeKrauthgamer} analyze a model
  in which the adversary is more powerful,
  and use it to show that Turner's algorithm~\cite{Turner}
  for approximating the bandwidth performs well 
  on semi-random inputs.
They also improve Turner's analysis.
Feige and Kilian~\cite{FeigeKilian}
  present polynomial-time algorithms that
  recover large independent sets, 
  $k$-colorings, and optimal bisections
  in semi-random graphs.
They also demonstrate that significantly better
  results would lead to surprising
  collapses of complexity classes.

\subsection{Our Results}\label{ssec:results}

We consider
  the maximum over $\zz$, $\orig{\yy}$,
  and $\vs{\aao}{1}{n}$ of the expected time taken
  by a two-phase shadow vertex simplex method to solve
 linear programming problems of the form
\begin{eqnarray}
 &  \mbox{maximize} & \zz^{T} \xx \nonumber  \\
 & \mbox{subject to} & \form{\aa _{i}}{\xx} \leq y _{i}, 
  \mbox{ for $1 \leq i \leq n$,} \label{eqn:lpEnumerated2}
\end{eqnarray} \index{zz@$\zz $}%
where each $\aa _{i}$ is a Gaussian random vector of standard deviation
  $\sigma \max_{i} \norm{(\orig{y}_{i}, \aao_{i})}$ centered at $\aao _{i}$,
  and each $y_{i}$ is a Gaussian random variable of  standard deviation
  $\sigma \max_{i} \norm{(\orig{y}_{i}, \aao_{i})}$ centered at $\orig{y} _{i}$.

We begin by considering the case in which
  $\yy = \oone $, $\norm{\aao _{i}} \leq 1$,
  and $\sigma < 1/3 \sqrt{d \ln n}$.
In this case, our first result, Theorem~\ref{thm:shadow}, says that
  for every vector
  $\tt $ the expected size of the {\em  shadow} of the polytope---the
  projection of the polytope defined
  by the equations (\ref{eqn:lpEnumerated2})  onto the plane
  spanned by $\tt $ and $\zz $---is polynomial in $n$, the dimension,
  and $1/\sigma $.
This result is the geometric foundation of our work, but
  it does not directly bound the running time of an algorithm,
  as the shadow relevant to the analysis of an algorithm
  depends on the perturbed program and cannot be specified
  beforehand as the vector $\tt$ must be.
In Section~\ref{sec:introSVM2phase}, we describe a two-phase 
  shadow-vertex simplex algorithm,
  and in Section~\ref{sec:phaseI} we
  use Theorem~\ref{thm:shadow} as a black box to show
  that it takes expected time polynomial in $n$, $d$,
  and $1/\sigma $ in the case described above.

Efforts have been made to analyze how much the solution of a linear
  program can change as its data is perturbed.
For an introduction to such analyses, 
  and an analysis of the complexity of interior point
  methods in terms of the resulting condition number,
  we refer the reader to
  the work of Renegar~\cite{RenegarFunc,RenegarCond,RenegarPert}.

\subsection{Intuition Through Condition Numbers}\label{sec:intuition}
For those already familiar with the simplex method and condition numbers,
  we include this section to provide some intuition for why our   
  results should be true.

Our analysis will exploit geometric properties
  of the condition number of a matrix, rather than of a
  linear program.
We start with the observation that if a corner of a polytope
  is specified by the equation $A_{I} \xx = \yy_{I}$,
  where $I$ is a $d$-set, then the condition number of
  the matrix $A_{I}$ provides a good measure of how far the corner
  is from being flat.
Moreover, it is relatively easy to show that if
  $A$ is subject to perturbation, then it is unlikely that
  $A_{I}$ has poor condition number.
So, it seems intuitive that if $A$ is perturbed, then most
  corners of the polytope should have angles bounded away
  from being flat.
This already provides some intuition as to why the simplex method
  should run quickly: one should make reasonable progress as
  one rounds a corner if it is not too flat.

There are two difficulties in making the above intuition rigorous:
  the first is that even if $A_{I}$ is well-conditioned for most
  sets $I$, it is not clear that $A_{I}$ will be well-conditioned
  for most sets $I$ that are bases of corners of the polytope.
The second difficulty is that even if most corners of the polytope
  have reasonable condition number, it is not clear that a simplex
  method will actually encounter many of these corners.
By analyzing the shadow vertex pivot rule, it is possible to resolve
  both of these difficulties.

The first advantage of studying the shadow vertex pivot rule is
  that its analysis comes down to studying the expected sizes
  of shadows of the polytope.
From the specification of the plane onto which the polytope will be projected,
  one obtains a characterization of all the corners that will be in
  the shadow, thereby avoiding the complication of an iterative
  characterization.
The second advantage is that these corners are specified by the
  property that they optimize a particular objective function,
  and using this property one can actually bound the probability
  that they are ill-conditioned.
While the results of Section~\ref{sec:shadow} are not stated in
  these terms, this is the intuition behind them.

Condition numbers also play a fundamental role in our
  analysis of the shadow-vertex algorithm.
The analysis of the algorithm differs from the mere analysis
  of the sizes of shadows in that, in the study of an algorithm,
  the plane onto which the polytope is projected depends upon
  the polytope itself.
This correlation of the plane with the polytope complicates
  the analysis, but is also resolved through the help
  of condition numbers.
In our analysis, we view the perturbation as the composition
  of two perturbations, where the second is small relative to the first.
We show that our choice of the plane onto which we
  project the shadow is well-conditioned with high
  probability after the first perturbation.
That is, we show that the second perturbation is unlikely
  to substantially change the plane onto which we project,
  and therefore unlikely to substantially change the shadow.
Thus,  it suffices to measure the expected size of the
  shadow obtained after the second perturbation onto the
  plane that would have been chosen after just the first
  perturbation.

The technical lemma that enables this analysis, Lemma~\ref{lem:MGC},
  is a concentration result that proves that it is highly
  unlikely that almost all of the minors of a random
  matrix have poor condition number.
This analysis also enables us to show that it is highly
  unlikely that we will need a large ``big-$M$''
  in phase I of our algorithm.

We note that the condition numbers of the $A_{I}$s
  have been studied before in the complexity of
  linear programming algorithms.
The condition number $\bar{\chi}_{A}$
  of Vavasis and Ye~\cite{VavasisYe} measures
  the condition number of the worst sub-matrix $A_{I}$,
  and their algorithm runs in time proportional
  to $\ln (\bar{\chi }_{A})$.
Todd, Tun{\c{c}}el, and Ye~\cite{ToddTuncelYe} have shown 
  that for a Gaussian random matrix the expectation
  of $\ln (\bar{\chi }_{A})$ is $O (\min (d \ln n, n))$.
That is, they show that it is unlikely that any $A_{I}$
  is exponentially ill-conditioned.
It is relatively simple to apply the techniques of
  Section~\ref{sec:phaseIManyGood} to obtain a similar
  result in the smoothed case.
We wonder whether our concentration result that it
  is exponentially unlikely that many $A_{I}$
  are even polynomially ill-conditioned could
  be used to obtain a better smoothed analysis
  of the Vavasis-Ye algorithm.

\subsection{Discussion}\label{sec:introDiscussion}

One can debate whether the definition of
  \textit{polynomial smoothed complexity}
  should be that an algorithm have complexity polynomial in $1/\sigma $
  or $\log (1/\sigma )$.
We believe that the choice of being polynomial in $1/\sigma $
  will prove more useful as the other definition is too strong
  and quite similar
  to the notion of being polynomial in the worst case.
In particular, one can convert any algorithm for linear programming
  whose smoothed complexity
  is polynomial in $d$, $n$ and $\log (1/\sigma) $
  into an algorithm whose worst-case complexity is polynomial in $d$,
  $n$, and $L$.
That said, one should certainly prefer complexity bounds that are
  lower as a function of $1/\sigma$, $d$ and $n$.

We also remark that a simple examination of the 
  constructions that provide exponential lower bounds
  for various pivot 
  rules~\cite{KleeMinty,Murty,GoldfarbSit,Goldfarb,AvisChvatal,Jeroslow}  
  reveals that none of these pivot rules
  have smoothed complexity polynomial in $n$ and
  sub-polynomial in $1/\sigma $.
That is, these constructions are unaffected by exponentially
  small perturbations.

% Local Variables: ***
% TeX-master:"shadow.tex" ***
% End: ***

\section{Notation and Mathematical Preliminaries}\label{sec:note}

In this section, we define the notation that will be used in the paper.
We will also review some background from mathematics 
  and derive a few simple statements that we will need.
The reader should probably skim this section now, and save
  a more detailed examination for when the relevant
  material is referenced.

\begin{itemize}
\item $[n]$ denotes the set of integers between 1 and $n$,
  and $\binom{[n]}{k}$ denotes the subsets of 
  $[n]$ of size $k$.

\item Subsets of $[n]$ are denoted by the capital Roman letters $I, J,L, K$.
  $\calM$ will denote a subset of integers, and $\calK$ will denote
  a set of subsets of $[n]$.

\item Subsets of $\Reals{?}$ are denoted by the capital  Roman letters
   $A, B, P, Q, R, S, T, U, V$.
\item Vectors in $\Reals{?}$ are denoted by bold lower-case Roman letters,
  such as $\aa_{i}, \aao_{i}, \aat_{i}$, $\bb_{i}, \cc _{i}$, $\dd_{i}, \hh$,
  $\tt ,\qq ,\zz , \yy$.

\item Whenever a vector, say $\aa \in \Reals{d}$ is present, its components
  will be denoted by lower-case Roman letters with subscripts, such as
  $\vs{a}{1}{d}$.

\item Whenever a collection of vectors, such as $\vs{\aa}{1}{n}$,
  are present, the similar bold upper-case letter, such as $\AA $,
  will denote the matrix of these vectors.
  For $I \in \binom{[n]}{k}$, $\AA _{I}$ will denote the matrix
  of those $\aa _{i}$ for which $i \in I$.

\item Matrices are denoted by bold upper-case Roman letters,
  such as $\AA, \AAo, \AAt , \BB , \MM $ and $\RR _{\oomega}$.

\item $\sphere{d-1}$ denotes the unit sphere in $\Reals{d}$.
  \index{Sigma@$\sphere{d}$}%

\item Vectors in $\sphere{?}$ will be denoted by
  bold Greek letters, such as $\oomega , \ppsi , \ttau$.

\item Generally speaking, univariate quantities with scale,
  such as lengths or heights, will be represented by
  lower case Roman letters such as $c$, $h$, $l$, $r$, $s$, and $t$.
  The principal exceptions are that $\kappa$ and $M$
  will also denote such quantities.

\item Quantities without scale, such as the ratios of quantities
  with scale or affine coordinates, will be represented by
  lower case Greek letters such as 
  $\alpha , \beta , \lambda , \xi , \zeta$.
  $\aalpha$ will denote a vector of such quantities
  such as $(\vs{\alpha }{1}{d})$.

\item Density functions are denoted by lower case Greek letters such as
   $\mu$ and $\nu $.

\item The standard deviations of Gaussian random variables
  are denoted by lower-case Greek letters such as
  $\sigma, \tau $ and $\rho $.

\item Indicator random variables are denoted by upper case Roman
  letters, such as $A$, $B$, $E$, $F$, $V$, $W$, $X$, $Y$, and $Z$

\item Functions into the reals or integers will be denoted
  by calligraphic upper-case letters, such as
  $\calF, \calG, \calS^{+}, \calS', \calT$.

\item Functions into $\Reals{?}$ are denoted
  by upper-case Greek letters, such as
   $\Phi_{\epsilon },  \Upsilon , \Psi$.

\item $\form{\xx }{\yy }$ denotes the inner 
  product of vectors $\xx $ and $\yy $.\index{<@$\form{\xx }{\yy }$}

\item 
For vectors $\oomega $ and $\zz$, we let
  $\angle{\oomega , \zz}$ denote the angle
  between these vectors at the origin.\index{angle@$\angle{\oomega , \zz}$}%

\item The logarithm base 2 is written $\lg$
  and the natural logarithm is written $\ln$.

\item The probability of an event $A$ is written $\prob{}{A}$,
  and the expectation of a variable $X$ is written $\expec{}{X}$.

\item The indicator random variable for an event $A$
  is written $\ind{A}$.

\end{itemize}

\subsection{Geometric Definitions}

For the following definitions, we let 
  $\vs{\aa }{1}{k}$ denote a set of vectors in $\Reals{d}$.

\begin{itemize}

\item $\Span{\vs{\aa }{1}{k}}$ denotes the subspace spanned
  by $\vs{\aa }{1}{k}$. \index{span@$\Span{}$}%

\item $\aff{\vs{\aa }{1}{k}}$ denotes the hyperplane that is the
  affine span of $\vs{\aa }{1}{k}$: \index{aff@$\aff{}$}
 the set of points $\sum _{i} \alpha _{i} \aa _{i}$,
  where $\sum _{i} \alpha _{i} = 1$,
  for all $i$.

\item $ \convhull{\vs{\aa }{1}{k}} $ denotes the convex hull
  of $\vs{\aa }{1}{k}$. \index{convh@$\convhull{}$}%

\item $ \cone{\vs{\aa }{1}{k}} $ denotes the 
  positive cone through $\vs{\aa}{1}{k}$:
  the set of points $\sum_{i} \alpha _{i} \aa _{i}$,
  for $\alpha _{i} \geq 0$.
 \index{cone@$\cone{}$}%

\item $\simp{\vs{\aa }{1}{d}}$ denotes the simplex 
  $\convhull{\vs{\aa }{1}{d}}$. \index{>@$\simp{}$}%

\end{itemize}

For a linear program specified by $\vs{\aa}{1}{n}$,
  $\yy$ and $\zz$, we will say that the linear program
  is in \textit{general position} if \index{general position@general position}
\begin{itemize}
\item The points $\vs{\aa}{1}{n}$ are in general position with respect to $\yy$,
  which means that for all $I \subset \binom{[n]}{d}$
  and $\xx = \AA_{I}^{-1} \yy _{I}$, and all $j \not \in I$,
  $\form{\aa _{j}}{\xx} \not = y_{j}$.

\item For all $I \subset \binom{[n]}{d-1}$,
  $\zz \not \in \cone{\AA _{I}}$.
\end{itemize}
Furthermore, we will say that the linear program
  is in \textit{general position with respect to a vector $\tt$}
  if the set of $\lambda $ for which there exists
  an $I \in \binom{[n]}{d-1}$ such that
\[
  (1-\lambda )\tt + \lambda \zz \in \cone{\AA _{I}}
\]
is finite and does not contain $0$.

\subsection{Vector and Matrix Norms}

The material of this section is principally used
  in Sections~\ref{sec:introSVM2phase} 
  and~\ref{sec:phaseIManyGood}.
The following definitions and propositions are standard, and may
  be found in standard texts on Numerical Linear Algebra.
  
\begin{definition}[Vector Norms]
For a vector $\xx$,
  we define
\begin{itemize}
\item $\norm{\xx} = \sqrt{\sum_{i} x_{i}^{2}}$.
\item $\norm{\xx}_{1} = \sum_{i} \abs{x_{i}}$.
\item $\norm{\xx}_{\infty} = \max_{i} \abs{x_{i}}$.
\end{itemize}
\end{definition}

\begin{proposition}[Vectors norms]\label{pro:vecNorms}
For a vector $\xx \in \Reals{d}$, 
\[
  \norm{\xx} \leq \norm{\xx}_{1} \leq \sqrt{d} \norm{\xx}.
\]
\end{proposition}

\begin{definition}[Matrix norm]\label{def:matrixNorms}
For a matrix $\AA$,
  we define
\[
  \norm{\AA} \defeq  \max_{\xx} \norm{\AA \xx} / \norm{x}.
\]
\end{definition}

\begin{proposition}[Properties of matrix norm]\label{pro:matrixNorms}
For  $d$-by-$d$ matrices $\AA$ and $\BB$, and a $d$-vector $\xx$,
\begin{enumerate}
\item  $\norm{\AA \xx} \leq \norm{\AA} \norm{\xx}$. \label{enu:matrixNormsAx}
\item  $\norm{\AA \BB} \leq \norm{\AA} \norm{\BB}$.
\item  $\norm{\AA} = \norm{\AA^{T}}$. \label{enu:matrixNormsTrans}
\item $\norm{\AA} \leq \sqrt{d} \max_{i} \norm{\aa_{i}}$,
  where $\AA = (\vs{\aa}{1}{d})$. \label{enu:matrixNormsMax}
\item $\det{\AA} \leq \norm{\AA}^{d}$. \label{enu:matrixNormsDet}
\end{enumerate}
\end{proposition}

\begin{definition}[$\smin{}$]\label{def:smin}
For a matrix $\AA$,
  we define
\[
  \smin{\AA} \defeq  \norm{\AA^{-1}}^{-1}.
\]
\end{definition}\index{smin@$\smin{}$}\index{smin@$\smin{}$}%
We recall that $\smin{\AA}$ is the smallest singular value of
  the matrix $\AA$, and that it is not a norm.
\index{smallest singular value@smallest singular value}%

\begin{proposition}[Properties of $\smin{}$]\label{pro:smin}
For  $d$-by-$d$ matrices $\AA$ and $\BB$, 
\begin{enumerate}
\item  $\smin{\AA} = \min_{\xx} \norm{\AA \xx} / \norm{\xx}$.
  \label{enu:matrixNormsSmallest}
\item  $\smin{\BB} \geq \smin{\AA} - \norm{\AA - \BB}$
  \label{enu:matrixNormsSminDiff}.
\end{enumerate}
\end{proposition}

\subsection{Probability}

For an event, $A$, we let $\ind{A}$ denote the indicator random variable
  for the event.
We generally describe random variables by their density functions.
If $\xx$ has density $\mu$, then
\[
  \prob{}{A (\xx)} \defeq  \int \ind{A (\xx)} \mu (\xx) \diff{\xx }.
\]
If $B$ is another event, then
\[
  \prob{B}{A (\xx)} 
\defeq 
  \prob{}{A (\xx) \big| B (\xx )}
\defeq 
  \frac{\int \ind{B (\xx )}\ind{A (\xx)} \mu (\xx) \diff{\xx }}
       {\int \ind{B (\xx )} \mu (\xx) \diff{\xx }}.
\]
In a context where multiple densities are present, we will use
  use the notation $\prob{\mu }{A (\xx)}$ to indicate the probability
  of $A$ when $\xx$ is distributed according to $\mu $.

In many situations, we will not know the density $\mu $ of a random variable $\xx $,
  but rather a function $\nu$ such that $\nu (\xx) = c \mu (\xx)$
  for some constant $c$.
In this case, we will say that $\xx$ 
  \textit{has density proportional to $\nu$}.

The following Propositions and Lemmas will play a prominent role in 
  the proofs in this paper. 
The only one of these which might not be intuitively obvious
  is Lemma~\ref{lem:comb}.

\begin{proposition}[Average $\leq$ maximum]\label{pro:favorite}
Let $\mu (x,y)$ be a density function, and
 let $x$ and $y$ be distributed according to $\mu (x,y)$.
If $A (x,y)$ is an event and
  $X (x,y)$ is random variable,
  then
\begin{align*}
  \prob{x,y}{A (x,y)}
 & \leq \max _{x} \prob{y}{A (x,y)}, \text{ and}\\
  \expec{x,y}{X (x,y)}
 & \leq \max _{x} \expec{y}{X (x,y)},
\end{align*}
where in the right-hand terms, $y$ is distributed according
  to the induced distribution $\mu (x,y)$.
\end{proposition}

\begin{proposition}[Expectation on sub-domain]\label{pro:condProb}
Let $\xx $ be a random variable and
  $A (\xx )$ an event.
Let $P$ be  a measurable subset of the domain of $\xx $.
Then,
\[
  \prob{\xx \in P}{A (\xx )}
  \leq 
  \prob{}{A (\xx )} / \prob{}{\xx \in P}.
\]
\end{proposition}
\begin{proof}
By the definition of conditional probability,
\begin{align*}
  \prob{\xx \in P}{A (\xx )}
& = 
  \prob{}{A (\xx ) | \xx \in P}\\
& = 
  \prob{}{A (\xx ) \text{ and } \xx \in P}
  / \prob{}{\xx \in P}, & \text{by Bayes' rule,}\\
& \leq 
  \prob{}{A (\xx )}
  / \prob{}{\xx \in P}.
\end{align*}
\end{proof}

\begin{lemma}[Comparing expectations]\label{lem:probXYA}
Let $X$ and $Y$ be non-negative random variables
  and $A$ an event satisfying
(1) $X \leq k$,
 (2)
 $\prob{}{A} \geq 1 - \epsilon $,
 and (3) there exists a constant $c$
 such that
 $\expec{}{X | A} \leq c \expec{}{Y | A}$.
Then,
\[
  \expec{}{X} \leq c \expec{}{Y} + \epsilon k.
\]
\end{lemma}
\begin{proof}
\begin{align*}
 \expec{}{X} 
& = 
 \expec{}{X | A} \prob{}{A} +
 \expec{}{X | \mathbf{not} (A)} \prob{}{\mathbf{not} (A)}\\
& \leq 
 c \expec{}{Y | A} \prob{}{A} +
 \epsilon k\\
& \leq 
 c \expec{}{Y} +
 \epsilon k,
\end{align*}
by Proposition~\ref{pro:condProb}.
%%Let 
%%\begin{align*}
%%  X_{0} & = \left\{
%%       \begin{array}{ll}
%%       X & \text{if $A$},\\
%%       0 & \text{otherwise}
%%        \end{array}
%%           \right.\\
%%  X_{1} & = \left\{
%%       \begin{array}{ll}
%%       0 & \text{if $A$},\\
%%       k & \text{otherwise}
%%        \end{array}
%%           \right.\\
%%\end{align*}
%%Then $X \leq X_{0} + X_{1}$, so 
%% $\expec{}{X} \leq \expec{}{X_{0}} + \expec{}{X_{1}}$
%%Moreover, 
%%\[
%%   \expec{}{X_{0}} \leq c \expec{}{Y},
%%\]
%%and
%%\[
%%  \expec{}{X_{1}} \leq \epsilon k,
%%\]
%%from which the lemma follows.
\end{proof}

\begin{lemma}[Similar distributions]\label{lem:approximate}
Let $X$ be a non-negative random variable such that $X \leq k$. 
Let $\nu $ and $\mu  $ be density functions for which
  there exists a set $S$ such that
(1) $\prob{\nu}{S} >1- \epsilon $ and
(2) there exists a constant $c \geq 1$ such 
  that for all $a\in S$, 
$ \nu (a) \leq c \mu (a)$.
Then,
\[
\expec{\nu}{X (a)}\leq 
  c \expec{\mu}{X (a)}+k \epsilon .
\]
\end{lemma}
\begin{proof}
We write
\begin{align*}
  \expec{\nu }{X}
& = \int _{a \in S} X (a) \nu (a) \diff{a}
  + \int _{a \not \in S} X (a) \nu (a) \diff{a}\\
& \leq c \int _{a \in S} X (a) \mu (a) \diff{a}
  + k \epsilon \\
& \leq c \int _{a} X (a) \mu (a) \diff{a}
  + k \epsilon \\
& = c \expec{\mu }{X}
  + k \epsilon.
\end{align*}
\end{proof}

\begin{lemma}[Combination lemma]\label{lem:comb}
Let $x$ and $y$ be random variables
  distributed according to $\mu (x,y)$.
Let $\calF(x)$ and $\calG (x,y)$ be non-negative functions
  and  $\alpha $ and $\beta $ be constants such that
\begin{itemize}
\item $\forall \epsilon \geq 0$, $\prob{x,y}{\calF(x) \leq  \epsilon } \leq  \alpha \epsilon $, and
\item $\forall \epsilon \geq 0$, $\max _{x} 
  \prob{y}{\calG (x,y) \leq  \epsilon } \leq  (\beta \epsilon )^{2}$,
\end{itemize}
where in the second line $y$ is distributed according to the induced
  density $\mu (x,y)$.
Then
\[
  \prob{x,y}{\calF(x) \calG (x,y) \leq  \epsilon } \leq  4 \alpha \beta \epsilon.
\]
\end{lemma}
\begin{proof}
Consider any $x$ and $y$ for which $\calF(x) \calG (x,y) \leq  \epsilon $.
If $i$ is the integer for which
\[
2^{i} \beta \epsilon < \calF (x) \leq 2^{i+1} \beta \epsilon,
\]
then $\calG (x,y) \leq 2^{-i} / \beta$.
Thus, $\calF(x) \calG (x,y) \leq  \epsilon $, implies that either
  $\calF(x) \leq  2 \beta \epsilon $, or there exists an integer $i \geq 1$
  for which
\[
  \calF (x) \leq  2^{i+1}\beta \epsilon 
  \qquad \mbox{ and } \qquad 
  \calG (x,y) \leq 2^{-i} / \beta.
\]
So, we obtain the bound
\begin{eqnarray*}
 \prob{x,y}{\calF(x) \calG (x,y) \leq  \epsilon }
& \leq &
 \prob{x,y}{\calF(x) \leq 2 \beta \epsilon }
 + \sum _{i \geq 1}
 \prob{x,y}{
    \calF (x) \leq  2^{i+1}\beta \epsilon 
     \mbox{ and }
    \calG (x,y) \leq 2^{-i} / \beta
  }\\
& \leq  & 
  2 \alpha \beta \epsilon 
+ \sum _{i \geq 1}
 \prob{x, y}{
    \calF (x) \leq  2^{i+1}\beta \epsilon 
  }
  \prob{x, y}{
    \calG (x,y) \leq 2^{-i} / \beta
  \big|
   \calF (x) \leq  2^{i+1}\beta \epsilon 
  }\\
& \leq  & 
  2 \alpha \beta \epsilon 
+ \sum _{i \geq 1}
 \prob{x, y}{
    \calF (x) \leq  2^{i+1}\beta \epsilon 
  }
  \max _{x}
  \prob{y}{
    \calG (x,y) \leq 2^{-i} / \beta
  }\\
 & \leq  &
  2 \alpha \beta \epsilon 
+ \sum _{i \geq 1}
   \left(2^{i+1} \alpha \beta \epsilon  \right)
   \left( 2^{-i} \right)^{2},
  \text{ by Proposition~\ref{pro:favorite},}
  \\
& = &
  2 \alpha \beta \epsilon 
+  \alpha \beta \epsilon 
   \sum _{i \geq 1}
   2^{1-i}
  \\
& = &
  4 \alpha \beta \epsilon.
\end{eqnarray*}
\end{proof}

As we have found this lemma very useful in our work, and
  we suspect others may as well, we state a more broadly
  applicable generalization.
It's proof is similar.

\begin{lemma}[Generalized combination lemma]\label{lem:genComb}
Let $x$ and $y$ be random variables
  distributed according to $\mu (x,y)$.
There exists a function $c (a,b)$ such that
 if $\calF (x)$ and $\calG (x,y)$ are non-negative functions
  and  $\alpha $, $\beta $, $a$ and $b$ are constants such that
\begin{itemize}
\item $\prob{x,y}{\calF (x) \leq  \epsilon } \leq  (\alpha \epsilon)^{a} $, and
\item $\max _{x} 
  \prob{y}{\calG (x,y) \leq  \epsilon } \leq  (\beta \epsilon )^{b}$,
\end{itemize}
where in the second line $y$ is distributed according to the induced
  density $\mu (x,y)$, then
\[
  \prob{x,y}{\calF (x) \calG (x,y) \leq  \epsilon } \leq  
  c (a,b) \alpha \beta \epsilon^{\min (a,b)} 
  \lg (1/\epsilon)^{\ind{a = b}},
\]
where $\ind{a = b}$ is $1$ if $a = b$ and $0$ otherwise.
\end{lemma}

\begin{lemma}[Almost polynomial densities]\label{lem:minMaxIntegral}
Let $k > 0$ and let $t$ be a non-negative
  random variable 
  with density proportional to
  $\mu (t) t^{k}$ such that, for some
  $t_{0} > 0$,
\[
  \frac{
  \max _{0 \leq t \leq t_{0}} \mu  (t)
}{
  \min _{0 \leq t \leq t_{0}} \mu  (t)
} \leq c.
\]
Then,
\[
  \prob{}{t < \epsilon } < c (\epsilon / t_{0})^{k+1}.
\]
\end{lemma}
\begin{proof}
For $\epsilon \geq t_{0}$, the lemma is vacuously true.  
Assuming $\epsilon < t_{0}$,
\begin{eqnarray*}
  \prob{}{t < \epsilon }
& \leq &
\frac{
  \prob{}{t < \epsilon }
}{
  \prob{}{t < t_{0} }
}
\\
& = &
\frac{
  \int _{t = 0}^{\epsilon } \mu (t) t^{k} \diff{t}
}{
  \int _{t = 0}^{t_{0}}  \mu (t) t^{k} \diff{t}
}
\\
& \leq  &
\frac{ 
  \max _{0 \leq t \leq t_{0}}  \mu (t) \int _{t = 0}^{\epsilon } t^{k} \diff{t}
}{
  \min _{0 \leq t \leq t_{0}}  \mu (t) \int _{t = 0}^{t_{0}}  t^{k} \diff{t}
}
\\
& \leq  &
c
\frac{ 
 \epsilon ^{k+1} / (k+1)
}{
 t_{0} ^{k+1} / (k+1)
}\\
& = &
  c (\epsilon / t_{0})^{k+1}.
\end{eqnarray*}
\end{proof}

\subsection{Gaussian Random Vectors}

For the convenience of the reader, we recall some standard facts
  about Gaussian random variables and vectors. 
These may be found in \cite[VII.1]{Feller1} and \cite[III.6]{Feller2}.
We then draw some corollaries of these facts and derive some
  lemmas that we will need later in the paper.

We first recall that a univariate Gaussian distribution with mean 0
  and standard deviation $\sigma$ has density
\[
  \frac{1}{\sqrt{2 \pi} \sigma} e^{-a^{2}/ 2 \sigma^{2}},
\]
and that a Gaussian random vector in $\Reals{d}$ centered at 
  a point $\aao $
  with covariance matrix $\MM $ has density \index{covariance matrix@covariance matrix}\index{M@$\MM$}%
\[
  \frac{1}{\left(\sqrt{2 \pi } \right)^{d} \mathrm{det} (\MM )}
   e^{- (\aa - \aao) ^{T} \MM ^{-1} (\aa -\aao) / 2}.
\]
For positive-definite $\MM $,
  there exists a basis in
  which the density can be written
\[
  \prod_{i=1}^{d} 
  \frac{1}{\sqrt{2 \pi} \sigma_{i}} e^{-a_{i}^{2} / 2 \sigma_{i}^{2}},
\]
where $\sigma_{1}^{2} \leq \cdots \leq \sigma_{d}^{2}$ are the
  eigenvalues of $\MM $.
When all the eigenvalues of $\MM $ are the same and equal to
  $\sigma$, then we will refer to the density as a 
\index{Gaussian, distribution of standard deviation $\sigma$}%
  \textit{Gaussian distribution of standard deviation $\sigma$}.

\begin{proposition}[Additivity of Gaussians]\label{pro:gaussSum}
If $\aa_{1} $ is a Gaussian random vector
  with covariance matrix $\MM _{1}$ centered at a point
  $\aao_{1}$ and 
  $\aa_{2} $ is a Gaussian random vector 
  with covariance matrix $\MM _{2}$ centered at a point
  $\aao_{2}$, then
  $\aa _{1} + \aa _{2}$ is the Gaussian random vector
  with covariance matrix $\MM _{1} + \MM  _{2}$
  centered at $\aao _{1} + \aao _{2}$.
\end{proposition}

\begin{lemma}[Smoothness of Gaussians]\label{lem:smoothGauss}
Let $\mu (\xx)$ be a Gaussian distribution
 of standard deviation $\sigma$ centered at a point $\orig{\aa}$.
Let $k \geq 1$, let 
  $\dist{\xx}{\orig{\aa}} \leq k$
  and let $\dist{\xx }{\yy} < \epsilon \leq k$.
Then,
\[
  \frac{\mu (\yy)}{\mu (\xx )} 
  \geq 
   e^{-3 k \epsilon / 2 \sigma ^{2}}.
\]
\end{lemma}
\begin{proof}
By translating $\orig{\aa}$, $\xx$ and $\yy$, we may assume
  $\orig{\aa} = 0$ and $\norm{\xx} \leq k$.
We then have
\begin{align*}
  \frac{\mu (\yy)}{\mu (\xx )} 
& =
   e^{- (\norm{\yy}^{2} - \norm{\xx}^{2})/2 \sigma^{2}}\\
& \geq 
   e^{- (2 \epsilon \norm{\xx } + \epsilon^{2})/ 2 \sigma^{2}},
  & \text{as $\norm{\yy} \leq \norm{\xx} + \epsilon $}
\\
& \geq 
   e^{- (2 \epsilon k + \epsilon^{2})/ 2 \sigma^{2}},
  & \text{as $\norm{\xx} \leq k$}
\\
& \geq 
   e^{- 3 \epsilon k /2 \sigma^{2}}
  & \text{as $\epsilon  \leq k$}.
\end{align*}
\end{proof}

\begin{proposition}[Restrictions of Gaussians]\label{pro:gaussPlane}
Let $\mu$ be a Gaussian distribution
  of standard deviation $\sigma$
  centered at a point $\orig{\aa}$.
Let $\vv$ be any vector and $r$ any real.
Then, the induced distribution
\[
  \mu (\xx | \vv^{T} \xx = r)
\]
is a Gaussian distribution of standard deviation $\sigma$ centered at 
  the projection of $\orig{\aa}$ onto the plane 
  $\setof{\xx : \vv^{T} \xx = r}$.
\end{proposition}

\begin{proposition}[Gaussian measure of halfspaces]\label{pro:gauss1d}
Let $\oomega$ be any unit vector in $\Reals{d}$ and $r$ any real.
Then,
\[
\left(\frac{1}{\sqrt{2 \pi} \sigma } \right)^{d}
\int _{\gg}
   \ind{\form{\oomega }{\gg }
              \leq r}
   e^{-\norm{\gg}^{2}/ 2 \sigma^{2}}
   \diff{\gg}
 = 
\frac{1}{\sqrt{2 \pi} \sigma }
 \int_{t = -\infty }
     ^{t = r}
   e^{-t^{2}/ 2 \sigma^{2}}
   \diff{t}
\]
\end{proposition}
\begin{proof}
Immediate if one expresses the Gaussian density in a basis
  containing $\oomega $.
\end{proof}

The distribution of the square of the norm of a Gaussian random vector
  is the Chi-Square distribution.  
We use the following weak bound on the Chi-Square distribution,
  which follows from Equality (26.4.8) of~\cite{AbramowitzStegun}.

\begin{proposition}[Chi-Square bound]\label{pro:chiSquare}
Let $\xx$ be a Gaussian random vector in $\Reals{d}$
  of standard deviation $\sigma$
  centered at the origin.
Then,
\begin{equation}\label{eqn:chiSquare}
  \prob{}{\norm{\xx} \geq k \sigma }
  \leq 
     \frac{\left(k^{2} \right)^{d/2 - 1} e^{-k^{2}/2}}
          {2^{d/2 - 1} \Gamma (\frac{d}{2})}.
\end{equation}
\end{proposition}

From this, we derive
\begin{corollary}[A chi-square bound]\label{cor:chiSquare}
Let $\xx$ be a Gaussian random vector in $\Reals{d}$ 
  of standard deviation $\sigma$
  centered at the origin.
Then, for $n \geq 3$
\[
  \prob{}{\norm{\xx} \geq 3 \sqrt{d \ln n} \sigma }
  \leq 
     n^{-2.9 d}.
\]
Moreover, if $n > d \geq 3$, and
  $\vs{\xx}{1}{n}$ are such vectors, then
\[
 \prob{}{\max _{i}\norm{\xx_{i}} \geq 3 \sqrt{d \ln n} \sigma }
  \leq 
     n^{-2.9 d + 1} 
  \leq 0.0015 \binom{n}{d}^{-1}.
\]
\end{corollary}
\begin{proof}
For $\alpha =3\sqrt{\ln n} \sigma $
   we can apply Stirling's formula~\cite{AbramowitzStegun} to
  \eqref{eqn:chiSquare} to find
\begin{align*}
  \prob{}{\norm{\xx} \geq \alpha \sqrt{d}}
 &
 \leq 
     \frac{(\alpha^{2} d)^{d/2 - 1} e^{-\alpha^{2} d /2} e^{d/2} \sqrt{d/2}}
          {2^{d/2 - 1} (d/2)^{d/2} \sqrt{2 \pi }}\\
 & = 
   \left( \alpha^{2} \right)^{d/2 - 1}  e^{- (\alpha^{2} - 1) d / 2}
     \frac{d^{d/2 - 1} \sqrt{d}}
          {2^{d/2 - 1} (d/2)^{d/2} 2 \sqrt{\pi }}\\
 & = 
   \left( \alpha^{2} \right)^{d/2 - 1}  e^{- (\alpha^{2} - 1) d / 2}
     \frac{1}
          {\sqrt{d \pi }}\\
 & \leq 
  \left(\alpha^{2} \right)^{d/2}
  e^{- (\alpha^{2} - 1) d / 2}\\
 & =
   e^{- (\alpha^{2} - \ln (\alpha^{2}) - 1) d / 2}\\
 & \leq 
   e^{- 2.9 d \ln n}\\
 & = 
   n^{-2.9 d},
\end{align*}
as
\[
  (\alpha^{2} - \ln (\alpha^{2}) - 1)
 =
  9 \ln (n) - \ln (9 \ln n) - 1
\geq 
  \ln (n) (9 - \ln 9 - 1)
\geq
  5.8 \ln (n).
\]
\end{proof}

We also prove it is unlikely that a Gaussian random variable
  has small norm.

\begin{proposition}[Gaussian near point or plane]\label{pro:gaussianNear}
Let $\xx$ be a $d$-dimensional Gaussian random vector of
  standard deviation $\sigma$ centered anywhere.
Then,
\begin{enumerate}
\item  For any point $\pp$, $  \prob{}{\dist{\xx}{\pp} \leq \epsilon} \leq 
   \left(\min \left(1,\sqrt{e/d}\right) (\epsilon / \sigma )\right)^{d}$, and

\item  For a plane $H$ of dimension $h$,
$\prob{}{\dist{\xx}{H} \leq \epsilon} \leq 
   (\epsilon / \sigma )^{d-h}$.
\end{enumerate}
\end{proposition}
\begin{proof}
Let $\orig{\xx}$ be the center of the Gaussian distribution,
  and let $B_{\epsilon} (\pp)$ denote the ball of radius $\epsilon$
  around $\pp$.
Recall that the volume of $B_{\epsilon } (\pp)$ is
\[
   \frac{2 \pi ^{d/2} \epsilon ^{d}}{d \Gamma (d/2)}.
\]
To prove part $(a)$, we bound the probability that 
  $\dist{\xx }{\pp} \leq \epsilon $
  by
\begin{eqnarray*}
  \left(\frac{1}{\sqrt{2 \pi} \sigma } \right)^{d}
  \int _{\xx \in B_{\epsilon } (\pp)}
    e^{-\norm{(\xx - \orig{\xx}) }^{2} / 2 \sigma^{2}} \diff{\xx}
 \leq 
  \left(\frac{1}{\sqrt{2 \pi} \sigma } \right)^{d}
  \left(\frac{2 \pi ^{d/2} \epsilon ^{d}}{d \Gamma (d/2)} \right) 
 = 
  \left(\frac{\epsilon}{\sigma} \right)^{d}
  \frac{
  2 
}{
  d 2^{d/2} \Gamma (d/2)
   }.
%\\
%& \leq & 
%  (\epsilon / \sigma ) ^{d}.
\end{eqnarray*}

By Proposition \ref{pro:Gamma}, we have for $d\geq 3$
\begin{eqnarray*}
\frac{
  2 
}{
  d 2^{d/2} \Gamma (d/2)
   }
\leq  (e/d)^{d/2}.
\end{eqnarray*}
Combining with the fact that $2/ (d 2^{d/2} \Gamma (d/2)) \leq  1$ 
  for all $d\geq 1$, we establish (a).

To prove part $(b)$, we consider a basis
  in which $d-h$ vectors are perpendicular to $H$,
  and apply part $(a)$ to the components of $\xx$
  in the span of those basis vectors.
\end{proof}

\begin{proposition}[Gamma Inequality]\label{pro:Gamma} For $d\geq 3$
\begin{eqnarray*}
\frac{
  2 
}{
  d 2^{d/2} \Gamma (d/2)
   }
& \leq &
(e/d)^{d/2} 
\end{eqnarray*}
\end{proposition}
\begin{proof}
For $d \geq  3$, we apply the inequality
$\Gamma (x+1)\geq  \sqrt{2\pi}\sqrt{x} (x/e)^{x}$ to show
\begin{eqnarray*}
\frac{
  2 
}{
  d 2^{d/2} \Gamma (d/2)
   }
& \leq &
\frac{
  2
}{
  d 2^{d/2} \sqrt{2\pi} \sqrt{(d-2)/2}  }
\left(\frac{2e}{d-2} \right)^{(d-2)/2}\\
& = & 
\left( \frac{
   e^{(d-2)/2}
}{
  d^{d/2} \sqrt{2\pi} \sqrt{(d-2)/2} }\right)
\left(\frac{d}{d-2}\right)^{(d-2)/2}
\\ 
&\leq &  (e/d)^{d/2},
\end{eqnarray*}
where the last inequality used the inequalities
   $1+2/ (d-2) \leq e^{2/ (d-2)}$ and
  $\sqrt{2\pi }\sqrt{(d-1)/2} > 1$ when $d\geq 3$.
\end{proof}

\begin{proposition}[Non-central Gaussian near the origin]\label{pro:noncentralgaussianNear}
For any integer $d\geq 3$, let $\xx$ be a $d$-dimensional Gaussian random vector of
  standard deviation $\sigma$ centered at $\orig{\xx} $.
Then, for $\epsilon \leq 1/ (\sqrt{2}e)$
\[
  \prob{}{\norm{\xx} \leq
\left(\sqrt{\norm{\orig{\xx}}^{2} + d\sigma^{2}}\right)\epsilon} \leq 
   \left(\sqrt{2e} \epsilon \right)^{d}
\]
.
\end{proposition}
\begin{proof}
Let $\lambda  = \norm{\orig{\xx}}$.
We divide the analysis into two cases: (1) $\lambda \leq
\sqrt{d}\sigma $, and (2) $\lambda \geq \sqrt{d}\sigma $.

For $\lambda  \leq  \sqrt{d} \sigma$,
\begin{eqnarray*}
 \prob{}{\norm{\xx} \leq(\sqrt{\lambda^{2} + d\sigma^{2}})\epsilon}  & \leq &  
 \prob{}{\norm{\xx} \leq(\sqrt{2d}\sigma )\epsilon} \leq (\sqrt{2e}\epsilon)^{d},
\end{eqnarray*}
by Part $(a)$ of Lemma \ref{pro:gaussianNear}.

For $\lambda  >  \sqrt{d} \sigma$,
  let $B_{r}$ be the ball of radius $r$ around the origin.
Applying the assumption  $\epsilon \leq 1/(\sqrt{2}e) $
     and letting $\lambda= c \sqrt{d}\sigma $ for $c\geq  1$, we have
\begin{eqnarray*}
 \prob{}{\norm{\xx} \leq(\sqrt{\lambda^{2} + d\sigma^{2}})\epsilon}  & \leq &  
 \prob{}{\norm{\xx} \leq(\sqrt{2}\lambda )\epsilon}\\
  & = &
  \left(\frac{1}{\sqrt{2 \pi}\sigma } \right)^{d}
  \int _{\xx \in B_{\sqrt{2}\epsilon\lambda}}
    e^{-\norm{(\xx - \orig{\xx}) }^{2} / 2\sigma^{2}} \diff{\xx}\\
& \leq &
  \left(\frac{1}{\sqrt{2 \pi} \sigma } \right)^{d}
  \left(\frac{2 \pi ^{d/2}}{d \Gamma (d/2)} \right)
(\sqrt{2}\epsilon\lambda ) ^{d} e^{- (1-1/e)^{2}\lambda^{2}/2\sigma^{2}}
 \\
& \leq  &
(\sqrt{2e}\epsilon)^{d}
\frac{\lambda^{d}}{d^{d/2}\sigma^{d}}e^{-
(1-1/e)^{2}\lambda^{2}/2\sigma^{2}} \\
& = &
(\sqrt{2e}\epsilon)^{d}
e^{d (\ln c - c^{2} (1-1/e)^{2}/2)}\\
 & \leq  & (\sqrt{2e}\epsilon)^{d},
\end{eqnarray*}
where the second inequality holds because $\epsilon \leq 1/
(\sqrt{2}e)$ and for any point 
  $\xx \in B_{\sqrt{2}\epsilon\lambda}$,
\[
e^{-\norm{(\xx - \orig{\xx}) }^{2} / 2\sigma^{2}} \leq e^{-
(1-\sqrt{2}\epsilon)^{2}\lambda^{2}/2\sigma^{2}} \leq 
e^{- (1-1/e)^{2}\lambda^{2}/2\sigma^{2}},
\]
the third inequality follows from Proposition \ref{pro:Gamma},
and the last inequality holds because one can prove for any $c\geq 1$, 
$\ln c - c^{2} (1-1/e)^{2}/2 < 0$.
\end{proof}

Bounds such as the following on the tails of Gaussian
  distributions are standard (see, for example~\cite[Section VII.1]{Feller1})
\begin{proposition}[Gaussian tail bound]\label{pro:ourFeller}
\[
   \left(\frac{\sigma }{x} \right)
   \frac{e^{-x^{2}/2 \sigma^{2}}}{\sqrt{2 \pi }}
  \geq 
 \frac{1}{\sqrt{2 \pi } \sigma } 
   \int _{t=x}^{\infty }  e^{-t^{2}/2 \sigma^{2}} \diff{t}
  \geq 
   \left(\frac{\sigma }{x} - \frac{\sigma ^{3}}{x^{3}} \right)
   \frac{e^{-x^{2}/2 \sigma^{2}}}{\sqrt{2 \pi }}.
\]
\end{proposition}

Using this, we prove:

\begin{lemma}[Comparing Gaussian tails]\label{lem:gaussRelTail}
Let $\sigma \leq 1$ and let
\[
\mu (t) = \frac{1}{\sqrt{2 \pi } \sigma } e^{-t^{2}/2\sigma^{2}}.
\]
Then, for $x \leq 2$ and $\abs{x-y} \leq  \epsilon $,
\begin{equation}\label{eqn:gaussRelTailOrig}
   \frac{\int _{t=y}^{\infty } \mu (t) \diff{t}}
        {\int _{t=x}^{\infty } \mu (t) \diff{t}}
  \geq 1 - \frac{8 \epsilon }{3 \sigma^{2}}.
\end{equation}
\end{lemma}
\begin{proof}
If $y < x$, the ratio is greater than 1 and the lemma is trivially
  true.
Assuming $y \geq x$, the ratio is minimized when
  $y = x + \epsilon $.
In this case, the lemma will follow from
\begin{equation}\label{eqn:gaussRelTail}
   \frac{\int _{t=x}^{x + \epsilon  } \mu (t) \diff{t}}
        {\int _{t=x}^{\infty } \mu (t) \diff{t}}
  \leq \frac{8 \epsilon }{3 \sigma^{2}}.
\end{equation}
It follows from part $(b)$ of Proposition \ref{pro:monotonicityGaussian}
  that the left-hand ratio in (\ref{eqn:gaussRelTail})
  is monotonically increasing in $x$, and
  therefore is maximized when 
  $x$ is maximized at $2$.
For $x = 2$, we apply Proposition~\ref{pro:ourFeller} to show
\[
  \frac{1}{\sqrt{2 \pi } \sigma }
  \int _{t=x}^{\infty } \mu (t) \diff{t}
  \geq 
   \left(\frac{\sigma }{2} - \frac{\sigma ^{3}}{8} \right)
   \frac{e^{-2/\sigma^{2}}}{\sqrt{2 \pi }}
  \geq 
   \frac{3\sigma e^{-2/\sigma^{2}}}{8 \sqrt{2 \pi }}.
\]
We then combine this bound with
\[
  \frac{1}{\sqrt{2 \pi } \sigma }
  \int _{t=x}^{x + \epsilon  } \mu (t) \diff{t}
  \leq 
    \frac{\epsilon e^{-2/\sigma^{2}}}{\sqrt{2 \pi } \sigma },
\]
to obtain
\begin{equation*}
   \frac{\int _{t=x}^{x + \epsilon  } \mu (t) \diff{t}}
        {\int _{t=x}^{\infty } \mu (t) \diff{t}}
 \leq 
  \left(    \frac{\epsilon e^{-2/\sigma^{2}}}{\sqrt{2 \pi } \sigma } \right)
  \left(    \frac{8 \sqrt{2 \pi }} {3\sigma e^{-2/\sigma^{2}}}\right)
 = 
  \frac{8 \epsilon }{3 \sigma^{2}}.
\end{equation*}
\end{proof}

\begin{proposition}[Monotonicity of Gaussian density]\label{pro:monotonicityGaussian}
Let
\[
\mu (t) = \frac{1}{\sqrt{2 \pi } \sigma } e^{-t^{2}/2\sigma^{2}}.
\]

\begin{enumerate}
\item For all $a>0$, $\mu (x)/\mu (x+a)$ is monotonically increasing
  in $x$;
\item The following ratio is monotonically increasing in $x$
\[
\frac{\mu (x)}{\int _{t=x}^{\infty } \mu (t) \diff{t}}
\]
\end{enumerate}
\end{proposition}
\begin{proof}
Part $(a)$ follows from 
\[
\frac{\mu (x)}{\mu (x+a)} = e^{(2ax + a^{2})/2\sigma^{2}},
\]
and that $e^{2ax}$ is monotonically increasing in $x$.

To prove part $(b)$ note that for all $a > 0$
\[
\frac{\int _{t=x}^{\infty } \mu (t) \diff{t}}{\mu (x)}
= 
\frac{\int _{t=0}^{\infty } \mu (x+t) \diff{t}}{\mu (x)}
\geq \frac{\int _{t=0}^{\infty } \mu (x+a + t) \diff{t}}{\mu (x+a)}
= \frac{\int _{t=x+a}^{\infty } \mu (t) \diff{t}}{\mu (x+a)},
\]
where the inequality follows from part $(a)$.
\end{proof}

\subsection{Changes of Variables}
The main proof technique used in Section~\ref{sec:shadow}
  is change of variables.
For the reader's convenience, we recall how
  a change of variables affects probability distributions.

\begin{proposition}[Change of variables]\label{pro:cov}
Let $\yy$ be a random variable 
  distributed according to density $\mu $.
If $\yy = \Phi (\xx)$, then $\xx$ has density
\[
  \mu (\Phi (\xx)) \jacob{\Phi (\xx)}{\xx}.
\]
\end{proposition}
Recall that $\jacob{\yy}{\xx}$ is the Jacobian of the change
  of variables.

We now introduce the fundamental change of variables used in this
  paper.
Let $\vs{\aa }{1}{d}$ be linearly independent points in $\Reals{d}$.
We will represent
  these points by specifying the plane passing through them and their
  positions on that plane.
Many studies of the convex hulls of random point sets have used
  this change of variables
  (for example, see~\cite{RenyiSulanke1,RenyiSulanke2,Efron,Miles}).
We specify the plane containing $\vs{\aa}{1}{d}$ by
  $\oomega $ and $r$, where \index{ome@$\oomega $}\index{r@$r$}%
  $\norm{\oomega} = 1$, $r \geq 0$ and
  $\form{\oomega }{\aa _{i}} = r$
  for all $i$.
We will not concern ourselves with the issue that
 $\oomega $ is ill-defined if the $\vs{\aa}{1}{d}$ are
  affinely dependent,
 as this is an event of probability zero.
To specify the positions of $\vs{\aa}{1}{d}$ on the plane
  specified by $(\oomega , r)$,
  we must choose a coordinate system for that plane.
To choose a canonical set of coordinates for 
  each $(d-1)$-dimensional hyperplane specified by $(\oomega, r)$,
  we first fix a reference unit vector in $\Reals{d}$, say $\qq $, and an
  arbitrary coordinatization of the subspace 
  orthogonal to $\qq $.
For any $\oomega \not = - \qq $, we let
\[
  \RR_{\oomega }
\] \index{R@$\RR_{\oomega} $}%
denote the linear transformation that rotates
  $\qq$ to $\oomega $ in the two-dimensional subspace through
  $\qq $ and $\oomega $ and that is the identity 
  in the orthogonal subspace.
Using $\RR_{\oomega }$, we can map points specified
  in the $d-1$ dimensional hyperplane specified by 
  $r$ and $\oomega $ to $\Reals{d}$ by
\[
  \aa _{i} = \RR_{\oomega } \bb _{i} + r \oomega,
\] \index{bb@$\bb _{i}$}%
where $\bb _{i}$ is viewed both as a vector in 
 $\Reals{d-1}$ and as an element of the subspace
  orthogonal to $\qq$.
We will not concern ourselves with the fact that 
  this map is not well defined if
  $\qq = - \oomega $,
  as the set of $\vs{\aa}{1}{d}$ that result in this coincidence
  has measure zero.

The Jacobian of this change of variables is computed by a famous theorem
  of integral geometry due to Blaschke~\cite{Blaschke} 
  (for more modern treatments, see~\cite{Miles} or~\cite[12.24]{Santalo}),
  and actually depends only marginally on the coordinatizations
  of the hyperplanes.

\begin{theorem}[Blaschke]\label{thm:blaschke}
For variables
  $\vs{\bb}{1}{d}$  taking values in $\Reals{d-1}$,
  $\oomega \in \sphere{d-1}$ and $r \in \Reals{}$,
 let 
\begin{align*}
 (\vs{\aa}{1}{d}) & 
  = \left(\RR_{\oomega } \bb_{1} + r \oomega , \ldots ,
     \RR_{\oomega } \bb_{d} + r \oomega  \right)
\end{align*}
The Jacobian of this map is
\[
\Jacob{\vs{\aa}{1}{d}}{\oomega , r, \vs{\bb}{1}{d}}
= 
(d-1)!\vol{\simp{\vs{\bb}{1}{d}}}.
\]
That is, 
\[
  \diff{\aa_{1}} \dotsc \diff{\aa_{d}} = 
(d-1)! \vol{\simp{\vs{\bb}{1}{d}}}
   \diff{\oomega } \diff{r} \diff{\bb_{1}} \dotsc  \diff{\bb_{d}}
\]
\end{theorem}

We will also find it useful to specify the plane by
 $\oomega $ and $s$, \index{s@$s$} where 
  $\form{s \qq}{\oomega } = r$, so that
  $s \qq$ lies on the plane specified by $\oomega $
  and $r$.
We will also arrange our coordinate system so
  that the origin on this plane
  lies at $s \qq$.

\begin{corollary}[Blaschke with $s$]\label{cor:blaschke}
For variables 
  $\vs{\bb}{1}{d}$ taking values in $\Reals{d-1}$,
  $\oomega \in \sphere{d-1}$ and $s \in \Reals{}$,
 let 
\begin{align*}
 (\vs{\aa}{1}{d}) & 
  =
   \left(\RR_{\oomega } \bb_{1} + s \qq, \ldots ,
     \RR_{\oomega } \bb_{d} + s \qq \right)
\end{align*}
The Jacobian of this map is
\[
\Jacob{\vs{\aa}{1}{d}}{\oomega , s, \vs{\bb}{1}{d}}
 = (d-1)!\form{\oomega }{\qq} \vol{\simp{\vs{\bb}{1}{d}}}.
\]
\end{corollary}
\begin{proof}
So that we can apply Theorem~\ref{thm:blaschke}, we will decompose
  the map into three simpler maps:
\begin{align*}
(\vs{\bb}{1}{d}, s, \oomega)
& \mapsto
\left(\bb _{1} + \RR_{\oomega}^{-1} (s \qq - r \oomega), \dots ,
  \bb _{d} + \RR_{\oomega}^{-1} (s \qq - r \oomega),
s, \oomega \right)\\
& \mapsto
\left(\bb _{1} + \RR_{\oomega}^{-1} (s \qq - r \oomega), \dots ,
  \bb _{d} + \RR_{\oomega}^{-1} (s \qq - r \oomega),
r, \oomega \right)\\
& \mapsto
\left(R_{\oomega} \left(\bb _{1} + \RR_{\oomega}^{-1} (s \qq - r \oomega) \right) + r \oomega, \dots ,
R_{\oomega} \left(\bb _{d} + \RR_{\oomega}^{-1} (s \qq - r \oomega) \right) + r \oomega \right)\\
& =
\left(R_{\oomega} \bb _{1} + s \qq, \dots ,
     R_{\oomega} \bb _{d} + s \qq \right)
\end{align*}
As $s \qq - r \oomega$ is orthogonal to $\oomega$,
  $\RR_{\oomega}^{-1}(s \qq - r \oomega)$ can be interpreted as a vector   
  in the $d-1$ dimensional space in which $\vs{\bb}{1}{d}$
  lie.
So, the first map is just a translation, and its Jacobian
  is $1$.
The Jacobian of the second map is
\[
\jacobn{r}{s} =  \form{\qq}{\oomega}.
\]
Finally, we note 
\[
  \vol{\bb _{1} + \RR_{\oomega}^{-1} (s \qq - r \oomega), \dots ,
  \bb _{d} + \RR_{\oomega}^{-1} (s \qq - r \oomega)}
 = \vol{\vs{\bb}{1}{d}},
\]
 and that the third map is one described in 
   Theorem~\ref{thm:blaschke}.
\end{proof}

In Section~\ref{sec:angle}, we will need to
  represent $\oomega $ by
  $c = \form{\oomega }{\qq}$ and $\ppsi \in \sphere{d-2}$,   
  where $\ppsi $ gives the location of $\oomega $ in the
  cross-section of $\sphere{d-1}$ for which 
  $\form{\oomega }{\qq} = c$. \index{c@$c$}\index{c@$\ppsi $}%
Formally, the map can be defined in a coordinate system
  with first coordinate $\qq$ by
\[
  \oomega = (c, \ppsi \sqrt{1-c^{2}}).
\]
For this change of variables, we have:

\begin{proposition}[Latitude and longitude]\label{pro:covC}
The Jacobian of the change of variables from 
  $\oomega $ to $(c, \ppsi) $ is
\[
\Jacob{\oomega }{c , \ppsi} = 
   (1-c^{2})^{(d-3)/2}.
\]
\end{proposition}
\begin{proof}
We begin by changing
  $\oomega$ to $(\theta , \ppsi)$, where $\theta$ is the angle
  between $\oomega$ and $\qq$, and
 $\ppsi$ represents the position of $\oomega$ in the $d-2$ dimensional
  sphere of radius $\sin (\theta)$ of points at angle $\theta$ to
  $\qq$.
To compute the Jacobian of this change of variables, we choose a local
  coordinate system on $\sphere{d-1}$ at $\oomega$ by taking
  the great circle through $\oomega$ and $\qq$, and then
  an arbitrary coordinatization of the great $d-2$ dimensional sphere
  through $\oomega$ orthogonal to the great circle.
In this coordinate system, $\theta$ is the position of $\oomega$ along
  the first great circle.
As the $d-2$ dimensional sphere 
  of points at angle $\theta$ to $\qq$
  is orthogonal to the great
  circle at $\oomega$, the coordinates in $\ppsi$ can be mapped
  orthogonally into the coordinates of the great $d-2$ dimensional
  sphere--the only difference being the radii of the sub-spheres.
Thus, 
\[
\Jacob{\oomega }{\theta , \ppsi}
  = 
  \sin (\theta)^{d-2}
\]
If we now let $c = \cos (\theta)$, then we find
\[
\Jacob{\oomega }{c, \ppsi }
= 
\Jacob{\oomega }{\theta , \ppsi }
\Jacob{\theta  }{c }
 = 
  \left(\sqrt{1-c^{2}} \right)^{d-2} \frac{1}{\sqrt{1-c^{2}}}
 = 
  \left(\sqrt{1-c^{2}} \right)^{d-3}.
\]

\end{proof}

% Local Variables: ***
% TeX-master:"shadow.tex" ***
% End: ***

%\newpage
\section{The Shadow Vertex Method}\label{sec:introSVM}

In this section, we will review the shadow vertex method
  and formally state the two-phase method analyzed in this paper.
We will begin by motivating the method.
In Section~\ref{sec:introSVMphase2}, we will explain how the method works assuming
  a feasible vertex is known.
In Section~\ref{sec:introSVMpolar}, we present a polar perspective on the method, from
  which our analysis is most natural.
We then present a complete two-phase method in Section~\ref{sec:introSVM2phase}.
For a more complete exposition of the Shadow Vertex Method, we refer the
  reader to~\cite[Chapter 1]{Borg82}.

The shadow-vertex simplex method is motivated by the observation that
  the simplex method is very simple in two-dimensions:
  the set of feasible points form a (possibly open) polygon, and
  the simplex method merely walks along the exterior of the polygon.
The shadow-vertex method lifts the simplicity of the simplex
  method in two dimensions to higher dimensions.
Let $\zz$ be the objective function of a linear program
  and let $\tt$ be an objective function optimized by $\xx$, a vertex
  of the polytope of feasible points for the linear program.
The shadow-vertex method considers the \emph{shadow} of the
  polytope---the projection of the
  polytope onto the plane spanned by $\zz$ and $\tt$.
One can verify that 
\begin{itemize}
\item [(1)] this shadow is a (possibly open) polygon,
\item [(2)] each vertex of the polygon is the image of a vertex
  of the polytope,
\item [(3)] each edge of the polygon is the image of an edge between
  two adjacent vertices of the polytope,
\item [(4)] the projection of $\xx $ onto the plane is a vertex
  of the polygon, and
\item [(5)] the projection of the vertex optimizing $\zz$ onto the plane
  is a vertex of the polygon.
\end{itemize}
Thus, if one 
  walks along the vertices of the polygon starting
  from the image of $\xx$, and keeps
  track of the vertices' pre-images on the polytope, then one will eventually
  encounter the vertex of the polytope optimizing $\zz$.
Given one vertex of the polytope that maps to a vertex of the polygon,
  it is easy to find the vertex of the polytope that maps to the
  next vertex of the polygon: fact $(3)$ implies that it must be
  a neighbor of the vertex on the polytope;
 moreover, for a linear program that is 
  in general position with respect to $\tt$, there will be $d$ such vertices.
Thus, the method will be efficient provided that the shadow polygon does
  not have too many vertices.
This is the motivation for the shadow vertex method.

\begin{figure}[h]
\centering\epsfig{figure=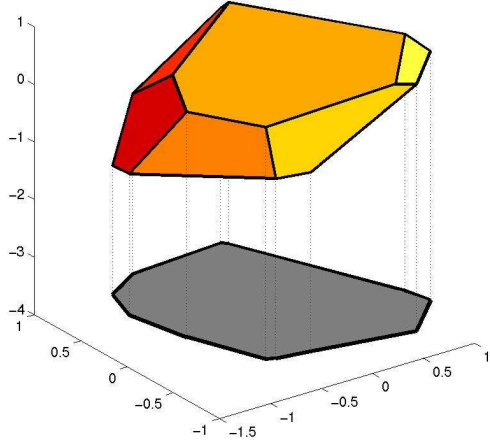,height=3in}
\caption{A shadow of a polytope}
\end{figure}

\subsection{Formal Description}\label{sec:introSVMphase2}

Our description of the shadow vertex simplex method will be
  facilitated by the following definition:
\begin{definition}[optVert]\index{optv@$\optVert{\zz}{\vs{\aa}{1}{n}; \yy }$}%
Given vectors $\zz$, $\vs{\aa}{1}{n}$ in $\Reals{d}$ and
  $\yy \in \Reals{n}$, we define
$\optVert{\zz}{\vs{\aa}{1}{n}; \yy }$ to be the set of
  $\xx$ solving
\begin{align*}
  \mbox{maximize} \qquad  & \zz^{T} \xx \nonumber  \\
  \mbox{subject to} \qquad & \form{\aa _{i}}{\xx} \leq y _{i}, 
  \mbox{ for $1 \leq i \leq n$}.
\end{align*}
If there are no such $\xx$, either because the program is
  unbounded or infeasible, we let $\optVert{\zz}{\vs{\aa}{1}{n}; \yy }$
  be $\emptyset$.
When $\vs{\aa}{1}{n}$ and $\yy$ are understood, we will 
  use the notation $\optvert{\zz }$.
\end{definition}
We note that, for linear programs in general position,
  $\optvert{\zz}$ will either be empty or contain one vertex.

Using this definition, we will give a description
  of the shadow vertex method assuming that
  a vertex $\xx_{0}$ and a vector $\tt$ are known
  for which $\optvert{\tt} = \xx_{0}$.
An algorithm that works without this assumption
  will be described in Section~\ref{sec:introSVM2phase}.
Given $\tt$ and $\zz$, we define objective functions
  interpolating between the two by
\[ 
  \qq_{\lambda} = (1-\lambda) \tt + \lambda \zz.
\] \index{qq@$\qq_{\lambda}$}\index{lam@$\lambda $}%
The shadow-vertex method will proceed by varying
  $\lambda$ from $0$ to $1$, and tracking
  $\optvert{\qq_{\lambda}}$.
We will denote the vertices encountered by
  $\xx_{0}, \xx_{1}, \dots, \xx_{k}$, and we will
  set
  $\lambda_{i}$ so that
  $\xx_{i} \in  \optvert{\qq_{\lambda}}$ for
  $\lambda \in [\lambda_{i}, \lambda_{i+1}]$.

As our main motivation for presenting the primal algorithm
  is to develop intuition in the reader, we will not dwell
  on issues of degeneracy in its description.
We will present a polar version of this algorithm with a proof
  of correctness in the next section.

\vskip .2in
\fbox{
\begin{minipage}{5in}
\noindent \textbf{primal shadow-vertex method}\\
Input: $\vs{\aa}{1}{n}$, $\yy$, $\zz$, and
  $\xx_{0}$ and $\tt$ satisfying
  $\setof{\xx_{0} } =  \optVert{\tt}{\vs{\aa}{1}{n}; \yy}$.
\begin{enumerate}
\item [(1)] Set  $\lambda_{0} = 0$, and $i = 0$.
\item [(2)] Set $\lambda_{1}$ to be maximal such that
  $\setof{\xx_{0}} = \optvert{\qq_{\lambda}}$ for
  $\lambda \in [\lambda_{0}, \lambda_{1}]$.
\item [(3)] while $\lambda_{i+1} < 1$,
\begin{enumerate}
\item [(a)] Set $i = i + 1$.
\item [(b)] Find an $\xx_{i}$ for which there exists a
  $\lambda_{i+1} > \lambda_{i}$ 
  such that
  $\xx_{i} \in  \optvert{\qq_{\lambda}}$ for
  $\lambda \in [\lambda_{i}, \lambda_{i+1}]$.
  If no such $\xx_{i}$ exists, return \textit{unbounded}.
\item [(c)] Let $\lambda_{i+1}$ be maximal such that
  $\xx_{i} \in  \optvert{\qq_{\lambda}}$ for
  $\lambda \in [\lambda_{i}, \lambda_{i+1}]$.
\end{enumerate}

\item [(4)] return $\xx_{i}$.

\end{enumerate}
\end{minipage}
}\index{primal shadow-vertex method@primal shadow-vertex method}\index{xx@$\xx _{i}$}%
\vskip .2in

Step $(b)$ of this algorithm deserves further explanation.
Assuming that the linear program is in general position
  with respect to $\tt $,
  each vertex $\xx_{i}$ will have exactly $d$ neighbors,
  and the vertex $\xx_{i+1}$ will be one of these~\cite[Lemma 1.3]{Borg82}.
Thus, the algorithm can be described as a simplex method.
While one could implement the method by examining these $d$ vertices
  in turn, more efficient implementations are possible.
For an efficient implementation of this algorithm in tableau form,
  we point the reader to the exposition 
  in~\cite[Section 1.3]{Borg82}.

\subsection{Polar Description}\label{sec:introSVMpolar}
Following Borgwardt~\cite{Borg82}, we will analyze the shadow
  vertex method from a polar perspective.
This polar perspective is natural provided that all
  $y_{i} > 0$.
In this section, we will describe a polar variant of the
  shadow-vertex method that works under this assumption.
In the next section, we will describe a two-phase shadow vertex method
  that uses this polar variant to solve linear programs with arbitrary
  $y_{i}$s.

While it is not strictly necessary for the results in this paper,
  we remind the reader that for a polytope 
  $P = \setof{\xx : \form{\xx}{\aa_{i}} \leq 1, \forall i}$,
 the polar of $P$ is $\setof{\yy : \form{\xx}{\yy} \leq 1, \forall \xx \in P}$.
An equivalent definition of the polar is $\convhull{\zzero , \vs{\aa}{1}{n}}$.
We remark that $P$ is bounded if and only if $\zzero$
  is in the interior of
  $\convhull{\vs{\aa}{1}{n}}$.
The polar motivates:

\begin{figure}[h]
 \noindent 
 \begin{minipage}{.3\linewidth}
 \centering\epsfig{figure=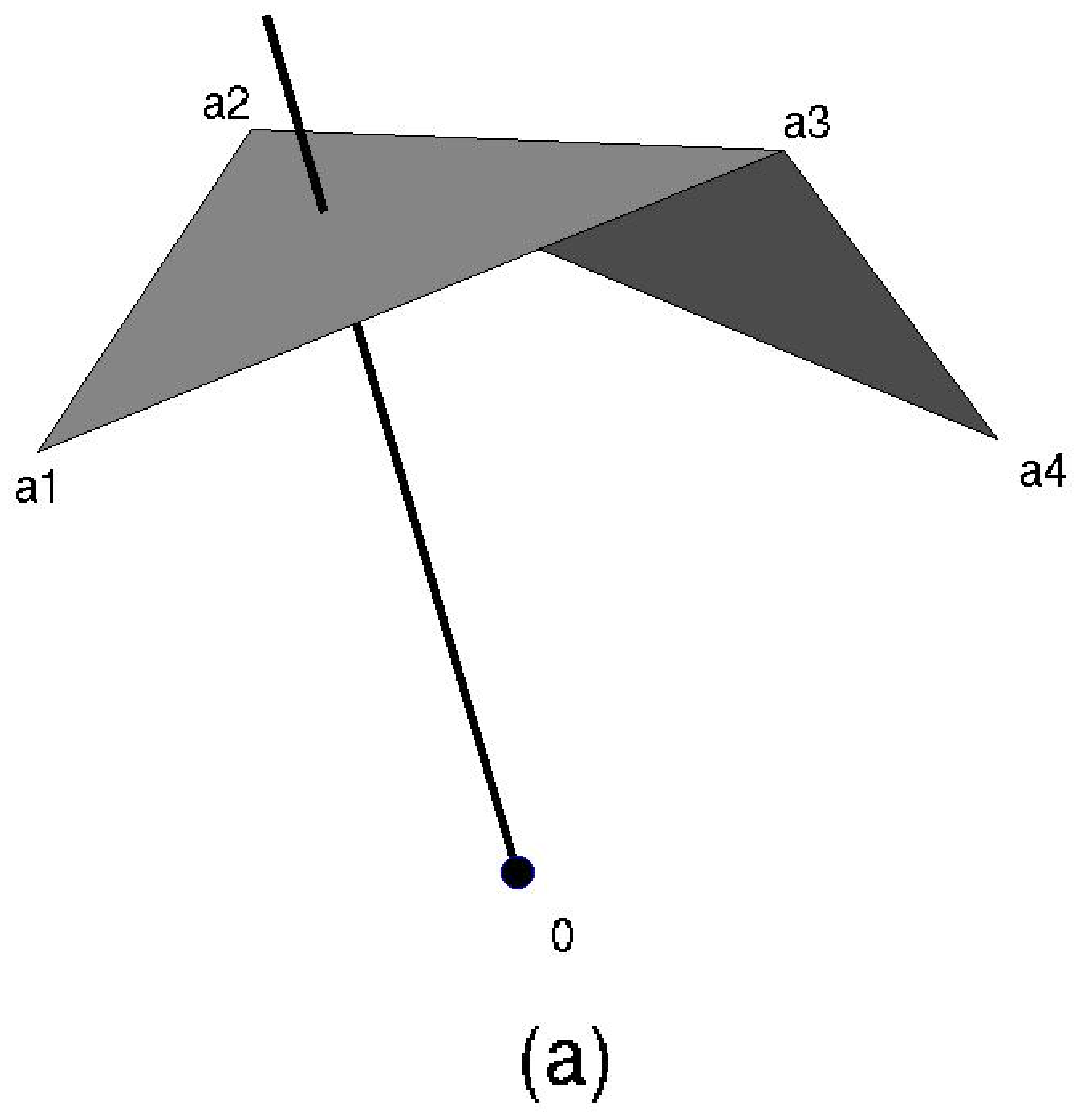,height=1.5in}
 \end{minipage}\hfill 
 \begin{minipage}{.3\linewidth}
 \centering\epsfig{figure=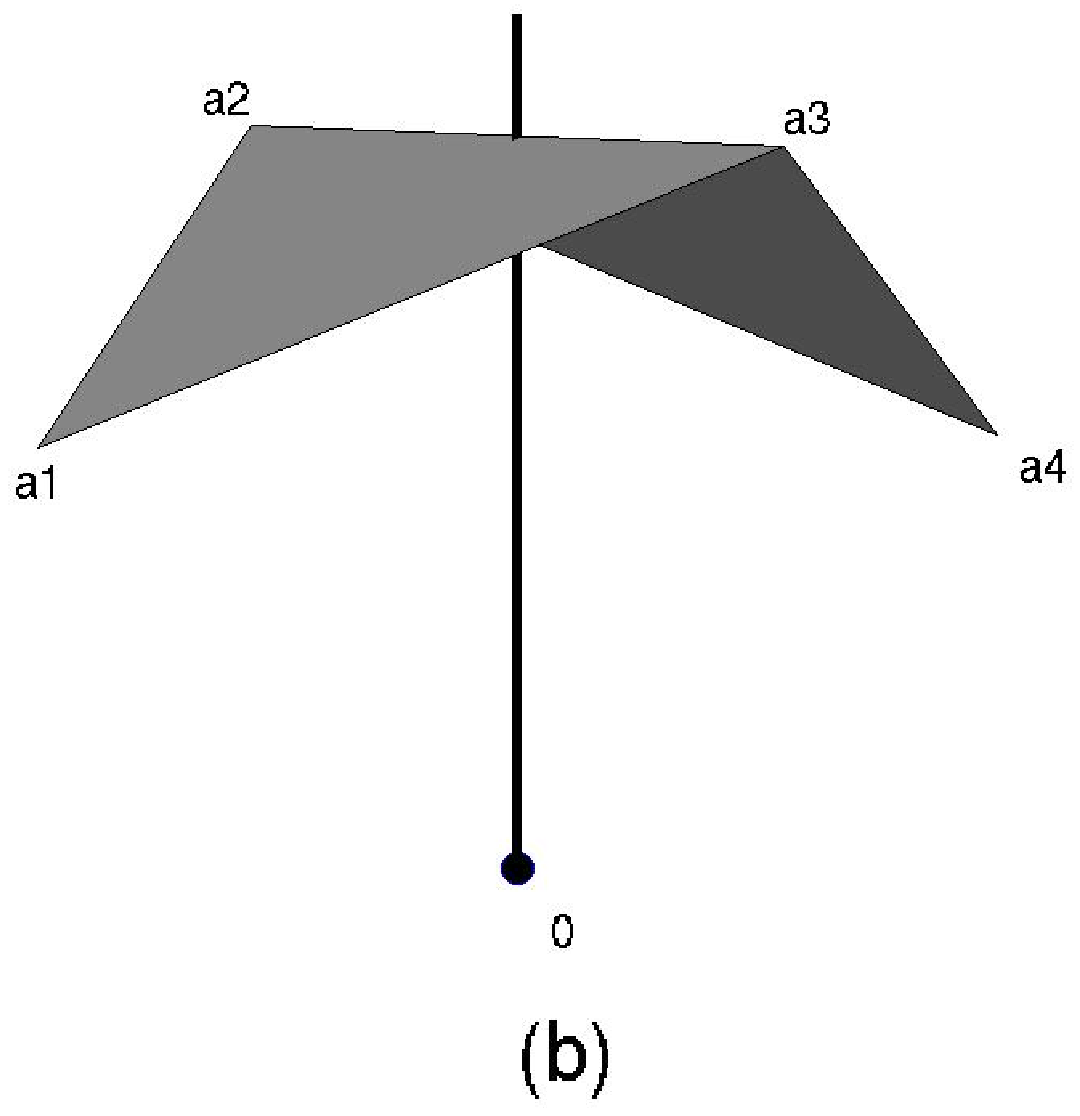,height=1.5in}
 \end{minipage}\hfill
 \begin{minipage}{.3\linewidth}
 \centering\epsfig{figure=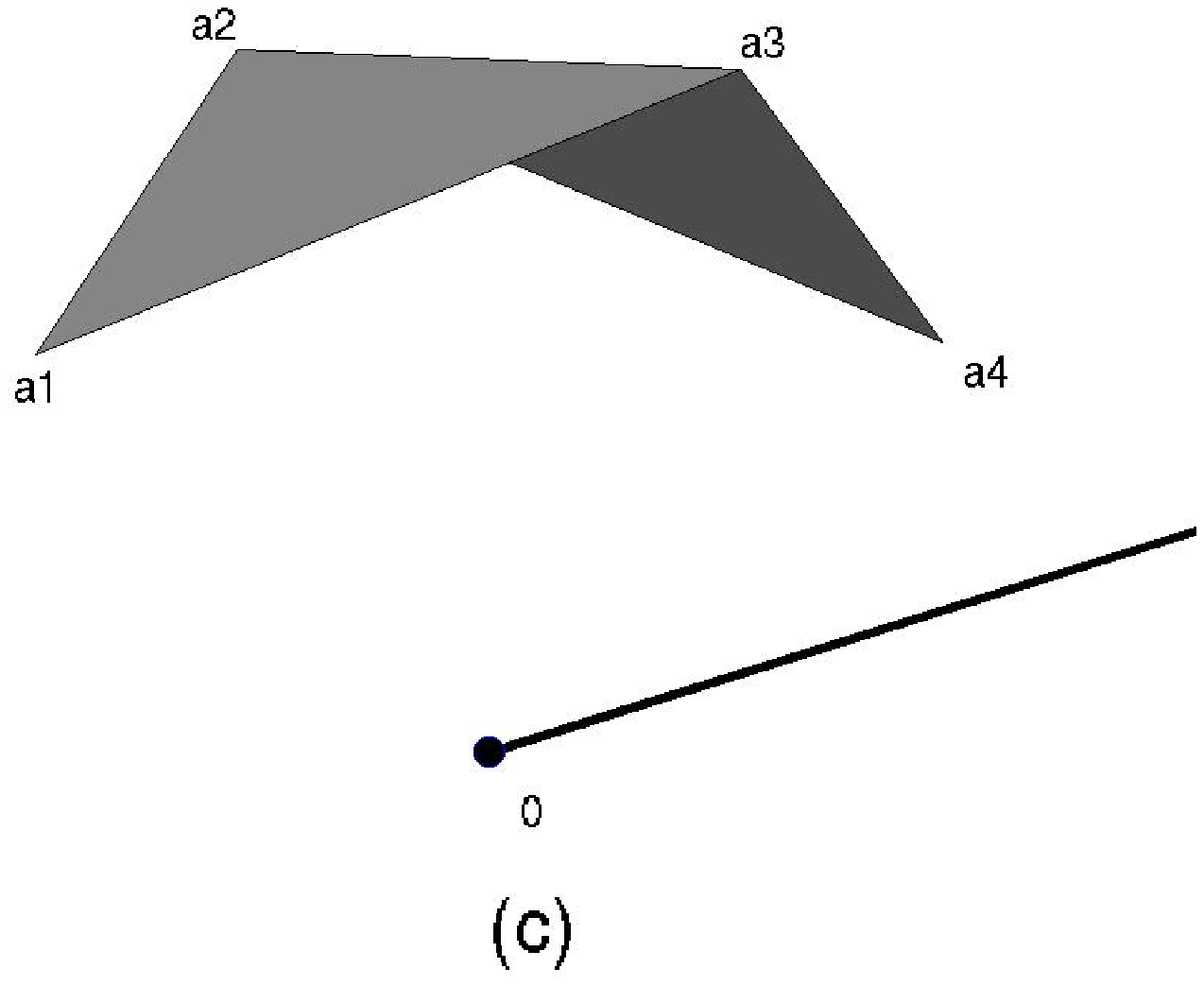,height=1.5in}
 \end{minipage}
\vspace{.1in}
\caption{In example (a), 
   $\mathrm{optSimp} = \setof{\setof{\aa_{1}, \aa_{2}, \aa_{3}}}$.
  In example (b),
   $\mathrm{optSimp} = \setof{\setof{\aa_{1}, \aa_{2}, \aa_{3}},
   \setof{\aa_{2}, \aa_{3}, \aa_{4}}}$.
  In example (c), 
   $\mathrm{optSimp} = \emptyset$,
}
\end{figure}

\begin{definition}[optSimp]\label{def:optSimp}\index{opts@$\optSimp{\zz}{\vs{\aa}{1}{n}}$}%
For $\zz$ and $\vs{\aa }{1}{n}$ in $\Reals{d}$
  and $\yy \in \Reals{n}$, $y_{i} > 0$,
  we let $\optSimp{\zz}{\vs{\aa}{1}{n} ; \yy}$
  denote the set of $I \in \binom{[n]}{d}$
 such that 
  $\AA _{I}$ has full rank,
   $\simp{(\aa_{i}/y_{i})_{i \in I}}$ is a facet
  of $\convhull{\zzero , \vsDiv{\aa}{y}{1}{n}}$
  and $\zz \in \cone{(\aa_{i})_{i \in I}}$.
When $\yy$ is understood to be $\oone$,
  we will use the notation
  $\optSimp{\zz}{\vs{\aa}{1}{n}}$
When $\vs{\aa}{1}{n}$ and $\yy$ are understood, we will 
  use the notation $\optsimp{\zz}$.
\end{definition}

We remark that for $\yy$, $\zz$ and $\vs{\aa}{1}{n}$
  in general position,
  $\optSimp{\zz}{\vs{\aa}{1}{n} ; \yy}$ will be
  the empty set or contain just one set of
  indices $I$.

The following proposition follows from the duality theory
  of linear programming:
\begin{proposition}[Duality]\label{pro:polar}
For $\vs{y}{1}{n} > 0$, 
  $I \in 
   \optSimp{\zz}{\vsDiv{\aa}{y}{1}{n}}$
  if and only if
  there exists an $\xx$ such that
  $\xx \in \optVert{\zz}{\vs{\aa}{1}{n} ; \yy}$
  and
  $\form{\xx}{\aa_{i}} = y_{i}$, 
  for $i \in I$.
\end{proposition}

We now state the polar shadow vertex method.

\vskip .2in
\fbox{

\begin{minipage}{5in}
\noindent \textbf{polar shadow-vertex method}\\
Input: 
  \begin{itemize}
\item $\vs{\aa}{1}{n}$,  $\zz$, and $\vs{y}{1}{n} > 0$,
\item  $I \in \binom{[n]}{d}$ and $\tt$ satisfying
  $I \in  \optSimp{\tt}{\vsDiv{\aa}{y}{1}{n}}$.
\end{itemize}
\begin{enumerate}
\item [(1)] Set $\lambda_{0} = 0$ and $i = 0$.
\item [(2)] Set $\lambda_{1}$ to be maximal such that for
  $\lambda \in [\lambda_{0}, \lambda_{1}]$,
\[
  I \in  \optSimp{\qq_{\lambda}}{\vsDiv{\aa}{y}{1}{n}}.
\]
\item [(3)] while $\lambda_{i+1} < 1$,
\begin{enumerate}
\item [(a)] Set $i = i + 1$.
\item [(b)] Find a $j$ and $k$ for which there exists a
  $\lambda_{i+1} > \lambda_{i}$ 
  such that
\[
I \cup \setof{j} - \setof{k}
 \in  \optSimp{\qq_{\lambda }}{\vsDiv{\aa}{y}{1}{n}}
\]
for
  $\lambda \in [\lambda_{i}, \lambda_{i+1}]$.
  If no such $j$ and $k$ exist, return \textit{unbounded}.

\item [(c)] Set $I = I \cup \setof{j} - \setof{k}$.

\item [(d)] Let $\lambda_{i+1}$ be maximal such that
  $I \in  \optSimp{\tt}{\vsDiv{\aa}{y}{1}{n}}$ for
  $\lambda \in [\lambda_{i}, \lambda_{i+1}]$.
\end{enumerate}

\item [(4)] return $I$.

\end{enumerate}
\end{minipage}
}\index{polar shadow-vertex method@polar shadow-vertex method}
\vskip .2in

The $\xx$ optimizing the linear program,
  namely $\optVert{\zz }{\vs{\aa}{1}{n}; \yy}$,
  is given by the equations $\form{\xx}{\aa_{i}} = y_{i}$,
  for  $i \in I$.

Borgwardt~\cite[Lemma 1.9]{Borg82} establishes that
  such $j$ and $k$ can be found in step $(b)$ if 
  there exists an $\epsilon$ for which
  $\optSimp{\qq_{\lambda_{i} + \epsilon}}{\vsDiv{\aa}{y}{1}{n}} \not = \emptyset$.
That the algorithm may conclude that the program is unbounded
  if a $j$ and $k$ cannot be found in step $(b)$ follows from:

\begin{proposition}[Detecting unbounded programs]\label{pro:introSVMunbounded}
If there is an $i$ and an $\epsilon > 0$ such that $\lambda_{i} + \epsilon < 1$ and
  $\optSimp{\qq_{\lambda_{i} + \epsilon}}{\vsDiv{\aa}{y}{1}{n}} = \emptyset$, then
  $\optSimp{\zz}{\vsDiv{\aa}{y}{1}{n}} = \emptyset$.
\end{proposition}
\begin{proof}
$\optSimp{\qq_{\lambda_{i} + \epsilon}}{\vsDiv{\aa}{y}{1}{n}} = \emptyset$
 if and only if
$\qq_{\lambda_{i} + \epsilon } \not \in  \cone{\vs{\aa}{1}{n}}$.
The proof now follows from the facts that 
  $\cone{\vs{\aa}{1}{n}}$ is a convex set and
  $\qq_{\lambda_{i} + \epsilon }$ 
  is a positive multiple of a
   convex combination of $\tt$ and $\zz$.
\end{proof}

The running time of the shadow-vertex method is bounded by the
  number of vertices in shadow of the polytope defined by
  the constraints of the linear program.
Formally, this is
\begin{definition}[Shadow] \index{shad@$\shadow{\tt, \zz}{\vs{\aa}{1}{n}; \yy}$}%
For independent vectors $\tt$ and $\zz$,
  $\vs{\aa}{1}{n}$ in $\Reals{d}$ and $\yy \in \Reals{n}$, $\yy > 0$,
\[
  \shadow{\tt, \zz}{\vs{\aa}{1}{n}; \yy}
 \defeq 
\union_{\qq \in \Span{\tt , \zz }} 
          \setof{\optSimp{\qq}{\vsDiv{\aa }{y}{1}{n}}}.
\]
If $\yy $ is understood to be $\oone$, we will just write
 $\shadow{\tt, \zz}{\vs{\aa}{1}{n}}$.
\end{definition}

\subsection{Two-Phase Method}\label{sec:introSVM2phase}

We now describe a two-phase shadow vertex method that solves 
  linear programs of form
\begin{align*}
  \mbox{maximize} \qquad  & \form{\zz}{\xx } \nonumber  \\
  \mbox{subject to} \qquad & \form{\aa _{i}}{\xx} \leq y _{i}, 
  \mbox{ for $1 \leq i \leq n$}.
 & (LP)
\end{align*}
%where $y_{i} \in [1,-1]$.
%To handle $y_{i}$ outside this range, one could either modify
%  the algorithm, or preprocess the input.
%We analyze the later approach in Corollary~\ref{cor:main}.

There are three issues that we must resolve before we can apply the
  polar shadow vertex method as described in Section~\ref{sec:introSVMpolar}
  to the solution of such programs:
\begin{enumerate}
\item [(1)] the method must know a feasible vertex of the linear program,
\item [(2)] the linear program might not even be feasible, and
\item [(3)] some $y_{i}$ might be non-positive.
\end{enumerate}
The first two issues are standard motivations for two-phase methods,
  while the third is motivated by the polar perspective from which we prefer
  to analyze the shadow vertex method.
We resolve these issues in two stages.
We first relax the constraints of $LP$ to construct a linear program
  $LP'$ such that
\begin{enumerate}
\item [$(a)$] the right-hand vector of the linear program is positive, and
\item [$(b)$] we know a feasible vertex of the linear program.
\end{enumerate}
After solving $LP'$, we construct another linear program, $LP^{+}$,
  in one higher dimension that interpolates between $LP$ and $LP'$.
$LP^{+}$ has properties $(a)$ and $(b)$, and we
  can use the shadow vertex method on $LP^{+}$ to transform the solution
  to $LP'$ into a solution of $LP$.

Our two-phase method first chooses
  a $d$-set $I$ to define the known feasible vertex
  of $LP'$.
The linear program $LP'$ is determined by $\AA$, $\zz$
  and the choice of $I$.
However, the magnitude of the right-hand entries in $LP'$
  depends upon $\smin{A_{I}}$.
To reduce the chance that these entries will need to be large,
  we examine several randomly chosen $d$-sets, and use the one
  maximizing $\mbox{\bf s}_{\textbf{min}}$.\index{I@$I$}%

The algorithm then sets
 
\begin{align*}
  M & = 2^{\ceiling{\lg \left(\max_{i} \norm{y_{i}, \aa_{i}} \right)} + 2}, \\
  \kappa & = 2^{\floor{\lg \left(\smin{\AA_{I}} \right)}}, \text{and}\\
  y_{i}' & =
  \begin{cases}
   M & \text{for $i \in I$}\\
   \sqrt{d}M^{2} / 4 \kappa  & \text{otherwise.}
  \end{cases}
\end{align*}\index{yp@$y'_{i}$}\index{M@$M$}\index{ka@$\kappa$}%
  
These define the program $LP'$:
\begin{align*}
  \mbox{maximize} \qquad  & \form{\zz}{\xx } \nonumber  \\
  \mbox{subject to} \qquad & \form{\aa _{i}}{\xx} \leq y'_{i}, 
  \mbox{ for $1 \leq i \leq n$}. 
 &  (LP')
\end{align*}

By Proposition~\ref{pro:initialVertex}, $\AA_{I}$ is a feasible basis for
  $LP'$, and optimizes any objective function of the form
  $\AA_{I} \aalpha$, for $\aalpha > 0$.
Our two-phase algorithm will solve $LP'$ by starting the polar
  shadow-vertex algorithm at the basis $I$ and the objective
  function $\AA_{I} \aalpha$ for a randomly chosen
  $\aalpha$ satisfying $\sum \alpha_{i} = 1$ and
  $\alpha_{i} \geq 1/d^{2}$, for all $i$.

\begin{proposition}[Initial simplex of $LP'$]\label{pro:initialVertex}
For any $\aalpha > 0$, $I = \optSimp{\AA_{I}\aalpha }{\vs{\aa}{1}{n}; \yyo }$.
\end{proposition}
\begin{proof}
Let $\xxo$ be the solution of the linear system 
\[
\form{\aa_{i}}{\xx'} = y'_{i}, \qquad \mbox{for}\qquad  i\in I.
\]
By Definition~\ref{def:matrixNorms} and Proposition
\ref{pro:matrixNorms} (\ref{enu:matrixNormsAx}), 
\[
  \norm{\xxo } 
\leq 
   \norm{\yyo_{I} } \norm{\AA_{I}^{-1}}
\leq 
    M \sqrt{d} \norm{\AA_{I}^{-1}} = M \sqrt{d} / \smin{\AA_{I}}.
\]
So, for all $i\not\in I $,
\[
   \form{\aa_{i}}{\xxo} 
\leq
 (  \max_{i} \norm{\aa_{i}})  M \sqrt{d} / \smin{\AA_{I}}
< 
    M^{2} \sqrt{d}  / 4\kappa 
    .
\]
Thus, for all $i \not \in I$,
\[
   \form{\aa_{i}}{\xxo } < y'_{i},
\]
and, by Definition~\ref{def:optSimp}, 
  $I = \optSimp{\AA_{I}\aalpha }{\vs{\aa}{1}{n}; \yyo }$.
\end{proof}

We will now define a linear program $LP^{+}$ that 
  interpolates between $LP'$ and $LP$.
This linear program will contain an extra variable $x_{0}$ and
  constraints of the form
\begin{equation*}
    \form{\aa_{i}}{\xx} \leq 
  \left(\frac{1+x_{0}}{2} \right) y_{i} + 
  \left(\frac{1-x_{0}}{2} \right) y'_{i},
\end{equation*}
and $-1 \leq x_{0} \leq 1$.
So, for $x_{0} = 1$ we see the original program $LP$ while
  for $x_{0} = -1$ we get $LP'$.
Formally, we let 
\begin{align*}
 \aap_{i} & = 
       \left\{\begin{array}{ll}
         \left((y'_{i} - y_{i})/2, \aa_{i} \right) 
               & \text{ for $1 \leq i \leq n$}\\
         \left(1, 0, \ldots , 0 \right) 
               & \text{ for $i = 0$}\\
         \left(-1, 0, \ldots , 0 \right) 
               & \text{ for $i = -1$}
    \end{array} \right.\\
  y^{+}_{i} & = 
      \left\{\begin{array}{ll}
         (y'_{i} + y_{i})/2
               & \text{ for $1 \leq i \leq n$}\\
         1 
               & \text{ for $i = 0$}\\
         1
               & \text{ for $i = -1$}
    \end{array} \right.
\\
  \zzp & =
   (1, 0, \ldots ,0), 
\end{align*} \index{aap@$\aap _{i}$}\index{yp@$y^{+}_{i}$}\index{zp@$\zzp $}%
and we define $LP^{+}$ by
\begin{align*}
  \mbox{maximize} \qquad  & \form{\zzp}{(x_{0}, \xx) } \nonumber  \\
  \mbox{subject to} \qquad & \form{\aap _{i}}{(x_{0}, \xx)} \leq y^{+}_{i}, 
  \mbox{ for $-1 \leq i \leq n$},
 & (LP^{+})
\end{align*}
and we set 
\[
  \yyp \defeq (\vs{y^{+}}{-1}{n}).
\]

By Proposition~\ref{pro:kappaM},
  $\sqrt{d} M / 4 \kappa  \geq 1$, so $y'_{i} \geq M$ and $y^{+}_{i} > 0$,
  for all $i$.
If $LP$ is infeasible, then the solution to $LP^{+}$
  will have $x_{0} < 1$.
If $LP$ is feasible, then the solution to $LP^{+}$
  will have  form $(1,\xx)$ where
  $\xx$ is a feasible point for $LP$.
If we use the shadow-vertex method to solve
  $LP^{+}$ starting from the appropriate initial vector,
  then $\xx$ will be an optimal solution to $LP$.

\begin{proposition}[relation of $M$ and $\kappa $]\label{pro:kappaM}
For $M$ and $\kappa$ as set by the algorithm, $\sqrt{d} M / 4 \kappa \geq 1$.
\end{proposition}
\begin{proof}
By definition, $\kappa \leq \smin{\AA_{I}}$.
On the other hand, 
$\smin{\AA_{I}} \leq \norm{\AA_{I}}
   \leq \sqrt{d} \max_{i} \norm{\aa_{i}}$,
  by Proposition~\ref{pro:matrixNorms} (\ref{enu:matrixNormsMax}).
Finally, $M \geq 4 \max_{i} \norm{\aa_{i}}$.
\end{proof}

We now state and prove the correctness of the two-phase shadow vertex method.
\vskip .2in
\fbox{

\begin{minipage}{5in}
\noindent \textbf{two-phase shadow-vertex method}\\
Input: $\AA = \left(\vs{\aa}{1}{n} \right)$, $\yy$, $\zz$. 
\begin{enumerate}

\item [(1)] Let $\calI = \setof{\vs{I}{1}{3nd \ln n}}$ be a collection
  of randomly chosen sets in $\binom{[n]}{d}$, and let
  $I \in \calI $ be the set maximizing $\smin{\AA _{I}}$.

\item [(2)] Set 
$M  = 2^{\ceiling{\lg \left(\max_{i} \norm{y_{i}, \aa_{i}} \right)} + 2}$
  and $\kappa  = 2^{\floor{\lg \left(\smin{\AA_{I}} \right)}}$.

\item [(3)] Set $  y_{i}'  =
  \begin{cases}
   M & \text{for $i \in I$}\\
   \sqrt{d}M^{2} / 4 \kappa  & \text{otherwise.}
  \end{cases}$.

\item [(4)] Choose $\aalpha$ uniformly at random from
   $\setof{\aalpha : \sum \alpha _{i} = 1 \text{ and } \alpha _{i} \geq 1/d^{2}}$.
  Set $\tto = \AA _{I} \aalpha$.\index{tt@$\tto $}%

\item [(5)] Let $J$\index{J@$J$}
  be the output of the polar shadow vertex algorithm on $LP'$
  on input $I$ and $\tto$.
  If $LP'$ is \textit{unbounded}, then return \textit{unbounded}.

\item [(6)] Let $\zeta   > 0$ be such that 
\[
\setof{-1} \cup J \in  \optSimp{(-\zeta , \zz)}{\vsDiv{\aap}{y^{+}}{-1}{n} }.
\]

\item [(7)] Let $K$\index{K@$K$}
  be the  output of the polar shadow vertex algorithm on $LP^{+}$
  on input $\setof{-1} \cup J$, $(-\zeta , \zz)$.
\item [(8)] Compute $(x_{0}, \xx)$ satisfying
  $\form{(x_{0}, \xx)}{\aap_{i}} = y_{i}$
  for $i \in K$.
\item [(9)] If $x_{0} < 1$, return \textit{infeasible}.
  Otherwise, return $\xx$.
\end{enumerate}
\end{minipage}
}\index{two-phase shadow-vertex method@two-phase shadow-vertex method}
\vskip .2in
The following propositions prove the correctness of the algorithm.

\begin{proposition}[Unbounded programs]\label{pro:twoPhaseUnbounded}
The following are equivalent
\begin{enumerate}
\item  $LP$ is unbounded;
\item  $LP'$ is unbounded;
\item  there exists a $1 > \lambda  > 0$
  such that 
  $\optSimp{\lambda (1, \zzero ) + (1-\lambda) (-\zeta , \zz)}
           {\vs{\aap}{-1}{n}; \yyp } =\emptyset $;
\item  for all $1 > \lambda  > 0$,
  $\optSimp{\lambda (1, \zzero ) + (1-\lambda) (-\zeta , \zz)}
           {\vs{\aap}{-1}{n}; \yyp } =\emptyset $.
\end{enumerate}
\end{proposition}

\begin{proposition}[Bounded programs]\label{pro:twoPhaseSolution}
If $LP'$ is bounded and has solution $J$, then
\begin{enumerate}
\item there exists $\zeta_{0}$ such that for all $\zeta > \zeta_{0}$,
  $\setof{-1} \cup J
    \in  \optSimp{(-\zeta , \zz)}{\vs{\aap}{-1}{n} ; \yyp }$, 
  \index{zez@$\zeta _{0}$}\index{ze@$\zeta $}%
\item  If $LP$ is feasible, then
  for $K' \in  \optSimp{\zz}{\vs{\aa}{1}{n}; \yy}$,
  there exists $\xi _{0}$ such that for all $\xi  > \xi _{0}$,
  $\setof{0} \cup K' \in  \optSimp{(\xi  , \zz)}{\vs{\aap}{-1}{n} ; \yyp }$, and
  \index{xiz@$\xi _{0}$}\index{xi@$\xi $}%
\item  if we use the shadow vertex method to solve
  $LP^{+}$ starting from $\setof{-1,J}$ and
  objective function $ (-\zeta , \zz)$, then the
  output of the algorithm will have form
  $\setof{0} \cup K'$, where $K'$ is a solution to $LP$.
\end{enumerate}
\end{proposition}

\begin{proof-of-proposition}{\ref{pro:twoPhaseUnbounded}}
$LP$ is unbounded if and only if there exists a vector $\vv$
  such that $\form{\zz}{\vv} > 0$ and 
  $\form{\aa_{i}}{\vv} \leq 0$ for all $i$.
The same holds for $LP'$, and establishes the equivalence of
  $(a)$ and $(b)$.
To show that $(a)$ or $(b)$ implies $(d)$, observe
\begin{align}
  \form{\lambda (1,\zzero) + (1-\lambda) (-\zeta , \zz  )}
       {(0, \vv)} & =
  (1-\lambda ) \form{\zz}{\vv} > 0, &  \label{eqn:twoPhaseUnbounded}\\
  \form{\aap_{i}}{(0, \vv)} & = \form{\aa_{i}}{\vv},
  \text{ for $i = 1, \ldots , n,$}
  \label{eqn:twoPhaseUnbounded2} \\
  \form{\aa_{0}^{+}}{(0, \vv)} & = 0,\text{ and} \nonumber
 \\
    \form{\aa_{-1}^{+}}{(0, \vv)} & = 0.\nonumber
\end{align}
To show that $(c)$ implies $(a)$ and $(b)$, note that
  $\aap_{0}$ and $\aap_{-1}$ are arranged so that
  if for some $v_{0}$ we have
\begin{equation*}
  \form{\aap_{i}}{(v_{0}, \vv)} \leq 0, \text{ for $-1 \leq i \leq n$},
\end{equation*}
then $v_{0} = 0$.
This identity allows us to apply \eqref{eqn:twoPhaseUnbounded}
  and \eqref{eqn:twoPhaseUnbounded2} to show $(c)$ implies $(a)$ and $(b)$. 
\end{proof-of-proposition}

\begin{proof-of-proposition}{\ref{pro:twoPhaseSolution}}
Let $J$ be the solution to $LP'$ and let
  $\xxo = \AA _{J}^{-1} \yy' _{J}$ be the corresponding 
  vertex.
We then have
\begin{align*}
   \form{\xx'}{\aa_{i}} & = y^{'}_{i},
   & \text{ for $i \in J$, and}\\
   \form{\xx'}{\aa_{i}} & \leq  y^{'}_{i},
   & \text{ for $i \not \in J$}.
\end{align*}

Therefore, it is clear that
\begin{align*}
   \form{(-1,\xx')}{\aap_{i}} & = y^{+}_{i},
   & \text{ for $i \in \setof{-1} \cup J$, and}\\
   \form{(-1,\xx')}{\aap_{i}} & \leq  y^{+}_{i},
   & \text{ for $i \not \in \setof{-1} \cup J$}.
\end{align*}
Thus, $\simp{\aap_{-1},(\aap_{i} )_{i \in J}}$ is a facet of $LP^{+}$.
To see that there exists a $\zeta_{0}$ such that
  it optimizes $(-\zeta , \zz)$ for all
  $\zeta  > \zeta_{0}$,
  first observe that there exist
  $\alpha _{i} > 0$, for $i \in J$, such that 
  $\sum_{i \in J} \alpha_{i} \aa_{i} = \zz$.
Now, let $(-\zeta_{0}, \zz) = \sum_{i \in J} \alpha_{i} \aap_{i}$.
For $\zeta  > \zeta_{0}$, we have
\[
  (-\zeta , \zz) = (\zeta - \zeta _{0}) \aap_{-1} + \sum_{i \in J} \alpha_{i} \aap_{i},
\]
which proves 
  $ (-\zeta , \zz) \in \cone{\aap_{-1},(\aap _{i})_{i \in J}}$ and
completes the proof of $(a)$.

The proof of $(b)$ is similar.

To prove part $(c)$, let $K$ be as
  in step $(7)$.
Then, there exists a $\lambda_{k}$ such that for all $\lambda \in (\lambda_{k},1)$,
\[
  K
    = \optSimp{(1-\lambda) (-\zeta , \zz) + \lambda \zzp }
              {\vs{\aap}{-1}{n}; \yyp}.
\]
Let $(x_{0}, \xx)$ satisfy
  $\form{(x_{0}, \xx)}{\aap_{i}} = y^{+}_{i}$, for $i \in K$.
Then, by Proposition~\ref{pro:polar}, 
\[
  (x_{0}, \xx )
    = \optVert{(1-\lambda) (-\zeta , \zz) + \lambda \zzp }
              {\vs{\aap}{-1}{n}; \yyp}.
\]
If $x_{0} < 1$, then LP was infeasible.
Otherwise, let $\xxs = \optVert{\zz}{\vs{\aa}{1}{n}; \yy}$.
By part $(b)$, there exists $\xi_{0}$ such that
  for all $\xi > \xi_{0}$,
\[
  (1 , \xxs )
    = \optVert{(\xi  , \zz)}
              {\vs{\aap}{-1}{n}; \yyp}.
\]
For $\xi = -\zeta + \lambda / (1-\lambda )$, we have
\[
  (\xi , \zz) = \frac{1}{1 - \lambda}
                \left((1-\lambda) (-\zeta , \zz) + \lambda \zzp \right).
\]
So, as $\lambda$ approaches 1,
$\xi = -\zeta + \lambda / (1-\lambda )$ goes to infinity and we have
\[
    \optVert{(1-\lambda) (-\zeta , \zz) + \lambda \zzp }
              {\vs{\aap}{-1}{n}; \yyp}
=
    \optVert{(\xi , \zz) }
              {\vs{\aap}{-1}{n}; \yyp},
\]
which implies $(x_{0}, \xx) = (1,\xxs  )$.
\end{proof-of-proposition}

Finally, we bound the number of steps taken in step (7)
  by the shadow size of a related polytope:

\begin{lemma}[Shadow path of $LP^{+}$]\label{pro:lp++}
Let $\vs{\aa^{+}}{-1}{n}$ and $\vs{y^{+}}{-1}{n}$ 
 be as defined in $LP^{+}$.
Let $\zeta > 0$ be such that 
  $\setof{-1}\cup J = 
  \optSimp{(-\zeta , \zz)}{\vsDiv{\aap}{y^{+}}{-1}{n} }$.
Then the number of simplex steps 
  made by the polar shadow vertex algorithm
  while solving $LP^{+}$ from
  initial basis
  $\setof{-1}\cup J$ and vector $(-\zeta , \zz)$ 
  is at most
\[
2 + \sizeof{\shadow{(0,\zz),\zzp}{\vsDiv{\aap}{y^{+}}{1}{n} }}.
\]
\end{lemma}
\begin{proof}
We will establish that $\setof{-1} \in I$
  for the first step only.
One can similarly prove that $\setof{0} \in I$
  is only true at termination.

Let $I \in \optSimp{\qq_{\lambda }}{\vsDiv{\aap}{y^{+}}{-1}{n}}$
  have form $\setof{-1} \cup L$.
As $\qq _{0} = \aap _{-1} \in \cone{\AA _{\setof{-1} \cup L}}$,
  and $\cone{\AA _{\setof{-1} \cup L}}$ is a convex set,
  we have 
  $\qq_{\lambda '}  \in \cone{\AA _{\setof{-1} \cup L}}$
  for all $0 \leq \lambda ' \leq \lambda $.
As $[\lambda _{i}, \lambda _{i+1}]$ is exactly the set of 
  $\lambda $ optimized by $\simp{\AA _{I}}$ in the $i$th step
  of the polar shadow vertex method,
  $I$ must be the initial set.

\end{proof}

\subsection{Discussion}
We also note that our analysis of the two-phase algorithm
  actually takes advantage of the fact that
  $\kappa$ and $M$ have been set to powers of two.
In particular, this fact is used to show that there are not
  too many likely choices for $\kappa$ and $M$.
For the reader who would like to drop this condition,
  we briefly explain how the argument of Section~\ref{sec:phaseI}
  could be modified to compensate:
  first, we could consider setting $\kappa$ and $M$
  to powers of $1 + 1/poly (n,d,1/\sigma )$.
  This would still result in a polynomially bounded number of choices for
  $\kappa$ and $M$.
One could then drop this assumption by observing that allowing
  $\kappa$ and $M$ to vary in a small range would
  not introduce too much dependency between the variables.

%For convenience, we will define
%\[
% LP^{+} (\AA,I, \hh ,\kappa) \defeq 
%  (\vs{\aap}{1}{n} ; \yyp),
%\]\index{LPp@$LP^{+} (\AA,I, \hh ,\kappa)$}
%\label{def:lp+}

%where $\vs{\aap}{1}{n}$ and $\yyp$ are formed from
%  $(\AA , I, \hh  ,\kappa)$
%  as described in this section.

% Local Variables: ***
% TeX-master:"shadow.tex" ***
% End: ***

%\newpage

\def\constAngGiven{\frac
    {9,371,990 \ n d^{2} \epsilon 
}{
    \sigma ^{6}
}
}

\def\constAngGivent{\frac
    {9,372,424 \ n d^{3} 
}{
    \sigma ^{6}
}
}

\def\constAngle{\frac
    {9,372,424 \ n d^{3} 
}{
    \sigma ^{6}
}
}

\def\constShadow{\frac
    {58,888,677 \ n d^{3}
}{
    \sigma ^{6}
}
}

\def\constFinal{\frac
    {58,888,678 \ n d^{3} 
}{
    \min \left(\sigma , 1/3\sqrt{d \ln n} \right)^{6}
}
}

\section{Shadow Size}\label{sec:shadow}
In this section, we bound the expected size of the shadow
  of the perturbation of a polytope onto a fixed plane.
This is the main geometric result of the paper.
The algorithmic results of this paper will rely on
  extensions of this theorem derived in
  Section~\ref{ssec:extension}.

\setcounter{theorem}{0}

\begin{theorem}[Shadow Size]\label{thm:shadow}
Let $d \geq 3$ and $n > d$.
Let $\zz$ and $\tt$ be independent vectors in $\Reals{d}$,
  and let
  $\vs{\mu }{1}{n}$ be Gaussian distributions 
  in $\Reals{d}$ of standard deviation
  $\sigma$
  centered at points each of norm at most $1$.
Then,
\begin{equation}\label{eqn:shadow}
  \expec{\vs{\aa}{1}{n}}
        {\sizeof{
  \shadow{\tt , \zz}{\vs{\aa}{1}{n}}
  }}
 \leq 
  \calD  (n, d, \sigma ),
\end{equation}\index{qq@$\qq$}%
where
\[
  \calD  (n,d, \sigma) = \constFinal,
\]\index{D@$\calD$}%
and $\vs{\aa}{1}{n}$ have density
  $\prod _{i=1}^{n} \mu _{i} (\aa_{i})$.
\end{theorem}

The proof of Theorem~\ref{thm:shadow}, will use
  the following definitions.

\begin{definition}[ang]\index{ang}
For a vector $\qq$ and a set $S$, we define
\[
  \angTo{\qq}{S}
 =
  \min _{\xx \in S}
  \angle{\qq , \xx },
\]
If $S$ is empty, we set $\angTo{\qq}{\emptyset} = \infty $.
\end{definition}

\begin{definition}[$\mathbf{ang}_{\qq}$]\index{ang@$\mathbf{ang}_{\qq}$}%
For a vector $\qq$ and points $\vs{\aa}{1}{n}$ in $\Reals{d}$,
  we define
\[
 \angZ{\qq}{\vs{\aa}{1}{n}} = 
  \angTo{\qq}{\partial \simp{\optSimp{\qq}{\vs{\aa}{1}{n}}}},
\]
where $\partial \simp{\optSimp{\qq}{\vs{\aa}{1}{n}}}$
  denotes
  the boundary of the simplex $\simp{\optSimp{\qq}{\vs{\aa}{1}{n}}}$.
\end{definition}

These definitions are arranged so that
  if the ray through $\qq$ does not pierce the convex hull
  of $\vs{\aa}{1}{n}$, then 
  $\angZ{\qq}{\vs{\aa}{1}{n}} = \infty $.

In our proofs, we will make frequent use of the fact that it is very
  unlikely that a Gaussian random variable is far from its mean.
To capture this fact, we define:
\begin{definition}[P]
$P$ is the set of $(\vs{\aa}{1}{n})$ for which
  $\norm{\aa_{i}} \leq 2$, for all $i$.
\end{definition}\index{p@$P$}%
Applying a union bound to Corollary~\ref{cor:chiSquare}, we obtain
\begin{proposition}[Measure of P]\label{pro:chiSquareP}
\[
  \prob{}{(\vs{\aa }{1}{n}) \in P} 
 \geq 1 - n (n^{-2.9d})
  = 1 -   n^{-2.9 d + 1}.
\]
\end{proposition}

\begin{proof-of-theorem}{\ref{thm:shadow}}
We first observe that we can assume $\sigma \leq 1/3\sqrt{d \ln n}$:
  if $\sigma > 1/3\sqrt{d \ln n}$, then we can scale down all the
  data until $\sigma = 1/3\sqrt{d \ln n}$.
As this could only decrease the norms of the centers of the distributions,
  the theorem statement would be unaffected.

Assume without loss of generality that $\zz$ and $\tt$
  are orthogonal.
Let 
\begin{equation}\label{eqn:zTheta}
\qq _{\theta } = \zz \sin (\theta ) + \tt \cos (\theta ).
\end{equation}\index{qq@$\qq _{\theta }$}%
We discretize the problem by using the intuitively obvious fact,
  which we prove as Lemma~\ref{lem:limit}, that the left-hand
  of \eqref{eqn:shadow} equals
\[
 \lim_{m \rightarrow \infty} 
  \expec{\vs{\aa}{1}{n}}
        {\sizeof{\union_{\theta  \in 
            \setof{\frac{2 \pi }{m}, \frac{2 \cdot 2 \pi }{m}, \ldots,
                         \frac{m \cdot 2 \pi }{m}}} 
          \setof{\optSimp{\qq_{\theta}}{\vs{\aa }{1}{n}}}}}.
\]
Let $E_{i}$ denote the event\index{E@$E_{i}$}%
\[
\ind{ \optSimp{\qq_{2 \pi i/ m}}{\vs{\aa}{1}{n}} \not =
           \optSimp{\qq_{2 \pi ((i+1) \mod m) / m}}{\vs{\aa}{1}{n}}}.
\]
Then, for any $m \geq 2$ and for all $\vs{\aa}{1}{n}$,
\[
\sizeof{\union_{\theta  \in 
            \setof{\frac{2 \pi }{m}, \frac{2 \cdot 2 \pi }{m}, \ldots,
                         \frac{m \cdot 2 \pi }{m}}} 
          \setof{\optSimp{\qq_{\theta}}{\vs{\aa }{1}{n}}}}
= 
  \sum_{i=1}^{m} E_{i} (\vs{\aa}{1}{n}).
\]
We bound this sum by
\begin{align*}
 \expec{}{\sum _{i=1}^{m} E_{i}}
  & =
     \expec{P}{   \sum_{i} E_{i}} \prob{}{P}
 +   \expec{\bar{P}}{   \sum_{i} E_{i}} \prob{}{\bar{P}}\\
 & \leq 
     \expec{P}{   \sum_{i} E_{i}} 
 +  \binom{n}{d} n^{-2.9d + 1}\\
 & \leq 
     \expec{P}{   \sum_{i} E_{i}} 
 + 1
\end{align*}
Thus, we will focus on bounding 
  $\expec{P}{   \sum_{i} E_{i}} $.

Observing that $E_{i}$ 
  implies $ \ind{\angZ{\qq _{2 \pi i/ m}}{\vs{\aa}{1}{n}} \leq 2 \pi / m}$,
  and applying linearity of expectation, we obtain
\begin{align*}
\expec{P}{   \sum_{i} E_{i}}
 & =    \sum _{i=1}^{m} 
   \prob{P}{E_{i}}\\
& \leq 
   \sum _{i=1}^{m} 
   \prob{P}{\textbf{ang}_{\qq_{2 \pi i/m}} (\vs{\aa }{1}{n}) < \frac{2 \pi }{m}}
\\
& \leq 2 \pi \constAngGivent    & \text{by Lemma~\ref{lem:angWrapper},}\\
& \leq \constShadow .
\end{align*}
\end{proof-of-theorem}

\begin{lemma}[Discretization in limit]\label{lem:limit}
Let $\zz$ and $\tt$ be orthogonal vectors in $\Reals{d}$,
  and let
  $\vs{\mu }{1}{n}$ be non-degenerate Gaussian distributions.
Then,
\begin{multline}\label{eqn:lemLimit}
  \expec{\vs{\aa}{1}{n}}
        {\sizeof{\union_{\qq \in \Span{\zz , \tt }} 
          \setof{\optSimp{\qq}{\vs{\aa }{1}{n}}}}}
=\\
 \lim_{m \rightarrow \infty} 
  \expec{\vs{\aa}{1}{n}}
        {\sizeof{\union_{\theta  \in 
            \setof{\frac{2 \pi }{m}, \frac{2 \cdot 2 \pi }{m}, \ldots,
                         \frac{m \cdot 2 \pi }{m}}} 
          \setof{\optSimp{\qq_{\theta}}{\vs{\aa }{1}{n}}}}},
\end{multline}
where $\qq_{\theta}$ is as defined in \eqref{eqn:zTheta}.
\end{lemma}
\begin{proof}
For a $I \in \binom{[n]}{d}$, let
\[
 F_{I} (\vs{\aa }{1}{n})
=
 \int _{\theta }
  \ind{\optSimp{\qq _{\theta }}{\vs{\aa}{1}{n}} = 
    I}
  \diff{\theta }.
\]
The left and right hand sides of \eqref{eqn:lemLimit} 
  can differ only if there exists a
  $\delta > 0$ such that for all $\epsilon > 0$,
\[
  \prob{\vs{\aa}{1}{n}}
       {\exists I \Big|
        \begin{array}{l}
            I = 
               \optSimp{\qq _{\theta }}{\vs{\aa}{1}{n}}
               \text{ for some $\theta $, and}\\
            F_{I} (\vs{\aa }{1}{n})< \epsilon
        \end{array}}
  \geq 
\delta.
\]
As there are only finitely many choices for
  $I$,
  this would imply the existence of
  a $\delta '$ and
  a particular
  $I$ such that for all $\epsilon > 0$,
\begin{equation*}
  \prob{\vs{\aa}{1}{n}}
  {     \begin{array}{l}
            I = 
               \optSimp{\qq _{\theta }}{\vs{\aa}{1}{n}}
               \text{ for some $\theta $, and}\\
            F_{I} (\vs{\aa }{1}{n})< \epsilon
        \end{array} 
  }
  \geq 
\delta'.
\end{equation*}
As $ F_{I} (\vs{\aa }{1}{n}) =  F_{I}
  (\AA _{I})$
given that 
 $I = \optSimp{\qq _{\theta }}{\vs{\aa}{1}{n}}$
 for some $\theta $,
 this implies
 that for all $\epsilon > 0$,
\begin{equation}\label{eqn:lemLimitSimp}
  \prob{\vs{\aa}{1}{n}}
  {     \begin{array}{l}
            I = 
               \optSimp{\qq _{\theta }}{\AA_{I}}
               \text{ for some $\theta $, and}\\
            F_{I} (\AA_{I})< \epsilon
        \end{array} 
  }
  \geq 
\delta'.
\end{equation}
Note that $I = \optSimp{\qq _{\theta }}{\AA_{I}}$
  if and only if
  $\qq_{\theta} \in \cone{\AA_{I}}$.
Now, let 
\[
  G (\AA_{I}) 
 = 
 \int _{\theta }
   \ind{\qq _{\theta } \in \cone{\AA_{I}}}
  \left( \angTo{\qq _{\theta }}{\partial \simp{\AA_{I}}} / \pi   \right)
 \diff{\theta }.
\]
As $G (\AA_{I}) \leq F_{I} (\AA_{I})$,
 \eqref{eqn:lemLimitSimp} implies that for all $\epsilon > 0$
\begin{equation*}
  \prob{\vs{\aa}{1}{n}}
  {     \begin{array}{l}
            I = 
               \optSimp{\qq _{\theta }}{\vs{\aa}{1}{n}}
               \text{ for some $\theta $, and}\\
            G (\AA_{I})< \epsilon
        \end{array} 
  }
  \geq 
\delta'.
\end{equation*}
However, 
  $G$ is a continuous function, and therefore measurable,
  so this would imply
\[
  \prob{\vs{\aa}{1}{n}}
  {     \begin{array}{l}
            I = 
               \optSimp{\qq _{\theta }}{\vs{\aa}{1}{n}}
               \text{ for some $\theta $, and}\\
            G (\AA_{I}) = 0
        \end{array} 
  }
  \geq 
\delta',
\]
which is clearly false as the set of
  $\AA_{I}$ satisfying
\begin{itemize}
\item $G (\AA_{I}) = 0 $, and
\item $\exists \theta : \optSimp{\qq _{\theta }}{\vs{\aa}{1}{n}}
  = \setof{\AA_{I}}$
\end{itemize}
has co-dimension 1, and so has
  measure zero under the product distribution of non-degenerate
  Gaussians.
\end{proof}

\begin{lemma}[Angle bound]\label{lem:angWrapper}
Let $d \geq 3$ and $n > d$.
Let $\qq $ be any unit vector and
  let $\vs{\mu }{1}{n}$ be Gaussian measures 
  in $\Reals{d}$
  of standard deviation $\sigma  \leq 1/3 \sqrt{d \ln n}$
  centered at points of norm at most 1.
Then,
\begin{equation*}
  \prob{P}{\textbf{ang}_{\qq} (\vs{\aa }{1}{n}) < \epsilon }
  \leq 
  \constAngle \epsilon 
\end{equation*}
where $\vs{\aa}{1}{n}$ have density 
\[
  \prod_{i=1}^{n} \mu _{i} (\aa_{i}).
\]
\end{lemma}

The proof will make use of the following definition:
\begin{definition}[$P_{I}^{j}$]\index{P@$P_{I}^{j}$}%
For a $I \in \binom{[n]}{d}$
  and $j \in I$,
 we define $P_{I}^{j}$ to be the set
  of $\vs{\aa}{1}{d}$ satisfying
\begin{itemize}
\item [(1)] For all $\qq$, if $\optSimp{\qq}{\vs{\aa}{1}{n}} \neq \emptyset$, then
  $s \leq 2 $, where $s$ is the real number for which
  $s \qq \in \simp{\optSimp{\qq}{\vs{\aa}{1}{n}}}$,
\item [(2)] $\dist{\aa_{ i}}{\aa_{k}} \leq 4 $, for
  $i, k \in I - \setof{j}$,
\item [(3)] $\dist{\aa_{j}}
             {\aff{\AA_{I - \setof{j}} } } \leq 4$, and
\item [(4)] $\dist{\aa_{j} ^{\bot}}{\aa_{i}} \leq 4 $,
  for all $i \in I - \setof{j}$, where $\aa_{j} ^{\bot}$ is the
  orthogonal projection of
  $\aa_{j}$ onto $\aff{\AA_{I - \setof{j}}}$.
\end{itemize}
\end{definition}

\begin{proposition}[$P \subset P^{j}_{I}$]\label{pro:PinP}
For all $j, I$, $P \subset P^{j}_{I}$. 
\end{proposition}
\begin{proof}
Parts $(2)$, $(3)$, and $(4)$ follow immediately from
  the restrictions $\norm{\aa_{i}} \leq 2$.
To see why part $(1)$ is true, note that $s \qq$
  lies in the convex hull of $\vs{\aa}{1}{n}$,
  and so its norm, $s$, can be at most 
  $\max_{i} \norm{\aa_{i}} \leq 2$,
  for $(\vs{\aa}{1}{n}) \in P$.
\end{proof}

\begin{proof-of-lemma}{\ref{lem:angWrapper}}
Applying a union bound twice, we write
\begin{align*}
\lefteqn{\prob{P}{\textbf{ang}_{\qq} (\vs{\aa}{1}{n}) < \epsilon }}\\
%\prob{P}{\textbf{ang}_{\qq} (\vs{\aa}{1}{n}) < \epsilon } \\
 & \leq 
  \sum _{I}
   \prob{P}{
    \begin{array}{l}
      \optSimp{\qq}{\vs{\aa}{1}{n}} = I
      \text{ and }\\
      \textbf{ang}(\qq, \partial \simp{\AA_{I}})
         < \epsilon 
    \end{array}
}\\
 & \leq 
  \sum _{I} \sum _{j=1}^{d}
   \prob{P}{
\begin{array}{l}
 \optSimp{\qq}{\vs{\aa}{1}{n}} = I
      \text{ and }\\
      \textbf{ang}(\qq,  \simp{\AA_{I - \setof{j}}})
         < \epsilon 
\end{array}
}\\
& 
 \leq 
 \sum _{I} \sum _{j=1}^{d}
   \prob{P^{j}_{I}}{
 \begin{array}{l}
 \optSimp{\qq}{\vs{\aa}{1}{n}} = I
      \text{ and }\\
      \textbf{ang}(\qq,  \simp{\AA_{I - \setof{j}}})
         < \epsilon 
 \end{array}
 }
 \Big/
 \prob{P_{I}^{j}}{P}\\
\intertext{(by Proposition~\ref{pro:condProb})}
& 
 \leq 
 \sum _{I} \sum _{j=1}^{d}
   \prob{P^{j}_{I}}{
 \begin{array}{l}
 \optSimp{\qq}{\vs{\aa}{1}{n}} = I
      \text{ and }\\
      \textbf{ang}(\qq, \simp{\AA_{I - \setof{j}}})
         < \epsilon 
 \end{array}
 }
 \Big/
 \prob{}{P}\\
\intertext{(by $P \subset P_{I}^{j}$)}
& 
 \leq \frac{1}{1 - n^{-2.9d + 1}} 
 \sum _{I} \sum _{j=1}^{d}
   \prob{P^{j}_{I}}{
 \begin{array}{l}
 \optSimp{\qq}{\vs{\aa}{1}{n}} = I
      \text{ and }\\
      \textbf{ang}(\qq, \simp{\AA_{I - \setof{j}}})
         < \epsilon 
 \end{array}
 }
\\
\intertext{(by Proposition~\ref{pro:chiSquareP})}
& 
 \leq \frac{1}{1 - n^{-2.9d + 1}} 
 \sum _{j=1}^{d} \sum _{I} 
   \prob{P^{j}_{I}}{
 \begin{array}{l}
 \optSimp{\qq}{\vs{\aa}{1}{n}} = I
      \text{ and }\\
      \textbf{ang}(\qq, \simp{\AA_{I - \setof{j}}})
         < \epsilon,
 \end{array}
 },
\end{align*}
by changing the order of summation.

We now expand the inner summation using Bayes' rule
  to get
\begin{multline}
\begin{split}
\lefteqn{
 \sum _{I} 
   \prob{P^{j}_{I}}{
 \begin{array}{l}
 \optSimp{\qq}{\vs{\aa}{1}{n}} = I
      \text{ and }\\
      \textbf{ang}(\qq, \simp{\AA_{I - \setof{j}}})
         < \epsilon 
 \end{array}
 }
} \qquad \qquad \qquad  &\\
 \qquad  =
 \sum _{I} 
 &   \prob{P^{j}_{I}}{
    \optSimp{\qq}{\vs{\aa}{1}{n}} = I} \cdot 
 \\
 &  \prob{P^{j}_{I}}{
   \begin{array}{l}
      \textbf{ang}(\qq, \simp{\AA_{I - \setof{j}}})
         < \epsilon 
   \big|\\
  \qquad  \optSimp{\qq}{\vs{\aa}{1}{n}} = I
   \end{array}}
\end{split}\label{eqn:angWrapper}
\end{multline}
As $\optSimp{\qq}{\vs{\aa}{1}{n}}$ is a set of size zero or one
  with probability $1$,
\[
 \sum _{I} 
   \prob{}{
    \optSimp{\qq}{\vs{\aa}{1}{n}} = I}
  \leq 1;
\]
from which we derive
\begin{align*}
\lefteqn{
  \sum _{I} 
   \prob{P^{j}_{I}}{
    \optSimp{\qq}{\vs{\aa}{1}{n}} = I}
}\\
 &
   \leq 
 \sum _{I} 
   \prob{}{
    \optSimp{\qq}{\vs{\aa}{1}{n}} = I}
    \big/ 
    \prob{}{P^{j}_{I}}
\\
\intertext{(by Proposition~\ref{pro:condProb})}
 &
   \leq 
  \frac{1}{1 - n^{-2.9d + 1}}
 \sum _{I} 
   \prob{}{
    \optSimp{\qq}{\vs{\aa}{1}{n}} = I}
\\
\intertext{(by $P \subset P^{j}_{I}$ and Proposition~\ref{pro:chiSquareP})}
  & 
  \leq 
  \frac{1}{1 - n^{-2.9d + 1}}.
\end{align*}
So,
\begin{align*}
\eqref{eqn:angWrapper} 
\leq 
  \frac{1}{1 - n^{-2.9d + 1}}.
   \max _{I}
   \prob{P^{j}_{I}}{
   \begin{array}{l}
      \textbf{ang}(\qq, \simp{\AA_{I - \setof{j}}})
         < \epsilon 
   \big|\\
  \qquad  \optSimp{\qq}{\vs{\aa}{1}{n}} = I
   \end{array}}.
\end{align*}
Plugging this bound in to the first inequality derived in the proof,
  we obtain the bound of
\begin{align*}
\lefteqn{\prob{P}{\textbf{ang}_{\qq} (\vs{\aa}{1}{n}) < \epsilon }}\\
& \leq 
 \frac{d}{(1 - n^{-2.9d + 1})^{2}} 
   \max _{j, I}
   \prob{P^{j}_{I}}{
   \begin{array}{l}
      \textbf{ang}(\qq, \simp{\AA_{I - \setof{j}}})
         < \epsilon 
   \big|\\
  \qquad  \optSimp{\qq}{\vs{\aa}{1}{n}} = I
   \end{array}}
 \\
& \leq 
  d
  \constAngGivent \epsilon , \text{ by Lemma~\ref{lem:angGiven}, $d \geq 3$ and $n \geq d+1$,} \\
& =
  \constAngle \epsilon .
\end{align*}
\end{proof-of-lemma}

\begin{definition}[Q]\index{Q@$Q$}%
We define $Q$ to be the set of $(\vs{\bb }{1}{d}) \in \Reals{d-1}$
  satisfying
\begin{itemize}
\item [(1)] $\dist{\bb _{1}}{\aff{\vs{\bb }{2}{d}}} \leq 4 $,
\item [(2)] $\dist{\bb _{i}}{\bb _{j}} \leq 4 $ for all 
  $i, j \geq 2$, 
\item [(3)] $\dist{\bb _{1}^{\bot}}{\bb _{i}} \leq 4 $
  for all $i \geq 2$, where
  $\bb _{1}^{\bot}$
  is the orthogonal projection
  of $\bb _{1}$ onto $\aff{\vs{\bb}{2}{d}}$, and
\item [(4)] $\zzero \in \simp{\vs{\bb}{1}{d}}$.
\end{itemize}
\end{definition}

\begin{lemma}[Angle bound given optSimp]\label{lem:angGiven}
Let 
 $\vs{\mu }{1}{n}$ be Gaussian measures in $\Reals{d}$
  of standard deviation $\sigma  \leq 1 / 3 \sqrt{d \ln n}$
  centered at points of norm at most $1$.
Then
\begin{equation}\label{eqn:angGiven}
   \prob{P^{1}_{1,\ldots ,d}}{
   \begin{array}{l}
      \textbf{ang}(\qq, \simp{\vs{\aa }{2}{d}})
         < \epsilon 
   \big|\\
  \qquad  \optSimp{\qq}{\vs{\aa }{1}{n}} = \setof{1,\ldots ,d}
   \end{array}}
  \leq 
  \constAngGiven 
\end{equation}
where $\vs{\aa}{1}{n}$ have density 
\[
  \prod_{i=1}^{n} \mu _{i} (\aa_{i}).
\]
\end{lemma}
\begin{proof}
We begin by making the change of variables 
  from $\vs{\aa}{1}{d}$
  to $\oomega , s, \vs{\bb}{1}{d}$
  described in Corollary~\ref{cor:blaschke},
  and we
  recall that the Jacobian of this change of variables is 
\[
  (d-1)! \form{\oomega }{\qq} \vol{\simp{\vs{\bb}{1}{d}}}.
\]
As this change of variables is arranged so that
  $s \qq \in \simp{\vs{\aa}{1}{d}}$ if and only if
  $\zzero \in \simp{\vs{\bb}{1}{d}}$,
  the condition that
  $ \optSimp{\qq}{\vs{\aa }{1}{n}} = \setof{1,\ldots ,d}$
  can be expressed as
\[
  \ind{\zzero \in \simp{\vs{\bb}{1}{d}}}
  \prod_{j > d} \ind{\form{\oomega }{\aa_{j}} \leq \form{\oomega }{s \qq}}.
\]

%The point $\xx \in \aff{\vs{\aa}{2}{d}}$ minimizing
%  $\angle{s \qq , \xx }$ is the point minimizing
%  $\dist{s \qq}{\xx}$.
Let $\xx$ be any point on $\simp{\vs{\aa}{2}{d}}$.
Given that $s \qq \in \simp{\vs{\aa}{1}{d}}$,
  conditions $(3)$ and $(4)$ for membership in $P^{1}_{1,\ldots ,d}$
  imply that
\[
 \dist{s \qq}{\xx } 
\leq 
 \dist{\aa _{1}}{\xx}
\leq 
\sqrt{\dist{\aa _{1}}{\aff{\vs{\aa }{2}{d}}}^{2}
  + \dist{\aa ^{\bot}_{1}}{\xx }^{2}
}
  \leq 4\sqrt{2},
\]
where $\aa_{1} ^{\bot}$ is the
  orthogonal projection of
  $\aa_{1}$ onto $\aff{\vs{\aa}{2}{d}}$.
So, Lemma~\ref{lem:division} implies
\[
  \angTo{\qq}{ \simp{\vs{\aa}{2}{d}}}
 \geq 
  \frac{\dist{s \qq }{\aff{\vs{\aa}{2}{d}}} \form{\oomega }{\qq}}
       {2+4\sqrt{2} }
  = 
  \frac{\dist{\zzero }{\aff{\vs{\bb}{2}{d}}} \form{\oomega }{\qq}}
       {2+4\sqrt{2}}.
\]

Finally, observe that $(\vs{\aa}{1}{d}) \in P^{1}_{1,\ldots,d}$
  is equivalent to the conditions
  $(\vs{\bb}{1}{d}) \in Q$ and $s \leq 2 $, given
  that $\optSimp{\qq}{\vs{\aa }{1}{d}} = \setof{1,\ldots ,d}$.
Now, the left-hand side of \eqref{eqn:angGiven} can be bounded by
\begin{equation}\label{eqn:angGivenNew}
\prob{\substack{\oomega , s \leq 2 \\
                (\vs{\bb}{1}{d}) \in Q}}
     { \frac{\dist{\zzero }{\aff{\vs{\bb}{2}{d}}} \form{\oomega }{\qq}}      
            {2+4\sqrt{2} } 
        < \epsilon },
\end{equation}
where the variables have density proportional to
\[
  \form{\oomega }{\qq}
  \vol{\simp{\vs{\bb}{1}{d}}}
  \left(  
    \prod_{j > d} 
     \int _{\aa_{j}}
     \ind{\form{\oomega }{\aa_{j}} \leq s \form{\oomega }{\qq}}
     \mu _{j} (\aa_{j}) \diff{\aa_{j}}
  \right)
  \prod_{i=1}^{d} \mu _{i} (\RR_{\oomega }\bb _{i} + s \qq ).
\]
As Lemma~\ref{lem:distance} implies
\[
\prob{\substack{\oomega , s \leq 2 \\
                (\vs{\bb}{1}{d}) \in Q}}
     {\dist{\zzero }{\aff{\vs{\bb}{2}{d}}}
        < \epsilon }
  \leq 
  \frac{900 e^{2/3} d^{2} \epsilon }{\sigma ^{4}},
\]
and Lemma~\ref{lem:angle} implies
\[
 \max _{s \leq 2, \vs{\bb }{1}{d} \in Q}
 \prob{\oomega }{\form{\oomega }{\qq} < \epsilon } < 
  \left(\frac{340 n  \epsilon}{\sigma^{2} } \right)^{2},
\]
 we can apply Lemma~\ref{lem:comb} to prove
\[
 \eqref{eqn:angGivenNew} \leq 
4 \cdot (2+4\sqrt{2}) \cdot 
\left(\frac{900 e^{2/3} d^{2} }{\sigma ^{4}} \right)
  \left(\frac{340 n }{\sigma^{2} } \right) \epsilon 
\leq 
 \constAngGiven 
.
\]
\end{proof}

\begin{lemma}[Division into distance and angle]\label{lem:division}
Let $\xx $ be a vector, let $0 < s \leq 2$, and let
  $\qq$ and $\oomega $ be unit vectors satisfying
\begin{enumerate}
\item $\form{\oomega }{\xx - s \qq} = 0$, and
\item $\dist{\xx}{s \qq} \leq 4\sqrt{2} $.
\end{enumerate}
Then,
\[
  \angle{\qq , \xx } 
  \geq 
  \frac{\dist{\xx }{s \qq } \form{\oomega }{\qq} }{2+4\sqrt{2} }.
\]
\end{lemma}
\begin{proof}
Let $\rr = \xx - s \qq$.
Then, $(a)$ implies
\[
  \form{\oomega }{\qq}^{2} + \Form{\frac{\rr}{\norm{\rr}}}{\qq}^{2}
  \leq \norm{\qq} = 1;
\]
so,
\[
  \form{\rr}{\qq} \leq \sqrt{1-\form{\oomega }{\qq}^{2}}
  \norm{\rr}.
\]
Let $h$ be the distance from $\xx$ to the ray 
  through $\qq$.
Then, 
\[
  h^{2} + \form{\rr}{\qq}^{2} = \norm{\rr}^{2};
\]
so,
\[
  h \geq \form{\oomega }{\qq}\norm{\rr} =
   \form{\oomega }{\qq} \dist{\xx }{s \qq}
\]
Now,
\[
  \angle{\qq , \xx } 
\geq 
  \sin (\angle{\qq , \xx })
= 
  \frac{h}{\norm{\xx }}
\geq 
  \frac{h}{s + \dist{\xx }{s \qq}}
\geq 
  \frac{h}{2+4\sqrt{2} }
\geq 
  \frac{ \form{\oomega }{\qq} \dist{\xx }{s \qq} }{2+4\sqrt{2} }.
\]
\end{proof}

\subsection{Distance}\label{sec:dist}

The goal of this section is to prove it is unlikely that
  $\zzero $ is near $\partial \simp{\vs{\bb}{1}{d}}$.

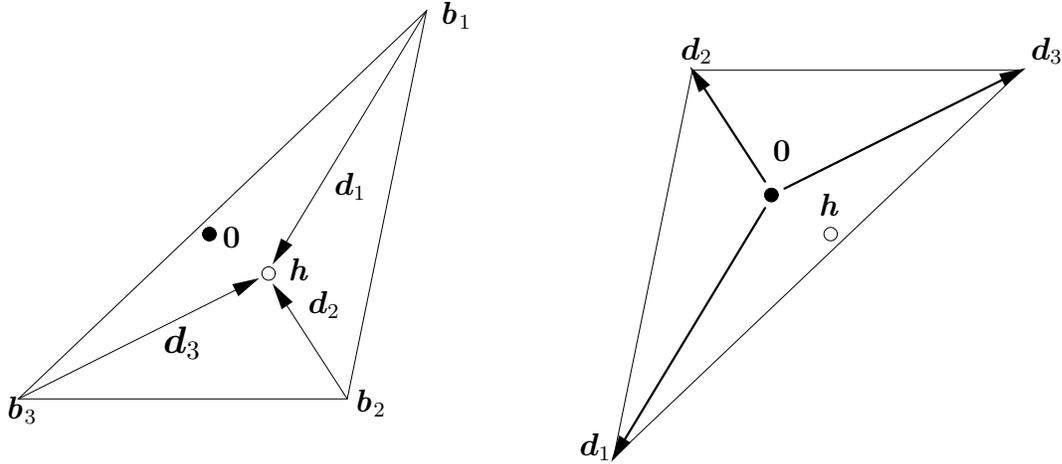
\begin{figure}[h]
\begin{center}
\setlength{\unitlength}{0.00083333in}
\begingroup\makeatletter\ifx\SetFigFont\undefined%
\gdef\SetFigFont#1#2#3#4#5{%
  \reset@font\fontsize{#1}{#2pt}%
  \fontfamily{#3}\fontseries{#4}\fontshape{#5}%
  \selectfont}%
\fi\endgroup%
{\renewcommand{\dashlinestretch}{30}
\begin{picture}(7272,2906)(0,-10)
\put(4780,1678){\blacken\ellipse{84}{84}}
\put(4780,1678){\ellipse{84}{84}}
\put(5149,1433){\ellipse{84}{84}}
\path(6341,2459)(4287,2459)(3794,37)(6341,2459)
\blacken\thicklines
\path(3833.986,166.430)(3794.000,37.000)(3890.197,132.379)(3833.986,166.430)
\path(3794,37)(4739,1597)
\thinlines
\blacken\thicklines
\path(4386.076,2366.615)(4287.000,2459.000)(4330.935,2330.857)(4386.076,2366.615)
\path(4287,2459)(4739,1762)
\thinlines
\blacken\thicklines
\path(6238.102,2370.892)(6341.000,2459.000)(6208.743,2429.690)(6238.102,2370.892)
\path(6341,2459)(4861,1720)
\put(3581,53){\makebox(0,0)[lb]{\smash{{{\SetFigFont{12}{14.4}{\rmdefault}{\mddefault}{\updefault}$\ddelta_1$}}}}}
\put(4213,2526){\makebox(0,0)[lb]{\smash{{{\SetFigFont{12}{14.4}{\rmdefault}{\mddefault}{\updefault}$\ddelta_2$}}}}}
\put(4789,1893){\makebox(0,0)[lb]{\smash{{{\SetFigFont{12}{14.4}{\rmdefault}{\mddefault}{\updefault}$\zzero$}}}}}
\put(6398,2526){\makebox(0,0)[lb]{\smash{{{\SetFigFont{12}{14.4}{\rmdefault}{\mddefault}{\updefault}$\ddelta_3$}}}}}
\put(5076,1548){\makebox(0,0)[lb]{\smash{{{\SetFigFont{12}{14.4}{\rmdefault}{\mddefault}{\updefault}$\hh$}}}}}
\thinlines
\put(1643,1187){\ellipse{84}{84}}
\put(1273,1433){\blacken\ellipse{84}{84}}
\put(1273,1433){\ellipse{84}{84}}
\path(82,406)(2135,406)(2628,2829)(82,406)
\path(2628,2829)(1684,1269)
\blacken\thicklines
\path(1723.925,1398.449)(1684.000,1269.000)(1780.152,1364.424)(1723.925,1398.449)
\thinlines
\path(2135,406)(1684,1104)
\blacken\thicklines
\path(1782.922,1011.450)(1684.000,1104.000)(1727.722,975.784)(1782.922,1011.450)
\thinlines
\path(82,406)(1560,1146)
\blacken\thicklines
\path(1457.198,1057.781)(1560.000,1146.000)(1427.775,1116.547)(1457.198,1057.781)
\put(1765,1146){\makebox(0,0)[lb]{\smash{{{\SetFigFont{12}{14.4}{\rmdefault}{\mddefault}{\updefault}$\hh$}}}}}
\put(985,693){\makebox(0,0)[lb]{\smash{{{\SetFigFont{14}{16.8}{\rmdefault}{\mddefault}{\updefault}$\ddelta _3$}}}}}
\put(2053,1678){\makebox(0,0)[lb]{\smash{{{\SetFigFont{12}{14.4}{\rmdefault}{\mddefault}{\updefault}$\ddelta _1$}}}}}
\put(1889,940){\makebox(0,0)[lb]{\smash{{{\SetFigFont{12}{14.4}{\rmdefault}{\mddefault}{\updefault}$\ddelta_2$}}}}}
\put(1356,1351){\makebox(0,0)[lb]{\smash{{{\SetFigFont{12}{14.4}{\rmdefault}{\mddefault}{\updefault}$\zzero$}}}}}
\put(0,282){\makebox(0,0)[lb]{\smash{{{\SetFigFont{12}{14.4}{\rmdefault}{\mddefault}{\updefault}$\bb_3$}}}}}
\put(2177,324){\makebox(0,0)[lb]{\smash{{{\SetFigFont{12}{14.4}{\rmdefault}{\mddefault}{\updefault}$\bb_2$}}}}}
\put(2710,2747){\makebox(0,0)[lb]{\smash{{{\SetFigFont{12}{14.4}{\familydefault}{\mddefault}{\updefault}$\bb_1$}}}}}
\end{picture}
}
\end{center}
\caption{The change of variables in Lemma~\ref{lem:distanceInPlane}.}
\end{figure}

\begin{lemma}[Distance bound]\label{lem:distance}
Let $\qq$ be a unit vector
 and let 
 $\vs{\mu }{1}{n}$ be Gaussian measures in $\Reals{d}$
  of standard deviation $\sigma  \leq 1 / 3 \sqrt{d \ln n}$ 
  centered at points of norm at most $1$.
Then,
\begin{equation}\label{eqn:lemDistance}
\prob{\substack{\oomega , s \leq 2 \\
                (\vs{\bb}{1}{d}) \in Q}}
     {\dist{\zzero }{\aff{\vs{\bb}{2}{d}}}
        < \epsilon }
  \leq 
  \frac{900 e^{2/3} d^{2}  \epsilon }{\sigma ^{4}},
\end{equation}
where the variables have density proportional to
\[
  \form{\oomega }{\qq}
  \vol{\simp{\vs{\bb}{1}{d}}}
  \left(  
    \prod_{j > d} 
     \int _{\aa_{j}}
     \ind{\form{\oomega }{\aa_{j}} \leq s \form{\oomega }{\qq}}
     \mu _{j} (\aa_{j}) \diff{\aa_{j}}
  \right)
  \prod_{i=1}^{d} \mu _{i} (\RR_{\oomega }\bb _{i} + s \qq ).
\]
\end{lemma}
\begin{proof}
Note that if we fix $\oomega $ and $s$, then the first and
  third terms in the density become constant.
For any fixed plane specified by
  $(\oomega , s)$, 
  Proposition~\ref{pro:gaussPlane} tells us that
  the induced density
  on $\bb _{i}$ remains a Gaussian of standard deviation $\sigma $ and
  is centered at the projection of the center of $\mu_{i}$ onto the plane.
As the origin of this plane is the point $s \qq$, and $s \leq 2 $,
  these induced Gaussians have centers of norm at most $3$.
Thus, we can use Lemma~\ref{lem:distanceInPlane} to bound
  the left-hand side of
  \eqref{eqn:lemDistance} by
\[
  \max _{\oomega , s \leq 2 }
  \prob{(\vs{\bb}{1}{d}) \in Q}
     {\dist{\zzero }{\aff{\vs{\bb}{2}{d}}}
        < \epsilon }
  \leq 
  \frac{900 e^{2/3} d^{2}  \epsilon }{\sigma ^{4}}.
\]
\end{proof}

\begin{lemma}[Distance bound in plane]\label{lem:distanceInPlane}
Let 
 $\vs{\mu }{1}{d}$ be Gaussian measures 
  in $\Reals{d-1}$.
  of standard deviation $\sigma  \leq 1/3 \sqrt{d \ln n}$
  centered at points of norm at most $3$.
Then
\begin{equation}\label{eqn:lemDistanceInPlane}
  \prob{\vs{\bb}{1}{d} \in Q}{\dist{\zzero }{\aff{\vs{\bb}{2}{d}}} <\epsilon }
  \leq 
  \frac{900 e^{2/3} d^{2}  \epsilon }{\sigma ^{4}},
\end{equation}
where $\vs{\bb}{1}{d}$ have density proportional to
\[
  \vol{ \simp{\vs{\bb}{1}{d}}}
  \prod_{i=1}^{d} \mu _{i} (\bb _{i}).
\]
\end{lemma}

\begin{proof}
In Lemma~\ref{lem:aspectRatio}, we will prove 
  it is unlikely that $\bb _{1}$ is close to $\aff{\vs{\bb }{2}{d}}$.
We will exploit this fact by proving that
  it is unlikely that $\zzero $ is much
  closer than
  $\bb _{1}$ to $\aff{\vs{\bb }{2}{d}}$.
We do this by fixing the shape of $\simp{\vs{\bb}{1}{d}}$,
  and then considering slight translations of this simplex.
That is, we make a change of variables to 
\begin{eqnarray*}
  \hh  & = & \frac{1}{d} \sum_{i=1}^{d} \bb _{i}\\
  \dd _{i} & = & \hh - \bb _{i}, \mbox{ for $i \geq 2$}.
\end{eqnarray*}\index{h@$\hh$}\index{ddi@$\dd_{i}$}%
The vectors $\vs{\dd }{2}{d}$ specify the shape of the simplex,
  and $\hh $ specifies its location.
As this change of variables is a linear transformation, 
  its Jacobian is constant.
For convenience, we also define 
  $\dd _{1} = \hh- \bb _{1} = - \sum _{i\geq 2} \dd _{i}$.

It is easy to verify that
\begin{eqnarray*}
 \zzero \in \simp{\vs{\bb}{1}{d}}
 & \iff &
 \hh \in \simp{\vs{\dd}{1}{d}},\\
 \dist{\zzero }{\aff{\vs{\bb}{2}{d}}}
 & = &
 \dist{\hh }{\aff{\vs{\dd}{2}{d}}}, \\
 \dist{\bb_{1} }{\aff{\vs{\bb}{2}{d}}}
 & = &
 \dist{\dd_{1} }{\aff{\vs{\dd}{2}{d}}}, \text{ and }\\
 \vol{\simp{\vs{\bb}{1}{d}}} 
 & = &
 \vol{\simp{\vs{\dd}{1}{d}}}.
\end{eqnarray*}
Note that the relation between $\dd _{1}$
  and  $\vs{\dd }{2}{d}$ guarantees
   $\zzero \in \simp{\vs{\dd }{1}{d}}$
  for all  $\vs{\dd }{2}{d}$.
So, $(\vs{\bb}{1}{d}) \in Q$
 if and only if $(\vs{\dd }{1}{d}) \in Q$
  and $h \in \simp{\vs{\dd}{1}{d}}$.
As $\dd _{1}$ is a function of $\vs{\dd }{2}{d}$,
  we let $Q'$ be the set of $\vs{\dd }{2}{d}$\index{Qp@$Q'$}
  for which $(\vs{\dd }{1}{d}) \in Q$.

So, the left-hand side of (\ref{eqn:lemDistanceInPlane}) equals
\begin{equation*}
  \prob{\substack{
   (\vs{\dd}{2}{d}) \in Q'\\
    \hh \in \simp{\vs{\dd}{1}{d}}
  }}
       {\dist{\hh }{\aff{\vs{\dd }{2}{d}}} <\epsilon }
\end{equation*}
where $\hh , \vs{\dd }{2}{d}$ have density proportional to
\begin{equation}\label{eqn:distanceDensity}
  \vol{ \simp{\vs{\dd}{1}{d}}}
  \prod_{i=1}^{d} \mu _{i} (\hh - \dd _{i}).
\end{equation}

Similarly, Lemma~\ref{lem:aspectRatio} can be seen to imply
\begin{eqnarray}\label{eqn:distance1}
  \prob{\substack{
   (\vs{\dd}{2}{d}) \in Q'\\
    \hh \in \simp{\vs{\dd}{1}{d}}
  }}
  {\dist{\dd_{1} }{\aff{\vs{\dd}{2}{d}}} <\epsilon }
  \leq 
  \left(\frac{\epsilon 3 e^{2/3} d}{\sigma^{2}} \right)^{3}
  \leq 
  \left(\frac{\epsilon 3 e^{2/3} d}{\sigma^{2}} \right)^{2}
\end{eqnarray}
under density proportional to (\ref{eqn:distanceDensity}).
We take advantage of \eqref{eqn:distance1} by proving
\begin{eqnarray}\label{eqn:distance2}
  \max _{\vs{\dd }{2}{d}  \in Q'}
  \prob{\hh  \in \simp{\vs{\dd}{1}{d}}}
   {\frac{
     \dist{\hh }{\aff{\vs{\dd}{2}{d}}}
   }{
     \dist{\dd_{1} }{\aff{\vs{\dd}{2}{d}}}
   }
 < \epsilon }
 < 
  \frac{75 d  \epsilon }{\sigma^{2}},
\end{eqnarray}
where $\hh $ has density proportional to
\[
  \prod_{i=1}^{d} \mu _{i} (\hh - \dd _{i}).
\]

Before proving (\ref{eqn:distance2}), we 
  point out that using Lemma~\ref{lem:comb} to combine (\ref{eqn:distance1})
  and (\ref{eqn:distance2}), we obtain
\[
  \prob{\substack{
   (\vs{\dd}{2}{d}) \in Q'\\
    \hh \in \simp{\vs{\dd}{1}{d}}
  }}
  {\dist{\hh }{\aff{\vs{\dd}{2}{d}} <\epsilon }}
  \leq 
  \frac{900 e^{2/3} d^{2}  \epsilon }{\sigma ^{4}},
\]
from which the lemma follows.

To prove (\ref{eqn:distance2}), we let
\[
  U_{\epsilon } = \setof{\hh \in \simp{\vs{\dd}{1}{d}} :    
   \frac{\dist{\hh }{\aff{\vs{\dd}{2}{d}}}}
        {\dist{\dd_{1}}{\aff{\vs{\dd}{2}{d}}}} \geq 
   \epsilon },
\]\index{U@$U_{\epsilon }$}%
and we set
  $\nu (\hh ) =   \prod_{i=1}^{d} \mu _{i} (\hh - \dd _{i})$.
Under this notation, the probability in (\ref{eqn:distance2}) is
  equal to
\[
  (\nu (U_{0}) - \nu (U_{\epsilon })) / \nu (U_{0}).
\]
To bound this ratio, we construct an isomorphism 
  from $U_{0}$ to $U_{\epsilon }$.
The natural isomorphism, which we denote $\Phi _{\epsilon }$,\index{fhi@$\Phi _{\epsilon }$}%
   \ is the map that contracts the simplex by a
  factor of $(1-\epsilon )$ at $\dd _{1}$.
To use this isomorphism to compare the measures of the sets,
  we use
  the facts that for 
  $\vs{\dd }{1}{d} \in Q$ and $\hh \in \simp{\vs{\dd }{1}{d}}$,
\begin{enumerate}
\item  $\norm{\hh - \dd_{i}} \leq 
  \max _{i,j} \norm{\dd _{i} - \dd _{j}} \leq 4\sqrt{2} $,
  so the distance from $\hh -\dd_{i}$ to the center of its
  distribution is at most $\norm{\hh -\dd_i}+3 \leq 4\sqrt{2}+3$;
\item  $  
    \dist{\hh }{\Phi _{\epsilon } (\hh )}
 \leq 
   \epsilon \max _{i} \dist{\dd _{1}}{\dd _{i}}
 \leq 
   4\sqrt{2} \epsilon  $
\end{enumerate}
to apply Lemma~\ref{lem:smoothGauss}
  to show that for all $\hh \in \simp{\vs{\dd }{1}{d}}$,
\[
     \frac{\mu _{i} (\Phi _{\epsilon } (\hh) -\dd_{i})}
          {\mu _{i} (\hh - \dd_{i})}
 \geq 
   e^{- \frac{3\cdot 4\sqrt{2} (4\sqrt{2}+3)\epsilon}{2\sigma^{2}}}
 = 
   e^{-\frac{(48+18\sqrt{2})\epsilon}{\sigma^{2}}}.
\]

So,
\begin{equation}\label{eqn:distance3}
  \min _{\hh \in \simp{\vs{\dd }{1}{d}}}
  \frac{\nu (\Phi _{\epsilon } (\hh ))}{\nu (\hh )}
=
  \min _{\hh \in \simp{\vs{\dd }{1}{d}}}
  \prod_{i=1}^{d} 
     \frac{\mu _{i} (\Phi _{\epsilon } (\hh) -\dd_{i})}
          {\mu _{i} (\hh - \dd_{i})}
  \geq 
   e^{- \frac{(48+18\sqrt{2}) d \epsilon }{\sigma^{2}}}
  \geq 
    1 - \frac{(48+18\sqrt{2}) d  \epsilon }{\sigma^{2}}.
\end{equation}
As the Jacobian 
\[
 \abs{\frac{\partial \Phi _{\epsilon } (\hh )}{\partial \hh }}
  =
  (1-\epsilon )^{d}
 \geq 
  1 - d \epsilon,
\]
using the change of variables $\xx = \Phi_{\epsilon} (\hh)$ we can compute
\begin{equation}\label{eqn:distance4}
  \nu (U_{\epsilon }) 
  = \int _{\xx \in U_{\epsilon }} \nu (\xx ) \diff{\xx }
  = \int _{\hh \in U_{0}} \nu (\Phi _{\epsilon } (\hh ) )
     \abs{\frac{\partial \Phi _{\epsilon } (\hh )}{\partial \hh }}
     \diff{\hh }
  \geq (1 - d \epsilon ) \int _{\hh \in U_{0}} \nu (\Phi _{\epsilon } (\hh ) )
     \diff{\hh }.
\end{equation}
So,
\begin{align*}
  \frac{\nu (U_{\epsilon })}{\nu (U_{0})}
 &  \geq 
  \frac{
 (1 - d \epsilon ) 
  \int _{\hh \in U_{0}} \nu (\Phi _{\epsilon } (\hh ) ) \diff{\hh }
   }{
  \int _{\hh \in U_{0}} \nu (\hh ) \diff{\hh }
  } & \text{ by \eqref{eqn:distance4}}\\
&
\geq 
  (1 - d \epsilon ) 
  \left(
    \min _{\hh \in \simp{\vs{\dd }{1}{d}}}
    \frac{\nu (\Phi _{\epsilon } (\hh ))}{\nu (\hh )}
  \right)
  \frac{
  \int _{\hh \in U_{0}} \diff{\hh }
   }{
  \int _{\hh \in U_{0}} \diff{\hh }
  }\\
&
\geq 
  (1 - d\epsilon ) 
\left(    1 - \frac{(48+18\sqrt{2}) d \epsilon }{\sigma^{2}} \right)
  & \text{ by \eqref{eqn:distance3}}\\
&
  \geq 
    1 - \frac{75 d \epsilon }{\sigma^{2}},
 & \text{ as $\sigma \leq 1$.}
\end{align*}
\eqref{eqn:distance2} now follows from
  $(\nu (U_{0}) - \nu (U_{\epsilon })) / \nu (U_{0}) 
  < \frac{75 d  \epsilon }{\sigma^{2}}$.
\end{proof}

\begin{figure}[h]
\begin{center}
\setlength{\unitlength}{0.00083333in}
\begingroup\makeatletter\ifx\SetFigFont\undefined%
\gdef\SetFigFont#1#2#3#4#5{%
  \reset@font\fontsize{#1}{#2pt}%
  \fontfamily{#3}\fontseries{#4}\fontshape{#5}%
  \selectfont}%
\fi\endgroup%
{\renewcommand{\dashlinestretch}{30}
\begin{picture}(5034,4327)(0,-10)
\put(1979,1790){\ellipse{122}{122}}
\put(443,438){\ellipse{122}{122}}
\put(3085,438){\ellipse{122}{122}}
\put(4191,438){\ellipse{122}{122}}
\path(443,868)(3085,868)(4191,4063)(443,868)
\path(1979,1790)(1979,1237)
\blacken\thicklines
\path(1929.845,1433.620)(1979.000,1237.000)(2028.155,1433.620)(1929.845,1433.620)
\thinlines
\dashline{90.000}(1979,1790)(1979,868)
\path(3515,1790)(3638,1728)(3638,1483)
	(3761,1421)(3638,1360)(3638,991)(3515,930)
\path(4387,4063)(4510,3940)(4510,3019)
	(4633,2957)(4510,2896)(4510,1876)(4387,1814)
\path(12,438)(4990,438)
\dashline{60.000}(4191,4063)(4191,438)
\dashline{60.000}(3085,868)(3085,438)
\dashline{60.000}(443,868)(443,438)
\put(381,69){\makebox(0,0)[lb]{\smash{{{\SetFigFont{14}{16.8}{\rmdefault}{\mddefault}{\itdefault}$\cc_3$}}}}}
\put(2962,69){\makebox(0,0)[lb]{\smash{{{\SetFigFont{14}{16.8}{\rmdefault}{\mddefault}{\itdefault}$\cc_2$}}}}}
\put(4129,69){\makebox(0,0)[lb]{\smash{{{\SetFigFont{14}{16.8}{\rmdefault}{\mddefault}{\itdefault}$\cc_1$}}}}}
\put(197,1053){\makebox(0,0)[lb]{\smash{{{\SetFigFont{14}{16.8}{\rmdefault}{\mddefault}{\itdefault}$\bb_3$}}}}}
\put(2778,930){\makebox(0,0)[lb]{\smash{{{\SetFigFont{14}{16.8}{\rmdefault}{\mddefault}{\itdefault}$\bb_2$}}}}}
\put(2040,1483){\makebox(0,0)[lb]{\smash{{{\SetFigFont{14}{16.8}{\rmdefault}{\mddefault}{\itdefault}$\ttau$}}}}}
\put(3822,1237){\makebox(0,0)[lb]{\smash{{{\SetFigFont{14}{16.8}{\rmdefault}{\mddefault}{\itdefault}$t$}}}}}
\put(4744,2773){\makebox(0,0)[lb]{\smash{{{\SetFigFont{14}{16.8}{\rmdefault}{\mddefault}{\itdefault}$l$}}}}}
\put(3822,4125){\makebox(0,0)[lb]{\smash{{{\SetFigFont{14}{16.8}{\rmdefault}{\mddefault}{\itdefault}$\bb_1$}}}}}
\end{picture}
}
\end{center}
\caption{The change of variables in Lemma~\ref{lem:aspectRatio}.}
\end{figure}
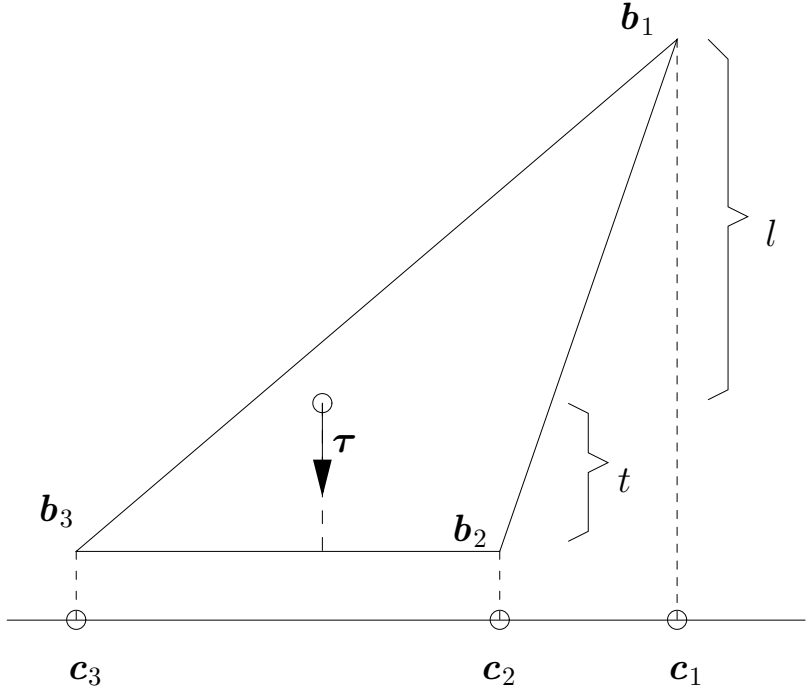

\begin{lemma}[Height of simplex]\label{lem:aspectRatio}
Let 
 $\vs{\mu }{1}{d}$ be Gaussian measures in $\Reals{d-1}$
  of standard deviation $\sigma  \leq 1 / 3 \sqrt{d \ln n}$
  centered at points of norm at most $3 $.
Then
\[
  \prob{\vs{\bb}{1}{d} \in Q}
       {\dist{\bb_{1} }{\aff{\vs{\bb}{2}{d}} <\epsilon }}
  \leq 
  \left(\frac{3 \epsilon  e^{2/3} d }{\sigma^{2}} \right)^{3}
\]
where $\vs{\bb}{1}{d}$ have  density proportional to
\[
  \vol{ \simp{\vs{\bb}{1}{d}}}
  \prod_{i=1}^{d} \mu _{i} (\bb _{i}).
\]
\end{lemma}
\begin{proof}
We begin with a simplifying change of variables.
As in Theorem~\ref{thm:blaschke}, 
  we let
\begin{align*}
 (\vs{\bb}{2}{d}) & 
  = \left(\RR_{\ttau } \cc_{2} + t \ttau , \ldots ,
     \RR_{\ttau } \cc_{d} + t \ttau  \right),
\end{align*}
where 
 $\ttau \in \sphere{d-2}$ and $t \geq 0$ specify  
  the plane through $\vs{\bb}{2}{d}$, and 
  $\vs{\cc }{2}{d} \in \Reals{d-2}$ 
  denote the local coordinates of these points
  on that plane.\index{tau@$\ttau $}\index{t@$t$}\index{cc@$\cc _{i}$}
Recall that the Jacobian of this change of variables
  is $\vol{\simp{\vs{\cc }{2}{d}}}$.
Let $l = - \form{\ttau }{\bb _{1}}$, and let $\cc _{1}$ denote
  the coordinates in $\Reals{d-2}$ of
  the projection of $\bb_{1}$ onto the plane specified by
  $\ttau $ and $t$.\index{l@$l$}%
\ Note that $l \geq 0$.
In this notation, we have
\[
  \dist{\bb_{1} }{\aff{\vs{\bb}{2}{d}}} = l + t.
\]
The Jacobian of the change from $\bb _{1}$ to $(l,\cc _{1})$
  is $1$ as the transformation is just an orthogonal change of coordinates.
The conditions for $(\vs{\bb }{1}{d}) \in Q$ translate into
  the conditions
\begin{enumerate}
\item $\dist{\cc _{i}}{\cc _{j}} \leq 4$ for all 
  $i \not = j$;
\item $(l + t) \leq 4 $; and
\item $\zzero \in \simp{\vs{\bb}{1}{d}}$.
\end{enumerate}
Let $R$ denote the set of $\vs{\cc}{1}{d}$ \index{R@$R$}
  satisfying the first condition.
As the lemma is vacuously true for $\epsilon \geq 4 $,
  we will drop the second condition and note that doing so
  cannot decrease the probability that $(t + l) < \epsilon $.
Thus, our goal is to bound
\begin{equation}\label{eqn:distR}
  \prob{ \ttau , t, l, (\vs{\cc}{1}{d}) \in R}
       {(l + t) < \epsilon },
\end{equation}
  where the variables have density proportional to%
\footnote{While we keep terms such as $\bb_{1}$ in the expression
 of the density, they should be interpreted as functions of
  $\ttau , t, l, \vs{\cc}{1}{d}$.}
\[
  \ind{\zzero \in \simp{\vs{\bb}{1}{d}}}
  \vol{ \simp{\vs{\bb}{1}{d}}}  
  \vol{ \simp{\vs{\cc}{2}{d}}} 
  \prod_{i=1}^{d} \mu _{i} (\bb _{i}).
\]
As $\vol{ \simp{\vs{\bb}{1}{d}}}  
  = (l+t) \vol{ \simp{\vs{\cc}{1}{d}}}/d$, this is the
  same as having density proportional to
\[
  (l + t)
  \ind{\zzero \in \simp{\vs{\bb}{1}{d}}}
  \vol{ \simp{\vs{\cc}{2}{d}}} ^{2}
  \prod_{i=1}^{d} \mu _{i} (\bb _{i}).
\]

Under a suitable system of coordinates, we can express
  $\bb _{1} = (-l, \cc _{1})$ and
  $\bb _{i} = (t, \cc _{i})$ for $i \geq 2$.
The key idea of this proof is that
  multiplying the first coordinates of these points by
  a constant does not change whether or not
  $\zzero \in \simp{\vs{\bb}{1}{d}}$;
 so,  we can determine  whether 
  $\zzero \in \simp{\vs{\bb}{1}{d}}$
  from the data $(l/t, \vs{\cc}{1}{d})$.
Thus, we will introduce a new variable
  $\alpha $, set $l = \alpha t$,\index{alp@$\alpha $}
  and let $S$ denote the set of\index{s@$S$}
  $(\alpha , \vs{\cc}{1}{d})$ for which
  $\zzero \in \simp{\vs{\bb}{1}{d}}$
  and $(\vs{\cc}{1}{d}) \in R$.
This change of variables from $l$  to $\alpha $ 
  incurs a Jacobian of $\frac{\partial l}{\partial \alpha } = t$, 
  so \eqref{eqn:distR} equals
\[
  \prob{ \ttau , t, (\alpha , \vs{\cc}{1}{d}) \in S}
       {(1 + \alpha )t < \epsilon },
\]
where the variables have density proportional to 
\[
   t^{2} (1 + \alpha )
  \vol{ \simp{\vs{\cc}{2}{d}}} ^{2}
   \mu _{1} (-\alpha t, \cc _{1})
  \prod_{i=2}^{d} \mu _{i} (t, \cc _{i}).
\]
We upper bound this probability by
\begin{eqnarray*}
 \max_{ \ttau , (\alpha , \vs{\cc}{1}{d}) \in S}
  \prob{t}{(1 + \alpha )t < \epsilon }
& \leq &
 \max_{ \ttau , (\alpha , \vs{\cc}{1}{d}) \in S}
  \prob{t}{\max (1,\alpha ) t < \epsilon },
\end{eqnarray*}
where $t$ has density proportional to
\[
  t^{2}
  \mu _{1} (-\alpha t, \cc _{1})
  \prod_{i=2}^{d} \mu _{i} (t, \cc _{i}).
\]
For $\vs{\cc}{1}{d}$ fixed, 
  the points $(-\alpha t,\cc_{1}),(t,\cc_{2}), \ldots, (t,\cc_{d})$
  become univariate Gaussians of standard deviation $\sigma $ 
  and mean of absolute value at most $3$.
Let $t_{0} = \sigma^{2}/ (3 \max (1,\alpha ) d )$.
Then, for $t$ in the range $[0,t_{0}]$,
  $-\alpha t$ is at most $3 + \alpha t_{0}$  
  from the mean of the first distribution and
  $t$ is at most $3 + t_{0}$ from the means of the other
  distributions.
We will now observe that if $t$ is restricted to
  a sufficiently small domain, then the densities
  of these Gaussians will have bounded variation.
In particular, Lemma~\ref{lem:smoothGauss} implies that
\begin{align*}
\frac{
  \max _{t \in [0, t_{0}]} 
  \mu _{1} (-\alpha t, \cc _{1})
  \prod_{i=2}^{d} \mu _{i} (t, \cc _{i})
}{
  \min _{t \in [0, t_{0}]} 
  \mu _{1} (-\alpha t, \cc _{1})
  \prod_{i=2}^{d} \mu _{i} (t, \cc _{i})
}
& \leq  
   e^{3 (3+ \alpha t_{0}) \alpha t_{0} / 2 \sigma^{2}}
  \prod_{i=2}^{d}
   e^{3 (3 +t_{0}) t_{0} / 2 \sigma^{2}}\\
& \leq 
   e^{9 \alpha t_{0}/ 2 \sigma^{2} }
  \left(  \prod_{i=2}^{d}
   e^{9 t_{0}/ 2 \sigma^{2}} \right)
\cdot 
   e^{3 (\alpha t_{0})^{2} / 2 \sigma^{2}}
 \left(  \prod_{i=2}^{d}
   e^{3 t_{0}^{2} / 2 \sigma^{2}} \right)\\
& \leq 
   e^{3 / 2 d }
  \left(  \prod_{i=2}^{d}
   e^{3 / 2 d} \right)
\cdot 
   e^{\sigma^{2}/6d^{2}}
 \left(  \prod_{i=2}^{d}
   e^{\sigma^{2}/6d^{2}} \right)\\
& \leq 
   e^{3 / 2}
\cdot 
   e^{1 / 6 d}
\\
& \leq  
   e^{2}.
\end{align*}
Thus, we can now apply Lemma~\ref{lem:minMaxIntegral}   
  to show that
\[
   \prob{t}{t < \epsilon }
  < 
 e^{2}
  \left( \frac{3 \epsilon (\max (1,\alpha ) d}{\sigma^{2}} \right)^{3},
\]
from which we conclude
\[
   \prob{t}{\max (1,\alpha ) t < \epsilon }
  < 
  \left(\frac{ 3 \epsilon  e^{2/3} d}{\sigma^{2}} \right)^{3}.
\]
\end{proof}

\subsection{Angle of $\qq$ to $\oomega$}\label{sec:angle}
\begin{lemma}[Angle of incidence]\label{lem:angle}
Let $d \geq 3$ and $n > d$.
Let $\vs{\mu }{1}{n}$ be 
  Gaussian densities in $\Reals{d}$ of standard deviation $\sigma $
  centered at points of norm
  at most 1 in $\Reals{d}$.
Let $s \leq 2$ and
  let $(\vs{\bb }{1}{d}) \in Q$.
Then,
\begin{equation}\label{eqn:angleMain}
 \prob{\oomega }{\form{\oomega }{\qq} < \epsilon } < 
  \left(\frac{340 \epsilon n }{\sigma^{2} } \right)^{2},
\end{equation}
where $\oomega $ has density proportional to
\[
  \form{\oomega }{\qq}
  \left(  
    \prod_{j > d} 
     \int _{\aa_{j}}
     \ind{\form{\oomega}{\aa_{j}} \leq 
       s \form{\oomega}{\qq}}
     \mu _{j} (\aa_{j}) \diff{\aa_{j}}
  \right)
  \prod_{i=1}^{d} \mu _{i} (\RR_{\oomega} \bb _{i} + s \qq ).
\]
\end{lemma}
\begin{proof}
First note that the conditions for $(\vs{\bb}{1}{d})$
  to be in $Q$ imply that for $1\leq i\leq d$,  $\bb _{i}$ has norm
  at most $\sqrt{(4)^{2} + (4)^{2}} 
  = 4 \sqrt{2} $ by properties $(1)$, $(3)$ and $(4)$
  of $Q$.

As in Proposition~\ref{pro:covC}, we change
  $\oomega $ to $(c, \ppsi )$, where
  $c = \form{\oomega }{\qq}$ and
  $\ppsi \in \sphere{d-2}$.\index{c@$\ppsi $}\index{c@$c$}%
The Jacobian of this change of variables is
\[
  (1-c^{2})^{(d-3)/2}.
\]
In these variables, the bound follows from
  Lemma~\ref{lem:angleC}.
\end{proof}

\begin{lemma}[Angle of incidence, II]\label{lem:angleC}
Let $d \geq 3$ and $n > d$.
Let $\vs{\mu }{d+1}{n}$ be 
  Gaussian densities in $\Reals{d}$ of standard deviation $\sigma $
  centered at points of norm
  at most 1 in $\Reals{d}$.
Let  $s \leq 2$,
  and let
  $\vs{\bb}{1}{d}$ each have norm at most $4 \sqrt{2} $.
Let $\ppsi \in \sphere{d-2}$.
Then
\[
  \prob{}{c < \epsilon } <   
  \left(\frac{340 \epsilon n }{\sigma^{2} } \right)^{2}
,
\]
where $c$ has density proportional to
\begin{equation}\label{eqn:angleCdensity}
  (1-c^{2})^{(d-3)/2} \cdot c \cdot
  \left(  
    \prod_{j > d} 
     \int _{\aa_{j}}
     \ind{\form{\oomega_{\ppsi, c}}{\aa_{j}} \leq 
       s \form{\oomega_{\ppsi, c} }{\qq}}
     \mu _{j} (\aa_{j}) \diff{\aa_{j}}
  \right)
  \prod_{i=1}^{d} \mu _{i} (\RR_{\oomega_{\ppsi, c}} \bb _{i} + s \qq )
\end{equation}
\end{lemma}
\begin{proof}
Let 
\begin{align*}
 \nu _{1} (c) & = (1-c^{2})^{(d-3)/2}, \\
 \nu _{2} (c) & = 
    \prod_{j > d} 
     \int _{\aa_{j}}
     \ind{\form{\oomega_{\ppsi, c}}{\aa_{j}} \leq 
       s \form{\oomega_{\ppsi, c} }{\qq}}
     \mu _{j} (\aa_{j}) \diff{\aa_{j}}, \text{ and }\\
 \nu _{3} (c) & =
  \prod_{i=1}^{d} \mu _{i} (\RR_{\oomega_{\ppsi, c}} \bb _{i} + s \qq ).
\end{align*}
Then, the density of $c$ is proportional to
\[
 \eqref{eqn:angleCdensity} = 
  c \cdot \nu _{1} (c) \nu _{2} (c) \nu _{3} (c).
\]

Let 
\begin{equation}\label{eqn:angleC0}
  c_{0} = \frac{\sigma^{2} }{240 n}.
\end{equation}
We will show that, for
 $c$ between $0$ and $c_{0}$, the density will vary by a factor
 no greater than 2.
We begin by letting $\theta _{0} = \pi /2 - \arccos (c_{0})$, and
  noticing that
  a simple plot of the $\arccos$ function reveals $c_{0} < 1/26$ implies
\begin{equation}\label{eqn:angleTheta0}
  \theta _{0} \leq 1.001 c_{0}.
\end{equation}
So, as $c$ varies in the range $[0,c_{0}]$,
  $\oomega_{\ppsi, c}$ travels in an arc of angle
  at most $\theta _{0}$ and therefore travels a distance
  at most $\theta _{0}$.
As $c = \form{\qq}{\oomega_{\ppsi, c}}$,
  we can apply Lemma~\ref{lem:angleUnderPlanes} to show
\begin{align}\label{eqn:angleC1}
\frac{
 \min _{0 \leq c \leq c_{0}}
  \nu _{2} (c)
}{
 \max _{0 \leq c \leq c_{0}}
  \nu _{2} (c)
}
 \geq 
  1 - \frac{8 n (1 + s) \theta _{0}}
       { 3 \sigma^{2}} 
 \geq 
  1 - \frac{24 n \theta _{0}}
       { 3 \sigma^{2}}
 \geq 
 1 - \frac{1.001 }
       {30}, 
\end{align}
by \eqref{eqn:angleC0} and \eqref{eqn:angleTheta0}.

We similarly note that as $c$ varies between $0$ and $c_{0}$,
  the point $\RR_{\oomega_{\ppsi, c}} \bb _{i} + s \qq$ moves
  a distance of at most 
\[
  \theta _{0}\norm{\bb _{i}} \leq 4 \sqrt{2} \theta _{0} .
\]
As this point is at distance at most 
\[
  1 + s + \norm{\bb _{i}} \leq 4\sqrt{2} + 3 
\]
from the center of $\mu _{i}$, Lemma~\ref{lem:smoothGauss} implies
\[
  \frac{
  \min _{0 \leq c \leq c_{0}}
  \mu _{i} (\RR_{\oomega_{\ppsi, c}} \bb _{i} + s \qq )
}{
  \max _{0 \leq c \leq c_{0}}
  \mu _{i} (\RR_{\oomega_{\ppsi, c}} \bb _{i} + s \qq )
}
 \geq 
  e^{-\left(3 (4 \sqrt{2} + 3 ) 4 \sqrt{2} \theta _{0} \right)/ 2 \sigma^{2}}
 \geq 
  e^{-147 \theta _{0}/\sigma^{2}}.
\]
So,
\begin{align}
\frac{
  \min _{0 \leq c \leq c_{0}}
  \nu _{3} (c)
}{
  \max _{0 \leq c \leq c_{0}}
  \nu _{3} (c)
}
 \geq 
     e^{-147 d  \theta _{0}/\sigma^{2}}
 \geq 
     e^{-148/240},
\label{eqn:angleC2}
\end{align}
by \eqref{eqn:angleC0} and \eqref{eqn:angleTheta0} and $d\leq  n$.

Finally, we note that 
\begin{equation}\label{eqn:angleC3}
1 \geq 
 \nu _{1} (c) 
= 
 (1-c^{2})^{(d-3)/2}
 \geq 
 (1 - 1/26d)^{(d-3)/2}
 \geq 
  \left(1 - \frac{1}{52} \right).
\end{equation}
So, combining equations \eqref{eqn:angleC1},
  \eqref{eqn:angleC2}, and \eqref{eqn:angleC3}, we obtain

\begin{equation*}
\frac{
\min _{0 \leq c \leq c_{0}}
 \nu _{1} (c) \nu _{2} (c) \nu _{3} (c)
}{
\max _{0 \leq c \leq c_{0}}
 \nu _{1} (c) \nu _{2} (c) \nu _{3} (c)
}
 \geq 
 \left(1 - \frac{1}{52} \right)
  e^{-\frac{148}{240}}
 \left( 1 - \frac{1.001 }{30} \right)
 \geq 
  1/2.
\end{equation*}

We conclude by using Lemma~\ref{lem:minMaxIntegral} to show
\[
  \prob{c}{c < \epsilon }
  \leq 
  2 (\epsilon / c_{0})^{2} 
= 
  2 \left(\frac{240 \epsilon n }{\sigma^{2} } \right)^{2}
\leq 
  \left(\frac{340 \epsilon n }{\sigma^{2} } \right)^{2}.
\]
\end{proof}

\begin{lemma}[Points under plane]\label{lem:angleUnderPlanes}
For $n > d$,
 let $\vs{\mu }{d+1}{n}$ be 
  Gaussian distributions in $\Reals{d}$ of standard deviation $\sigma $
  centered at points of norm at most 1.
Let $s \geq 0$ and let $\oomega _{1}$ 
  and $\oomega _{2}$ be unit vectors
  such that $\form{\oomega _{1}}{\qq}$
  and $\form{\oomega _{2}}{\qq}$  
  are non-negative.
Then, 
\[
\frac{
    \prod_{j > d} 
     \int _{\aa_{j}}
     \ind{\form{\oomega_{2} }{\aa_{j}} \leq s \form{\oomega_{2} }{\qq}}
     \mu _{j} (\aa_{j}) \diff{\aa_{j}}
}{
    \prod_{j > d} 
     \int _{\aa_{j}}
     \ind{\form{\oomega_{1} }{\aa_{j}} \leq s \form{\oomega_{1} }{\qq}}
     \mu _{j} (\aa_{j}) \diff{\aa_{j}}
}
  \geq 
 1 - \frac{8 n (1 + s) \norm{\oomega _{1} - \oomega _{2}}}
       { 3 \sigma^{2}}.
\]
\end{lemma}
\begin{proof}
As the integrals in the statement of the lemma are just
  the integrals of Gaussian measures over half-spaces, they can
  be reduced to univariate integrals.
If $\mu _{j}$ is centered at $\aao _{j}$, then
\begin{align*}
 \int _{\aa_{j}}
     \ind{\form{\oomega_{1} }{\aa_{j}} \leq s \form{\oomega_{1} }{\qq}}
     \mu _{j} (\aa_{j}) \diff{\aa_{j}}
& =
 \left(\frac{1}{\sqrt{2 \pi } \sigma } \right)^{d}
 \int _{\aa_{j}}
   \ind{\form{\oomega_{1} }{\aa_{j}} \leq s \form{\oomega_{1} }{\qq}}
   e^{-\norm{\aa_{j} - \aao_{j}}^{2}/ 2 \sigma^{2}} \diff{\aa_{j}}\\
& = 
 \left(\frac{1}{\sqrt{2 \pi } \sigma } \right)^{d}
 \int _{\gg_{j}}
   \ind{\form{\oomega_{1} }{\gg_{j} + \aao _{j}}
           \leq s \form{\oomega_{1} }{\qq}}
   e^{-\norm{\gg_{j}}^{2}/ 2 \sigma^{2}}
   \diff{\gg_{j}},\\
\intertext{(setting $\gg _{j} = \aa_{j} - \aao _{j}$)}
& = 
 \left(\frac{1}{\sqrt{2 \pi } \sigma } \right)^{d}
 \int _{\gg_{j}}
   \ind{\form{\oomega_{1} }{\gg _{j}}
              \leq \form{\oomega_{1} }{s \qq - \aao _{j}}}
   e^{-\norm{\gg_{j}}^{2}/ 2 \sigma^{2}}
   \diff{\gg_{j}}\\
& = 
 \frac{1}{\sqrt{2 \pi } \sigma }
 \int_{t = -\infty }
     ^{t = \form{\oomega_{1} }{s \qq - \aao _{j}}}
   e^{-t^{2}/ 2 \sigma^{2}}
   \diff{t}\\
\intertext{(by Proposition~\ref{pro:gauss1d})}
& = 
 \frac{1}{\sqrt{2 \pi } \sigma }
\int_{t = -\form{\oomega_{1} }{s \qq - \aao _{j}}}
     ^{t = \infty}
    e^{-t^{2}/ 2 \sigma^{2}}
    \diff{t}.
\end{align*}
As $\norm{\aao_{j}} \leq 1$,
we know
\begin{equation}\label{eqn:angleUnderPlanes1}
 -\form{\oomega_{1} }{s \qq - \aao_{j}}
\ = \
-\form{\oomega _{1}}{s \qq}
+\form{\oomega _{1}}{\aao _{j}}
\ \leq \
\form{\oomega _{1}}{\aao _{j}}
\  \leq \
 1
\end{equation}
Similarly, 
\begin{align}
  \big|  -\form{\oomega_{1} }{s \qq - \aao_{j}}
 +\form{\oomega_{2} }{s \qq - \aao_{j}} \big|
 &  = 
 \big| -\form{\oomega_{1} - \oomega _{2} }{s \qq - \aao_{j}} \big|
 \nonumber \\
& \leq 
  \norm{\oomega _{1} - \oomega _{2}}
  \norm{s \qq - \aao_{j}}\\
& \leq 
 \norm{\oomega _{1} - \oomega _{2}}  (s+1) . \label{eqn:angleUnderPlanes2}
\end{align}
Thus, by applying Lemma~\ref{lem:gaussRelTail} to  
  \eqref{eqn:angleUnderPlanes1} and
  \eqref{eqn:angleUnderPlanes2}, we obtain
\begin{align*}
\frac{
     \int _{\aa_{j}}
     \ind{\form{\oomega_{2} }{\aa_{j}} \leq s \form{\oomega_{2} }{\qq}}
     \mu _{j} (\aa_{j}) \diff{\aa_{j}}
}{
     \int _{\aa_{j}}
     \ind{\form{\oomega_{1} }{\aa_{j}} \leq s \form{\oomega_{1} }{\qq}}
     \mu _{j} (\aa_{j}) \diff{\aa_{j}}
}
& = 
\frac{
\int_{t = -\form{\oomega_{2} }{s \qq - \aao_{j}}}
     ^{t = \infty}
    e^{-t^{2}/ 2 \sigma^{2}}
    \diff{t}.
}{
\int_{t = -\form{\oomega_{1} }{s \qq - \aao_{j}}}
     ^{t = \infty}
    e^{-t^{2}/ 2 \sigma^{2}}
    \diff{t}.
}\\
&\geq 
  \left(1 - \frac{8(1+s) \norm{\oomega _{1} - \oomega _{2}}}
                 {3 \sigma^{2}} \right).
\end{align*}
Thus,
\begin{align*}
\frac{
    \prod_{j > d} 
     \int _{\aa_{j}}
     \ind{\form{\oomega_{2} }{\aa_{j}} \leq s \form{\oomega_{2} }{\qq}}
     \mu _{j} (\aa_{j}) \diff{\aa_{j}}
}{
    \prod_{j > d} 
     \int _{\aa_{j}}
     \ind{\form{\oomega_{1} }{\aa_{j}} \leq s \form{\oomega_{1} }{\qq}}
     \mu _{j} (\aa_{j}) \diff{\aa_{j}}
}
& \geq 
  \left(1 - \frac{8(1+s) \norm{\oomega _{1} - \oomega _{2}}}
                 {3 \sigma^{2}} \right)^{n - d}\\
& \geq 
  \left(1 - \frac{8 n (1+s) \norm{\oomega _{1} - \oomega _{2}}}
                 {3 \sigma^{2}} \right).
\end{align*}
\end{proof}

\subsection{Extending the shadow bound} \label{ssec:extension}

In this section, we relax the restrictions made in
  the statement of Theorem~\ref{thm:shadow}.
The extensions of Theorem~\ref{thm:shadow} are needed
  in the proof of Theorem~\ref{thm:twoPhaseMain}.

We begin by removing the restrictions on where the distributions
  are centered in the shadow bound.

\begin{corollary}[$\norm{\aa_{i}}$ free]\label{cor:freeA}
Let $\zz$ and $\tt $ be unit vectors and 
  let 
  $\vs{\aa}{1}{n}$ be Gaussian random vectors in $\Reals{d}$
  of standard deviation $\sigma  \leq 1/3 \sqrt{d \ln n}$
  centered at points
  $\vs{\aao}{1}{n}$.
Then,
\[
 \expec{}{\shadow{{\zz,\tt}}{\vs{\aa }{1}{n}}} \leq 
  \calD \left(d,n,\frac{\sigma }{\max \left(1, \max_{i}\norm{\aao } \right)} \right)
\]
where $\calD  (d,n, \sigma) $ is as given in Theorem \ref{thm:shadow}.
\end{corollary}
\begin{proof}
Let $k = \max_{i} \norm{\aao_{i}}$.
Assume without loss of generality that $k \geq 1$,
 and let
  $\bb_{i} = \aa_{i} / k$ for all $i$.
Then, $\bb_{i}$ is a Gaussian random variable of
  standard deviation $(\sigma/k)$ centered at a point of norm at most $1$.
So, Theorem~\ref{thm:shadow} implies
\[
 \expec{}{\shadow{{\zz,\tt}}{\vs{\bb }{1}{n}}} \leq 
  \calD \left(d,n,\frac{\sigma }{k} \right).
\]
On the other hand,
  the shadow of the polytope defined by the $\bb_{i}$s
  can be seen to be a dilation of the polytope defined
  by the $\aa_{i}$s: 
  the division of the $\bb_{i}$s by a factor
  of $k$ is equivalent to the multiplication of $\xx$ by $k$.
So, we may conclude that for all $\vs{\aa}{1}{n}$,
\[
 \sizeof{\shadow{{\zz,\tt}}{\vs{\aa }{1}{n}}} =
 \sizeof{\shadow{{\zz,\tt}}{\vs{\bb }{1}{n}}}.
\]
\end{proof}

\begin{corollary}[Gaussians free]\label{cor:gaussiansFree}
Let $\zz$ and $\tt $ be unit vectors and 
  let 
  $\vs{\aa}{1}{n}$ be Gaussian random vectors in $\Reals{d}$
  with covariance matrices $\vs{\MM}{1}{n}$
  centered at points $\vs{\aao}{1}{n}$, respectively.
If  the eigenvalues of each $\MM _{i}$ lie between
  $\sigma^{2}$ and $1 / 9 d \ln n$, then
\[
 \expec{}{\shadow{{\zz,\tt}}{\vs{\aa }{1}{n}}} \leq 
  \calD \left(d,n,\frac{\sigma }{1 + \max_{i}\norm{\aao }} \right) + 1
\]
where $\calD  (d,n, \sigma) $ is as given in Theorem \ref{thm:shadow}.
\end{corollary}
\begin{proof}
By Proposition~\ref{pro:gaussSum}, each $\aa_{i}$ can be expressed
  as 
\[
  \aa_{i} = \aao_{i} + \gg_{i} + \ggt_{i},
\]
where $\ggt_{i}$ is a Gaussian random vector of standard deviation
  $\sigma$ centered at the origin and
  $\gg_{i}$ is a Gaussian random vector centered at the origin
  with covariance matrix $\MM^{0}_{i} = \MM _{i} - \sigma^{2}I$, each of whose eigenvalues
  is at most $1/9 d \ln n$.
Let $\aat_{i} = \aao_{i} + \gg_{i}$.
If $\norm{\aat_{i}} \leq 1 + \norm{\aao_{i}}$, for all $i$, then we
  can apply Corollary~\ref{cor:freeA} to show
\[
 \expec{\vs{\ggt}{1}{n}}
  {\shadow{\zz,\tt}{\vs{\aa }{1}{n}}} 
\leq 
  \calD \left(d,n,\frac{\sigma }{\max \left(1, \max_{i}\norm{\aat } \right)} \right)
\leq 
  \calD \left(d,n,\frac{\sigma }{1 + \max_{i}\norm{\aao }} \right).
\]
On the other hand, Corollary~\ref{cor:chiSquare} implies
\[
  \prob{\vs{\gg}{1}{n}}
  {\exists i : \norm{\aat_{i}} > 1 + \norm{\aao_{i}}}
 \leq 0.0015 \binom{n}{d}^{-1}
\]
So, using Lemma~\ref{lem:probXYA} and
  $\shadow{\zz,\tt}{\vs{\aa }{1}{n}} \leq \binom{n}{d}$, we
  can show
\[
  \expec{\vs{\ggt}{1}{n}}
   {\expec{\vs{\gg}{1}{n}}{\shadow{{\zz,\tt}}{\vs{\aa }{1}{n}}}}
 \leq 
  \calD \left(d,n,\frac{\sigma }{1 + \max_{i}\norm{\aao }} \right) 
  + 
  1.
\]
from which the Corollary follows.
\end{proof}

\begin{corollary}[$y_{i}$ free]\label{cor:freeYi}
Let $\yy \in \Reals{n}$ be a positive vector.
Let $\zz$ and $\tt $ be unit vectors and 
  let 
  $\vs{\aa}{1}{n}$ be Gaussian random vectors in $\Reals{d}$
  with covariance matrices $\vs{\MM}{1}{n}$
  centered at points $\vs{\aao}{1}{n}$, respectively.
If  the eigenvalues of each $\MM _{i}$ lie between
  $\sigma^{2}$ and $1 / 9 d \ln n$, then
\[
 \expec{}{\shadow{{\zz,\tt}}{\vs{\aa }{1}{n}} ; \yy} \leq 
  \calD \left(d,n,\frac{\sigma }
  {(1 + \max _{i} \norm{\aao _{i}}) (\max_{i} y_{i}) / (\min_{i} y_{i}) } \right) + 1
\]
where $\calD  (d,n, \sigma) $ is as given in Theorem \ref{thm:shadow}.
\end{corollary}

\begin{proof}
Nothing in the statement is changed if we rescale the $y_{i}$s.
So, assume without loss of generality that $\min_{i} y_{i} = 1$.

Let $\bb_{i} = \aa_{i} / y_{i}$.
Then $\bb_{i}$ is a Gaussian random vector with covariance matrix
  $\MM_{i} / y_{i}^{2}$ 
  centered at a point of norm at most $\norm{\aa _{i}} / y_{i} \leq \norm{\aa _{i}}$.
Then, the eigenvalues of each $\MM_{i}$ lie between   
  $\sigma^{2} / y_{i}^{2}$ and 
  $1 / (9 d \ln n y_{i}^{2})  \leq 1 / 9 d \ln n$,
 so we may complete the proof by applying Corollary~\ref{cor:gaussiansFree}.
\end{proof}

% Local Variables: ***
% TeX-master:"shadow.tex" ***
% End: ***

%\newpage
\section{Smoothed Analysis of a Two-Phase Simplex Algorithm}
\label{sec:phaseI}

In this section, we will analyze the smoothed
   complexity of the {two-phase shadow-vertex simplex method}
   introduced in Section \ref{sec:introSVM2phase}.
The analysis of the algorithm will use as a black-box 
   the bound on the expected
   sizes of shadows proved in the previous section.
However, the analysis is not immediate from this bound.

\setcounter{theorem}{0}

The most obvious difficulty in applying the shadow bound
  to the analysis of an algorithm is that, in the statement of
  the shadow bound, the plane onto which the polytope was projected
  to form the shadow was fixed, and unrelated to the
  data defining the polytope.
However, in the analysis of the shadow-vertex algorithm, 
  the plane onto which the polytope is projected will
  necessarily depend upon data defining the linear program.
This is the dominant complication in the analysis of 
  the number of steps taken to solve $LP'$.

Another obstacle will stem from the fact that, in 
  the analysis of $LP^{+}$,
  we need to consider the expected sizes of shadows
  of the convex hulls of points
  of the form $\aap _{i}/y_{i}^{+}$,
  which do not have a Gaussian distribution.
In our analysis of $LP^{+}$, we essentially handle
  this complication by demonstrating that in almost every
  small region the distribution can be approximated by
  some Gaussian distribution.

The last issue we need to address is that if
  $\smin{\AA _{I}}$ is too small, then the resulting values
  for $y'_{i}$ and $y^{+}_{i}$ can be too large.
In Section~\ref{sec:phaseIManyGood}
  we resolve this problem by proving that one of $3nd\ln n$
  randomly chosen $I$ will have
  reasonable  $\smin{\AA_{I}}$
  with very high probability.
Having a reasonable $\smin{\AA_{I}}$ is also essential
  for the analysis of $LP'$.

As our two-phase shadow-vertex simplex algorithm is randomized,
  we will measure its expected complexity on each input.
For an input linear program specified by $\AA$, $\yy $ and $\zz$,
  we let
\[
  \calC (\AA , \yy ,\zz)
\]\index{C@$\calC$}%
denote the expected number of simplex steps taken by
  the algorithm on input
  $(\AA ,\yy ,\zz)$.
As this expectation is taken over the choices for
  $\calI $ and $\aalpha$,
  and can be divided into the number of steps taken
  to solve $LP^{+}$ and $LP'$,
  we introduce the functions
\[
  \calS'_{\zz} (\AA , \yy ,\calI ,\aalpha),
\]\index{Sprime@$\calS'$}%
to denote the number of simplex steps taken by the algorithm
  in step~(5) to solve $LP'$ for a given $\AA$, $\yy$, $\calI $
  and $\aalpha$,
and
\[
  \calSp_{\zz} (\AA , \yy ,\calI) + 2
\]\index{Splus@$\calSp$}%
to denote the number of simplex steps\footnote{
The seemingly odd appearance of $+2$ in this definition
  is explained by~\ref{pro:lp++}.
}
  taken by the 
  algorithm
  in step~(7) to solve $LP^{+}$ for a given $\AA$, $\yy$ and $\calI $.
We note that the complexity of the second phase does not depend
  upon $\aalpha$, however it does depend upon $\calI $ as $\calI$ affects
  the choice of $\kappa $ and $M$.
We have
\[
 \calC (\AA , \yy ,\zz)
 \leq  
  \expec{\calI , \aalpha}{\calS'_{\zz} (\AA , \yy ,\calI ,\aalpha)}
  + 
  \expec{\calI , \aalpha}{\calSp_{\zz} (\AA , \yy ,\calI ,\aalpha)}
 + 2.
\]

\begin{theorem}[Main]\label{thm:twoPhaseMain}
There exists a polynomial $\calP$ and a constant $\sigma_{0}$
  such that for every
  $n > d \geq 3$,
   $\orig{\AA} = [\vs{\orig{\aa}}{1}{n}]\in \Reals{n\times d}$,
   $\orig{\yy} \in \Reals{n}$ and $\zz \in\Reals{d}$,
 and $\sigma > 0$,
\[
\expec{\AA ,\yy }{\calC (\AA ,\yy , \zz)}
\leq  
\min \left(\calP (d, n, 1/\min (\sigma ,\sigma_{0})),
 \binom{n}{d} + \binom{n}{d+1} + 2 \right),
\]
where
  $\AA$ is a Gaussian random matrix centered at $\orig{\AA}$
  of standard deviation $\sigma \max _{i}\norm{(\orig{y}_{i}, \orig{\aa}_{i})}$,
  and $\yy$ is a Gaussian random vector centered at
  $\orig{\yy}$ of standard deviation 
  $\sigma \max _{i}\norm{(\orig{y}_{i}, \orig{\aa}_{i})}$.
\end{theorem}
\begin{proof}
We first observe that the behavior of the algorithm is
  unchanged if one multiplies $\AA$ and $\yy$ by a power of two.
That is,
\[
   \calC (\AA ,\yy ,\zz) =  \calC (2^{k}\AA ,2^{k}\yy , \zz),
\]
for any integer $k$.  
When $\AA$ and $\yy$ are Gaussian random variables
  centered at $\AAo$ and $\orig{\yy}$ of standard
  deviation $\sigma \max _{i}\norm{(\orig{y}_{i}, \orig{\aa}_{i})}$,
 $2^{k} \AA$ and $2^{k} \yy$ are Gaussian random variables
  centered at $2^{k} \AAo$ and $2^{k} \orig{\yy}$
  of standard deviation
  $\sigma \max _{i}\norm{(2^{k}\orig{y}_{i}, 2^{k}\orig{\aa}_{i})}$.
Accordingly, we may assume without loss of generality
  in our analysis that
  $\max _{i}\norm{(\orig{y}_{i}, \orig{\aa}_{i})} \in (1/2,1]$.

The Theorem now follows from
  Proposition~\ref{pro:trivial} and
  Lemmas~\ref{lem:LP'} and~\ref{lem:LP+}.

%Under this assumption, we set 
%  $\sigma ' = \sigma  \max _{i}\norm{(\orig{y}_{i}, \orig{\aa}_{i})}$,
%  and apply Lemma~\ref{lem:LP'} to 
%  show
%\[
%  \expec{\calI , \aalpha}{\calS'_{\zz} (\AA , \yy ,\calI ,\aalpha)}
% \leq \cdots (\sigma ')
%\]
%We then apply Lemma~\ref{lem:LP+}  to show
%\[
%  \expec{\calI , \aalpha}{\calSp_{\zz} (\AA , \yy ,\calI ,\aalpha)}
% \leq \cdots (\sigma ').
%\]
%Combining these two bounds, we obtain 
%\begin{align*}
%  \expec{\AA ,\yy }{\calC (\AA ,\yy , \zz)}
%& \leq 
%  \expec{\calI , \aalpha}{\calS'_{\zz} (\AA , \yy ,\calI ,\aalpha)}
% +
%  \expec{\calI , \aalpha}{\calSp_{\zz} (\AA , \yy ,\calI ,\aalpha)}
% + 2\\
%& \leq  \cdots 
%\end{align*}
\end{proof}

Before proceeding with the proof of Theorem~\ref{thm:twoPhaseMain},
  we state a trivial upper bound on $\calS'$ and $\calSp$:

\begin{proposition}[trivial shadow bounds]\label{pro:trivial}
For all $\AA$, $\yy$, $\zz$, $\calI$ and $\aalpha$:
\[
\calS'_{\zz} (\AA , \yy ,\calI ,\aalpha) \leq 
  \binom{n}{d}
\qquad \text{ and } \qquad 
\calSp_{\zz} (\AA , \yy ,\calI ,\aalpha) \leq 
  \binom{n}{d+1}.
\]
\end{proposition}
\begin{proof}
The bound on $\calS'$ follows from the fact that
  there are  $\binom{n}{d}$ $d$-subsets of $[n]$.
The bound on $\calSp$ follows from the observation in
  Lemma~\ref{pro:lp++} that the number of steps taken
  by the second phase is at most $2$ plus
  the number of $(d+1)$-subsets of $[n]$.
\end{proof}

% Local Variables: ***
% TeX-master:"shadow.tex" ***
% End: ***

\subsection{Many Good Choices}\label{sec:phaseIManyGood}

For a Gaussian random $d$-by-$d$ matrix $(\vs{\aa}{1}{d})$, it is
  possible to show that the probability that
  the smallest singular value of $(\vs{\aa}{1}{d})$
  is less than $\epsilon$
  is at most $O (d^{1/2} \epsilon)$.
In this section, we consider the 
  probability that almost all of the $d$-by-$d$
  minors of a $d$-by-$n$ matrix $(\vs{\aa}{1}{n})$
  have small singular value.
If the events for different minors were independent,
  then the proof would be straightforward.
However, distinct minors may have significant overlap.
While we believe stronger concentration results should be obtainable,
  we have only been able to prove:

\begin{lemma}[Many good choices]\label{lem:MGC}
For $n > d \geq 3$, 
  let $\vs{\aa}{1}{n}$ be Gaussian random variables in $\Reals{d}$ of
  standard deviation
  $\sigma$ centered at points of norm at most $1$.
Let $\AA = (\vs{\aa}{1}{n})$.
Then, we have
\[
\prob{\vs{\aa}{1}{n}}
     {
      \sum_{I \in \binom{[n]}{d}}
        \ind{\smin{\AA_{I}} \leq \kappa_{0}}
      \geq \left(1 - \frac{1}{n} \right)
     \binom{n}{d}
     }
  \leq  n^{-d} + n^{-n+d-1} + n^{-2.9d + 1},
\]
where
\begin{equation}\label{eqn:kappa0}
    \kappa_{0}  \defeq  \frac{\sigma \min (1, \sigma )}
                      {12 d^{2} n^{7} \sqrt{\ln n}}.
\end{equation}\index{kappa0@$\kappa_{0}$}%
\end{lemma}

In the analyses of $LP'$ and $LP^{+}$, we use the following
  consequence of Lemma~\ref{lem:MGC},
  whose 
  statement
  is facilitated by the following notation for a set of
  $d$-sets, $\calI $
\[
   \calI (\AA) \defeq  \mathrm{arg max}_{I \in \calI } \left(\smin{\AA _{I}} \right).
\]\index{I(A)@$\calI (\AA)$}%

\begin{corollary}[probability of small $\smin{\AA_{\calI (\AA)}}$]\label{cor:MGC}
For $n > d \geq 3$, 
  let $\vs{\aa}{1}{n}$ be Gaussian random variables in $\Reals{d}$ of
  standard deviation
  $\sigma$ centered at points of norm at most $1$, and
 let $\AA = (\vs{\aa}{1}{n})$.
For $\calI$ a set of $3 n d \ln n$ randomly chosen $d$-subsets of
  $[n]$,
\[
  \prob{\AA ,\calI}
       {\smin{\AA_{\calI (\AA)}} \leq \kappa _{0}}
 \leq 
   0.417 \binom{n}{d}^{-1}.
\]
\end{corollary}
\begin{proof}
\begin{align*}
\lefteqn{ \prob{\AA ,\calI}
       {\smin{\AA_{\calI (\AA)}} \leq \kappa _{0}}}\\
&   = 
  \prob{\AA ,\calI}
       {\forall I \in \calI : \smin{\AA_{I}} \leq \kappa _{0}}\\
& \leq 
  \prob{\AA}
       {      \sum_{I \in \binom{[n]}{d}}
        \ind{\smin{\AA_{I}} \leq \kappa_{0}}
      < \left(1 - \frac{1}{n} \right)
     \binom{n}{d}}\\
& \quad 
 +
  \prob{\calI, \AA }
  {\forall I \in \calI : \smin{\AA_{I}} \leq \kappa_{0}
   \Bigg|
    \sum_{I \in \binom{[n]}{d}}
        \ind{\smin{\AA_{I}} \leq \kappa_{0}}
      \geq \left(1 - \frac{1}{n} \right)
     \binom{n}{d}
  }\\
& \leq 
   n^{-d} + n^{-n+d-1} + n^{-2.9d + 1} +
  \left(1-\frac{1}{n}\right)^{\sizeof{\calI }},
  \text{ by Lemma~\ref{lem:MGC}}
\\
& \leq 
   n^{-d} + n^{-n+d-1} + n^{-2.9d + 1} + n^{-3d},
  \text{ as $\sizeof{\calI } = 3 n d \ln n$,}\\
& \leq 
  0.417 \binom{n}{d}^{-1},
\end{align*}
for $n > d \geq 3$.
\end{proof}

We also use the following corollary, which states that it
  is highly  unlikely that $\kappa$ falls outside the
  set $\calK$, which we now define:
\begin{equation}\label{eqn:defK}
  \calK 
= 
  \setof{
  2^{\floor{\lg (x)}} :
   \kappa_{0} \leq x \leq 
  \sqrt{d} + 3 d \sqrt{\ln n} \sigma 
 }.
\end{equation}\index{K@$\calK$}%

\begin{corollary}[probability of $\kappa$ in $\calK $]\label{cor:K}
For $n > d \geq 3$, 
  let $\vs{\aa}{1}{n}$ be Gaussian random variables in $\Reals{d}$ of standard deviation
  $\sigma$ centered at points of norm at most $1$, and
 let $\AA = (\vs{\aa}{1}{n})$.
For $\calI$ a set of $3 n d \ln n$ randomly chosen $d$-subsets of
  $[n]$,
\[
\prob{\AA ,\calI }
 {2^{\floor{\lg (\smin{\AA_{\calI (\AA)}})}} \not \in \calK }
\leq 
   0.42 \binom{n}{d}^{-1}.
\]
\end{corollary}
\begin{proof}

It follows from  Corollary~\ref{cor:MGC} that
\begin{align*}
  \prob{\AA ,\calI}
       {\smin{\AA_{\calI (\AA)}} \leq \kappa _{0}}
& \leq 
   0.417 \binom{n}{d}^{-1}.
\end{align*}
On the other hand, as
\[
\smin{\AA _{I}} 
\leq \norm{\AA _{I}} 
\leq \sqrt{d} \max _{i}\norm{\aa _{i}},
\]
\[
\prob{\AA, \calI }
  {\smin{\AA _{\calI (\AA)}} \geq 
  \sqrt{d}\left(1 + 3 \sqrt{d \ln n} \sigma  \right)
  }
\leq 
\prob{\AA}
     {\max _{i} \norm{\aa _{i}} \geq 1 + 3 \sqrt{d \ln n}\sigma}
   \leq 
  0.0015 \binom{n}{d}^{-1},
\]
by Corollary~\ref{cor:chiSquare}.
\end{proof}

\begin{proposition}[size of $\calK $]\label{pro:sizeK}
\[
\sizeof{\calK} \leq 9 \lg (n d / \min (\sigma ,1)).
\]
\end{proposition}

The rest of this section is devoted to the proof of  Lemma~\ref{lem:MGC}.
The key to the proof 
  is an examination of the relation between
  the events which we now define.
\begin{definition}
For $I \in \binom{[n]}{d}$, $K \in \binom{[n]}{d-1}$, and $j \not \in K$,
  we define the indicator random variables
\begin{align*}
   X_{I} & = \ind{\smin{\AA_{I}} \leq \kappa_{0}}, \text{ and }\\
   Y_{K}^{j} & = \ind{\dist{\aa_{j}}{\Span{A_{K}}} \leq  h_{0}},
\end{align*}\index{Xi@$X_{I}$}\index{Yk@$Y_{K}^{j}$}%
where 
\[
  h_{0} \defeq  \frac{\sigma}{4 n^{4}}.
\] \index{h0@$h_{0}$}%
\end{definition}

In Lemma~\ref{lem:MGCSumK}, we
  obtain a concentration result on the $Y_{K}^{j}$s using the fact that
  the $Y_{K}^{j}$  are independent for fixed $K$ and different $j$.
To relate this concentration result to the $X_{I}$s, we show
  in Lemma~\ref{lem:MGCRelateSums} 
  that when $X_{I}$ is true, it is probably the case that
  $Y_{I-\setof{j}}^{j}$ is true for most $j$.
%  $Y_{I-\setof{j}}^{j}$ is true for most $j$.

\begin{proof-of-lemma}{\ref{lem:MGC}}
The proof has two parts.
The first, and easier, part is Lemma~\ref{lem:MGCSumK}
  which implies
\[
  \prob{\vs{\aa}{1}{n}}
       {\sum_{K \in \binom{[n]}{d-1}}
       \sum_{j \not \in K} Y_{K}^{j} \leq  \ceiling{\frac{n-d-1}{2}} \binom{n}{d-1}}
    >
   1 - n^{-n+d-1}.
\]
To apply this fact, we use Lemma~\ref{lem:MGCRelateSums}, which implies
\[
 \prob{\vs{\aa}{1}{n}}
      {\sum_{K \in \binom{[n]}{d-1}}
       \sum_{j \not \in K} Y_{K}^{j} > \frac{d}{2} \sum_{I} X_{I}}
 >
   1 - n^{-d} - n^{-2.9d + 1}.
\]
Combining these two Lemmas, we obtain
\[
 \prob{\vs{\aa}{1}{n}}
      {
  \frac{d}{2} \sum_{I} X_{I}
   <  \ceiling{\frac{n-d-1}{2}} \binom{n}{d-1}
   } 
  \geq 1 -  n^{-d} -  n^{-n+d-1} - n^{-2.9d + 1}.
\]
Observing,
\begin{align*}
  \frac{d}{2} \sum_{I} X_{I}
   <  \ceiling{\frac{n-d-1}{2}} \binom{n}{d-1}
 \quad \implies \quad 
\sum_{I} X_{I}
& < 
 \frac{n-d}{d} \binom{n}{d-1}\\
&  =
 \frac{n-d}{n-d+1} \binom{n}{d}\\
& =
 \left(1 -  \frac{1}{n-d+1} \right) \binom{n}{d}\\
& \leq 
 \left(1 -  \frac{1}{n} \right) \binom{n}{d},
\end{align*}
we obtain
\[
\prob{\vs{\aa}{1}{n}}
     {\sum_{I} X_{I}
      \geq \left(1 - \frac{1}{n} \right) \binom{n}{d}}
  \leq  n^{-d} + n^{-n+d-1} + n^{-2.9d + 1}.
\]
\end{proof-of-lemma}

\begin{lemma}[Probability of $Y_{K}^{j}$]\label{lem:MGCProbK}
Under the conditions of
  Lemma~\ref{lem:MGC}, for all $K \in \binom{[n]}{d-1}$
  and $j \not \in K$,
\[
  \prob{\vs{\aa}{1}{n}}{Y_{K}^{j}} \leq \frac{h_{0}}{\sigma}.
\]
\end{lemma}
\begin{proof}
Follows from Proposition~\ref{pro:gaussianNear}.
\end{proof}

\begin{lemma}[Sum over $j$ of $Y_{K}^{j}$]\label{lem:MGCExistsK}
Under the conditions of
  Lemma~\ref{lem:MGC}, for all $K \in \binom{[n]}{d-1}$,
\[
  \prob{\vs{\aa}{1}{n}}
       {\sum_{j \not \in K} Y_{K}^{j} \geq \ceiling{(n-d+1)/2}}
  \leq 
  \left(\frac{4 h_{0}}{\sigma} \right)^{\ceiling{(n-d+1)/2}}
\]
\end{lemma}
\begin{proof}
Using the fact that for fixed $K$, the events
  $Y_{K}^{j}$ are independent, we compute
\begin{align*}
  \prob{\vs{\aa}{1}{n}}
       {\sum_{j \not \in K} Y_{K}^{j} \geq \ceiling{(n-d+1)/2}}
 & \leq 
 \sum_{J \in \binom{[n] - K}{\ceiling{(n-d+1)/2}}}
  \prob{\vs{\aa}{1}{n}}
       {\forall j \in J, Y_{K}^{j}}\\
 & =
 \sum_{J \in \binom{[n] - K}{\ceiling{(n-d+1)/2}}}
  \prod_{j \in J}
  \prob{\vs{\aa}{1}{n}}
       {Y_{K}^{j}}\\
 & \leq 
 \sum_{J \in \binom{[n] - K}{\ceiling{(n-d+1)/2}}}
  \left(\frac{h_{0}}{\sigma} \right)^{\ceiling{(n-d+1)/2}},
 &\text{ by Lemma~\ref{lem:MGCProbK}, }\\
 & \leq
  \left(\frac{4 h_{0}}{\sigma} \right)^{\ceiling{(n-d+1)/2}},
\end{align*}
as $\sizeof{\binom{[n] - K}{\ceiling{(n-d+1)/2}}} \leq 2^{\sizeof{[n] - K}} 
  = 2^{n-d+1}$.
\end{proof}

\begin{lemma}[Sum over $K$ and $j$ of $Y_{K}^{j}$]\label{lem:MGCSumK}
Under the conditions of Lemma~\ref{lem:MGC},
\[
  \prob{\vs{\aa}{1}{n}}
       {\sum_{K \in \binom{[n]}{d-1}}
       \sum_{j \not \in K} Y_{K}^{j} > \ceiling{\frac{n-d-1}{2}} \binom{n}{d-1}}
    \leq 
    n^{-n+d-1}.
\]
\end{lemma}
\begin{proof}
If $\sum_{K \in \binom{[n]}{d-1}}
     \sum_{j \not \in K} Y_{K}^{j} > \ceiling{\frac{n-d-1}{2}}
   \binom{n}{d-1}$,
 then there must exist a $K$ for which
 $\sum_{j \not \in K} Y_{K}^{j} > \ceiling{\frac{n-d-1}{2}}$,
 which implies for that $K$
\[
  \sum_{j \not \in K} Y_{K}^{j} 
  \geq  \ceiling{\frac{n-d-1}{2}} + 1
  =  \ceiling{\frac{n-d+1}{2}} .
\]
Using this trick, we compute
\begin{align*}
  \prob{\vs{\aa}{1}{n}}
       {\sum_{K \in \binom{[n]}{d-1}}
       \sum_{j \not \in K} Y_{K}^{j} \geq \ceiling{\frac{n-d-1}{2}} \binom{n}{d-1}}
& \leq 
  \prob{\vs{\aa}{1}{n}}
       {\exists K \in \binom{[n]}{d-1} : 
       \sum_{j \not \in K} Y_{K}^{j} \geq \ceiling{\frac{n-d+1}{2}}}\\
& \leq 
  \binom{n}{d-1}
  \prob{\vs{\aa}{1}{n}}
       {\sum_{j \not \in K} Y_{K}^{j} \geq \ceiling{\frac{n-d+1}{2}}}\\
& \leq 
  \binom{n}{d-1}
    \left(\frac{4 h_{0}}{\sigma} \right)^{\ceiling{(n-d+1)/2}}\\
\intertext{(by Lemma~\ref{lem:MGCExistsK})}
& = 
  \binom{n}{n - d+ 1}
    \left(\frac{4 h_{0}}{\sigma} \right)^{\ceiling{(n-d+1)/2}}\\
& \leq 
  n^{n-d+1}
    \left(\frac{1}{n^{4}} \right)^{\ceiling{(n-d+1)/2}}\\
& \leq 
  n^{-n+d-1}.
\end{align*}
\end{proof}

The other statement needed for the proof of Lemma~\ref{lem:MGC} is:

\begin{lemma}[Relating $X$s to $Y$s]\label{lem:MGCRelateSums}
Under the conditions of Lemma~\ref{lem:MGC}, 
\[
 \prob{\vs{\aa}{1}{n}}
      {\sum_{K \in \binom{[n]}{d-1}}
       \sum_{j \not \in K} Y_{K}^{j} \leq \frac{d}{2} \sum_{I} X_{I}}
 \leq 
    n^{-d} + n^{-2.9d + 1}
\]
\end{lemma}
\begin{proof}
Follows immediately from Lemmas~\ref{lem:MGCXandY} and~\ref{lem:MGCKrare}.
\end{proof}

\begin{lemma}[Geometric condition for bad $I$]\label{lem:MGCXandY}
If there exists a $d$-set $I$ such that
\[
X_{I} \quad \mbox{ and } \quad 
   \sum_{j\in I}Y_{I -\setof{j}}^{j} \leq d/2,
\]
then
there exists a set $L \subset I$, $\sizeof{L} = \floor{d/2 - 1}$
  and a $j_{0} \in I - L$ such that
\[
  \dist{\aa_{j_{0}}}{\Span{\AA _L}}
  \leq 
  \sqrt{d} \kappa_{0}
  \left(1 +  \ceiling{\frac{d}{2}}
            \frac{\max_{i} \norm{\aa_{i}}}{h_{0}} \right).
\]
\end{lemma} 
\begin{proof}
Let $I = \setof{\vs{i}{1}{d}}$.
By Proposition~\ref{pro:smin} (\ref{enu:matrixNormsSmallest}),
  $X_{I}$ implies the existence of $\vsb{u}{i}{1}{d}$,
  $\norm{(\vsb{u}{i}{1}{d})} = 1$, such that
\[
  \norm{\sum_{i \in I} u_{i} \aa_{i}} \leq  \kappa_{0}.
\]
On the other hand, $\sum_{j\in I}Y_{I -\setof{j}}^{j} \leq d/2$
  implies the existence of a $J \subset I$, $\sizeof{J} = \ceiling{d/2}$,
  such that
  $Y_{I - \setof{j}}^{j} = 0$ for all $j \in J$.
By Lemma~\ref{lem:MGCHeightCoeff}, this implies $\abs{u_{j}} <  \kappa_{0}/ h_{0}$
  for all $j \in J$.
As $\norm{(\vsb{u}{i}{1}{d})} = 1$ and $\kappa_{0} / h_{0}  \leq  1/\sqrt{d}$,
  there exists some $j_{0} \in I - J$ such that 
  $\abs{u_{j_{0}}} \geq 1 / \sqrt{d}$.
Setting $L = I - J - \setof{j_{0}}$, we compute
\begin{align*}
 \norm{\sum_{j \in  I} u_{j} \aa_{j}} \leq \kappa_{0}
& \quad  \implies \quad 
 \norm{u_{j_{0}} \aa_{j_{0}} + \sum_{j \in L} u_{j} \aa_{j} 
                 + \sum_{j \in J} u_{j} \aa_{j}} \leq \kappa_{0}\\
& \quad  \implies \quad 
 \norm{u_{j_{0}} \aa_{j_{0}} + \sum_{j \in L} u_{j} \aa_{j} }
        \leq \kappa_{0} + \norm{\sum_{j \in J} u_{j} \aa_{j}}\\
& \quad  \implies \quad 
 \norm{\aa_{j_{0}} + \sum_{j \in L} ( u_{j}/ u_{j_{0}} ) \aa_{j} }
    \leq (1/\abs{u_{j_{0}}})\left( \kappa_{0} + \norm{\sum_{j \in J} u_{j} \aa_{j}} \right)\\
& \quad  \implies \quad 
 \norm{\aa_{j_{0}} + \sum_{j \in L} ( u_{j}/ u_{j_{0}} ) \aa_{j} }
    \leq \sqrt{d} \left( \kappa_{0} + \sum_{j \in J} \abs{u_{j}} \norm{\aa_{j}} \right)\\
& \quad  \implies \quad 
 \dist{\aa_{j_{0}}}{\Span{A_{L}}}
    \leq \sqrt{d} \left(\kappa_{0} + 
   \ceiling{\frac{d}{2}}
   \frac{ \kappa_{0} \max_{i} \norm{\aa_{i}}}{ h_{0}} \right).
\end{align*}
\end{proof}

\begin{lemma}[Big height, small coefficient]\label{lem:MGCHeightCoeff}
Let $\vs{\aa}{1}{d}$ be vectors and $\uu$ be a unit vector
  such that
\[
  \norm{\sum_{i=1}^{d} u_{i} \aa_{i}}  \leq  \kappa_{0}.
\]
If $\dist{\aa_{j}}{\Span{\setof{\aa_{i}}_{ i \not = j}}} > h_{0}$,
  then
  $\abs{u_{j}} < \kappa_{0}/ h_{0}$.
\end{lemma}
\begin{proof}
We have
\begin{align*}
    \norm{\sum_{i=1}^{d} u_{i} \aa_{i}} \leq \kappa_{0}
& \quad  \implies \quad 
  \norm{u_{j} \aa_{j} + \sum_{i \not = j} u_{i} \aa_{i}} \leq \kappa_{0}\\
& \quad  \implies \quad 
  \norm{\aa_{j} + \sum_{i \not = j} (u_{i}/u_{j}) \aa_{i}} 
        \leq \kappa_{0} /\abs{u_{j}} \\
& \quad  \implies \quad 
  \dist{\aa_{j}}{\Span{\setof{\aa_{i}}_{i \not = j}}}
        \leq  \kappa_{0} /\abs{u_{j}},
\end{align*}
from which the lemma follows.
\end{proof}

\begin{lemma}[Probability of bad geometry]\label{lem:MGCKrare}
Under the conditions of Lemma~\ref{lem:MGC}, 
\[
 \prob{\vs{\aa}{1}{n}}
     {\begin{array}{l}
     \exists L \in \binom{[n]}{\floor{d/2 - 1}}, j_{0} \not \in L 
     \text{ such that}
     \\
      \qquad  \dist{\aa_{j_{0}}}{\Span{A_{L}}}
        \leq 
       \sqrt{d} \kappa_{0}
       \left(1 + 
       \ceiling{\frac{d}{2}}
       \frac{ \max_{i} \norm{\aa_{i}}}{ h_{0}} \right)
      \end{array}
   }
\leq 
  n^{-d} + n^{-2.9 d + 1}.
\]
\end{lemma}
\begin{proof}
We first note that
\begin{align}
\lefteqn{
 \prob{\vs{\aa}{1}{n}}
     {\begin{array}{l}
     \exists L \in \binom{[n]}{\floor{d/2 - 1}}, j_{0} \not \in L 
    \text{ such that}
     \\
      \qquad  \dist{\aa_{j_{0}}}{\Span{A_{L}}}
        \leq 
      \sqrt{d} \kappa_{0}
       \left(1 + 
       \ceiling{\frac{d}{2}}
       \frac{ \max_{i} \norm{\aa_{i}}}{ h_{0}} \right)
      \end{array}
   }
}  \nonumber \\
& \qquad \leq 
\prob{\vs{\aa}{1}{n}}
     {\begin{array}{l}
     \exists L \in \binom{[n]}{\floor{d/2 - 1}}, j_{0} \not \in L 
    \text{ such that}
     \\
      \qquad  \dist{\aa_{j_{0}}}{\Span{A_{L}}}
        \leq 
      \sqrt{d} \kappa_{0}
       \left(1 + 
       \ceiling{\frac{d}{2}}
        \frac{1 + 3 \sqrt{d \ln n} \sigma}{ h_{0}} \right)
      \end{array}
   }
 \label{eqn:lowSpan}\\
& \qquad \qquad +
\prob{\vs{\aa}{1}{n}}
     {\max_{i} \norm{\aa_{i}} > 1 + 3 \sqrt{d \ln n} \sigma}.
  \label{eqn:bigNorm}
\end{align}
We now apply Proposition~\ref{pro:gaussianNear} 
   to bound \eqref{eqn:lowSpan} by
\begin{multline}\label{eqn:MGCKrare}
\lefteqn{\sum_{L \in \binom{[n]}{\floor{d/2 - 1}}} \sum_{j_{0} \not \in L}
\prob{\vs{\aa}{1}{n}}
     {\dist{\aa_{j_{0}}}{\Span{A_{L}}}
     \leq 
      \sqrt{d} \kappa_{0}
       \left(1 + 
       \ceiling{\frac{d}{2}}
       \frac{1 + 3 \sqrt{d \ln n} \sigma}{h_{0}} \right)
     }} \\
\leq 
  \binom{n}{\floor{d/2 - 1}} (n - d/2 + 1)
   \left(\frac{\sqrt{d}\kappa_{0} }{ \sigma }
       \left(1 + 
       \ceiling{\frac{d}{2}}
       \frac{1 + 3 \sqrt{d \ln n} \sigma}{h_{0}} \right) 
  \right)^{d - \sizeof{L}}
\end{multline}
To simplify this expression, we note that
  $\ceiling{\frac{d}{2}} \leq \frac{2d}{3}$, for $d \geq 3$.
We then recall
\[
  \frac{\kappa _{0}}{h_{0}}
  =
   \frac{\min (\sigma , 1)}{3 d^{2} n^{3} \sqrt{\ln n}},
\]
and apply $d \geq 3$ to show
\begin{align*}
\frac{\sqrt{d}\kappa_{0} }{ \sigma }
       \left(1 + 
       \ceiling{\frac{d}{2}}
       \frac{1 + 3 \sqrt{d \ln n} \sigma}{h_{0}} \right) 
& \leq 
\frac{\sqrt{d} \kappa _{0}}{\sigma } +
\frac{\kappa _{0}}{h_{0}}
\left(
  \frac{2 d^{3/2}}{3 \sigma }
  +
  2 d^{2} \sqrt{\ln n}
 \right)\\
& \leq 
 \frac{1}{n^{3}}.
\end{align*}
So, we have 

\begin{align}
\eqref{eqn:MGCKrare} 
& \leq  
  \binom{n}{\floor{d/2 - 1}} (n - d/2 + 1)
   \left( \frac{1}{n^{3}}
  \right)^{\ceiling{d/2}}\\
& \leq 
  n^{\floor{d/2 - 1} + 1} n^{-3d/2} \\
& \leq  
  n^{-d}.
\end{align}

On the other hand, we can 
 use Corollary~\ref{cor:chiSquare}, to bound \eqref{eqn:bigNorm} 
  by $n^{-2.9 d +1}$.
\end{proof}

\subsubsection{Discussion}
It is natural to ask whether 
  one could avoid the complication of
  this section by  setting
  $I = \setof{1,\ldots ,d}$, or
  even choosing $I$ to be the best $d$-set in
  $\setof{1, \ldots , d+k}$ for some constant $k$.
It is possible to show that the probability that
  all $d$-by-$d$ minors of a perturbed $d$-by-$(d+k)$
  matrix have condition number at most $\epsilon$
  grows like $(\sqrt{d} \epsilon / \sigma)^{k}$.
Thus, the best of these sets would have
  reasonable condition number with polynomially
  high probability.
This bound would be sufficient to handle
  our concerns about the magnitude of
  $y'_{i}$.
The analysis 
  in Lemma~\ref{lem:comparisonPrime} might
  still be possible in this situation;
 however, it would require considering
  multiple possible splittings of the perturbation
  (for multiple values of $\tau_{1}$), and
 it is not clear whether such an analysis
  can be made rigorous.
Finally, it seems difficult in this situation
  to apply
  the trick in the proofs of
  Lemma~\ref{lem:LP+} and~\ref{lem:LP'} 
  of summing over all likely values
  for $\kappa $.
If the algorithm is given $\sigma$ as input,
  then it is possible to avoid the need for
  this trick (and an such an analysis appeared
  in an earlier draft of this paper).
However, we believe that it is preferable   
  for the algorithm to make sense without
  taking $\sigma$ as an input.

While choosing $I$ in such a simple 
  fashion could possibly simplify this
  section, albeit at the cost of complicating
  others, we feel that once
  Lemma~\ref{lem:MGC}  has been improved
  and the correct concentration bound
  has been obtained, this technique will
  provide the best bounds.

One of the anonymous referees pointed out that it should be possible
  to use the rank revealing QR factorization to find an $I$
  with almost maximal $\smin{\AA_{I}}$ (see~\cite{ChanHansen}).
While doing so seems to be the best choice algorithmically,
  it is not clear to us how we could analyze the smoothed
  complexity of the resulting two-phase algorithm.
The difficulty is that the assumption that a particular $I$
  was output by the rank revealing QR factorization
  would impose conditions on $\AA$ that we are currently not
  able to analyze.

% Local Variables: ***
% TeX-master:"shadow.tex" ***
% End: ***

\subsection{Bounding the shadow of $LP'$}
\label{sec:lp'}

Before beginning our analysis of the shadow of $LP'$, we
   define the set from which $\aalpha$ is chosen to be
   $A_{1/d^{2}}$, where we define
\begin{align*}
   A & = \setof{\aalpha : \form{\aalpha}{\oone } = 1}, \text{ and }\\
   A_{\delta}&= \setof{\aalpha: \form{\aalpha}{\oone}= 1
     \text{ and } \alpha_{i} \geq \delta , \forall i}.
\end{align*} \index{A@$A$} \index{Adel@$A_{\delta }$}

The principal obstacle to proving the bound for $LP'$ 
  is that Theorem~\ref{thm:shadow}
  requires one to specify the plane on which the shadow of the
  perturbed polytope will be measured before the perturbation is known,
 whereas 
 the shadow relevant to the analysis of $LP'$ depends on
  the perturbation---it is the shadow onto $\Span{\AA \aalpha ,\zz}$.
To overcome this obstacle, we prove in Lemma~\ref{lem:comparisonPrime}
  that if $\smin{\AAo_{\calI (A)}} \geq \kappa_{0}/2$, then
  the expected size of the shadow onto  $\Span{\AA \aalpha ,\zz}$
  is close to the expected size of the shadow onto
  $\Span{\AAo \aalphao ,\zz}$, where $\aalphao$ is chosen from $A_{0}$.
As this plane is independent of the perturbation, we can apply
  Theorem~\ref{thm:shadow} to bound the size of the shadow on this plane.
Unfortunately, $\AAo$ is arbitrary, so we cannot make any
  assumptions about $\smin{\AAo_{\calI (A)}}$.
Instead, we 
  decompose the perturbation into two parts, as in
  Corollary~\ref{cor:gaussiansFree}, and can then
  use Corollary~\ref{cor:MGC} to show that with high
  probability
  $\smin{\AAt_{\calI (A)}} \geq \kappa_{0}/2$.
We begin the proof with this decomposition, and build to
  the point at which we can apply Lemma~\ref{lem:comparisonPrime}.  

A secondary obstacle in the analysis is that
  $\kappa $ and $M$ are correlated with $\AA$ and $\yy$.
We overcome this obstacle by considering the sum of the
  expected sizes of the shadows when $\kappa $ and $M$
  are fixed to each of their likely values.
This analysis is facilitated by the notation
\[
\calT'_{\zz} (\AA, I, \aalpha , \kappa , M )
   \defeq
\sizeof{  \shadow{A_{I}\aalpha , \zz}{\vs{\aa}{1}{n}; \yy'}},
\quad
\text{where }
   y_{i}' =
   \begin{cases}
    M & \text{if $i \in I$}\\
    \sqrt{d}M^{2}/ 4 \kappa  & \text{otherwise.}
   \end{cases}
\]\index{Tprime@$\calT'$}%
We note that
\[
  \calS'_{\zz} (\AA, \yy , \calI , \aalpha )
  =
  \calT '_{\zz} \left(\AA ,\calI (\AA), \aalpha ,
      2^{\floor{\lg \smin{\AA_{\calI (\AA)}}}},
      2^{\ceiling{\lg (\max_{i} \norm{(y_{i},\aa_{i})})}+2} \right).
\]\index{Sprime@$\calS'$}%

\begin{lemma}[LP']\label{lem:LP'}
Let $d \geq 3$ and $n \geq d + 1$.
Let $\orig{\AA} = [\vs{\orig{\aa}}{1}{n}]\in \Reals{n\times d}$,
   $\orig{\yy} \in \Reals{n}$ and $\zz \in\Reals{d}$
  satisfy
  $\max_{i} \norm{(\orig{y}_{i}, \aao_{i})} \in (1/2,1]$.
For any $\sigma > 0$, let
  $\AA$ be a Gaussian random matrix centered at $\orig{\AA}$
  of standard deviation $\sigma$,
  and let $\yy$ by a Gaussian random vector centered at
  $\orig{\yy}$ of standard deviation $\sigma $.
Let $\aalpha$ be 
 chosen uniformly at random 
 from  $A_{1/d^{2}}$ and 
 let $\calI $ be a collection
 of $3nd\ln n$ randomly chosen $d$-subsets of $[n]$.
Then,
\[
\expec{\AA,\yy,\calI,\aalpha}{\calS'_{\zz} (\AA ,\yy ,\calI,\aalpha )} 
  = 326 nd (\ln n) \lg (dn/\min (1,\sigma ))\ \calD 
  \left(d, n, \frac{\min (1,\sigma^{4})}{12,960 d^{8.5}n^{14}\ln^{2.5} n}\right),
\]
where $\calD  (d,n, \sigma) $ is as given in 
 Theorem \ref{thm:shadow}.
\end{lemma}

\begin{proof}
Instead of treating $\AA$ as a perturbation of standard deviation $\sigma$
  of $\orig{\AA}$, 
  we will view $\AA$ as the result of applying a perturbation of
  standard deviation $\tau_{0}$ followed by a perturbation of standard
  deviation
  $\tau_{1}$, where $\tau_{0}^{2} + \tau_{1}^{2} = \sigma^{2}$.
Formally, we will let 
  $\GG$ be a Gaussian random matrix 
  of standard deviation $\tau_{0}$ centered at the origin,
  $\AAt = \AAo + \GG$,\index{Atilde@$\AAt$}
  $\GGt$ be a Gaussian random matrix\index{Gtilde@$\GGt$}
  of standard deviation $\tau_{1}$ centered at the origin,
  and
  $\AA = \AAt + \GGt$, where
\[
   \tau_{1}
 \defeq 
  \frac{\kappa _{0}}
       {6 d^{3} \sqrt{\ln n}}.
\]\index{tau1@$\tau_{1}$}%
and $\tau_{0}^{2} = \sigma^{2} - \tau_{1}^{2}$.\index{tau0@$\tau_{0}$}
We similarly decompose the perturbation to $\yy$
  into a perturbation of standard deviation $\tau _{0}$
  from which we obtain $\yyt $, and a perturbation
  of standard deviation $\tau _{1}$ from which we obtain
  $\yy$.\index{ytilde@$\yyt$}
We will let $\hht = \yy - \yyt $.\index{htilde@$\hht$}%

We can then apply Lemma~\ref{lem:sminAtilde} to show
\begin{equation}\label{eqn:sminAt}
\prob{\calI , \AAt , \GGt }{\smin{\AAt _{\calI (\AA)}}< \kappa _{0}/2}
< 
 0.42 \binom{n}{d}^{-1}.
\end{equation}

One difficulty in bounding the expectation of
  $\calT '$ is that its input parameters are correlated.
To resolve this difficulty,
  we will bound the expectation of $\calT '$ by the
  sum over the expectations obtained by
  substituting each of the likely choices for
  $\kappa $ and $M$.

In particular, we set
\[
\calM
=
 \setof{2^{\ceiling{\lg x}+2} :
  \left( \max _{i} \norm{(\yt_{i}, \aat _{i})} \right)
  - 3 \sqrt{d \ln n} \tau _{1}
 \leq 
  x
 \leq 
  \left( \max _{i} \norm{(\yt_{i}, \aat _{i})} \right)
  + 3 \sqrt{d \ln n} \tau _{1}
 }.
\]

We now define indicator random variables $V$, $W$, $X$, $Y$, and $Z$
  by
\begin{align*}
V & =   \ind{\sizeof{\calM } \leq 2},\\
W & =   \ind{\max _{i} \norm{(\yt _{i}, \aat _{i})} \leq 1 + 3 \sqrt{(d+1) \ln n} \sigma},\\
X & =   \ind{\smin{\AAt_{\calI (\AA)}} \geq \kappa_{0}/2},\\
Y & =    \ind{2^{\floor{\lg  \smin{\AA_{\calI (\AA)}}}} \in \calK  },
 \text{ and}\\
Z & =    \ind{2^{\ceiling{\lg \max_{i} \norm{(y_{i}, \aa_{i})}}+2} \in \calM },
\end{align*}
and then expand
\begin{multline}\label{eqn:lp'twoTerms}
\expec{\calI , A, \yy ,\aalpha }
      {\calS' (\AA ,\yy ,\calI , \aalpha )}\\
 = 
\expec{\calI , A, \yy ,\aalpha }
      {\calS' (\AA ,\yy ,\calI , \aalpha ) V W X Y Z
      }
+
\expec{\calI , A, \yy ,\aalpha }
      {\calS' (\AA ,\yy ,\calI , \aalpha ) (1 - V W X Y Z)
      }.
\end{multline}

From Corollary~\ref{cor:K}, we know
\begin{equation}\label{eqn:lp'probK}
  \prob{\AA, \calI }{\mathrm{not} (Y)} =    \prob{\AA ,\calI } 
       {2^{\floor{\lg  \smin{\AA_{\calI (\AA)}}}} \not \in \calK}
\leq 
 0.42 \binom{n}{d}^{-1}
 .
\end{equation}
Similarly, Corollary~\ref{cor:chiSquare} implies for any
  $\AAt$ and $\yyt $ and $n > d \geq 3$.
\begin{equation}\label{eqn:lp'probB}
  \prob{\GGt, \hht  ,\calI  }{\mathrm{not} (Z)} =  \prob{\GGt, \hht  ,\calI } 
       {2^{\ceiling{\lg \max_{i} \norm{(y_{i}, \aa_{i})}}+2} \not \in \calM}
\leq 
  0.0015 \binom{n}{d}^{-1}.
\end{equation}
From Corollary~\ref{cor:chiSquare} we have
\[
  \prob{\AAt, \yyt }{\mathrm{not} (W)} 
  \leq  
  n^{-2.9 (d+1) + 1}
  \leq 
  0.0015 \binom{n}{d}^{-1}.
\]
For $i_{0}$ an index for which $\norm{(y_{i_{0}},\aa _{i_{0}})} \geq 1/2$,
  Proposition~\ref{pro:noncentralgaussianNear}
  implies 
\[
  \prob{\AAt }{\mathrm{not} (V)} \leq 
  \prob{\aat_{i_{0}}, \yt _{i_{0}}}
  {\norm{(\yt _{i_{0}}, \aat _{i_{0}})} < 9 \sqrt{(d+1) \ln n} \tau _{1}}
  \leq 
      0.01 \binom{n}{d}^{-1}.
\]
By also applying inequality \eqref{eqn:sminAt} to bound
  the probability of  $\mathrm{not} (X)$, 
 we find
\[
\prob{\AA ,\yy ,\calI  }{(1 - V W X Y Z) = 1}
\leq 
  0.86 \binom{n}{d}^{-1}.
\]
As  
\[
  \calS' (\AA ,\yy ,\calI , \aalpha )
 \leq 
   \binom{n}{d},
  \text{ (by Proposition~\ref{pro:trivial})}
\]
the second term of  \eqref{eqn:lp'twoTerms} can be bounded by 1.

To bound the first term of \eqref{eqn:lp'twoTerms},
  we note
\begin{multline}\label{eqn:lp'2}
\expec{\calI , A, \yy ,\aalpha }
      {\calS' (\AA ,\yy ,\calI , \aalpha ) V W X Y Z
      }\\
 \leq 
\expec{\calI, \AAt , \yyt }{V W 
 \sum_{\kappa \in \calK , M \in \calM}
 \expec{\GGt , \hht , \aalpha }{
  \calT ' \left(\AA ,\calI (\AA), \aalpha ,
     \kappa , M \right)
  X W }
}
\end{multline}
Moreover,
\begin{align*}
\lefteqn{\expec{\GGt , \hht , \aalpha }{
  \calT ' \left(\AA ,\calI (\AA), \aalpha ,
     \kappa , M \right)
 X W  
}}\\
& = 
 \expec{\GGt , \hht , \aalpha }{
 \sum_{I \in \calI }  
   \calT ' \left(\AA ,I, \aalpha ,
     \kappa , M \right)
   W \ind{\smin{\AAt _{I}} \geq \kappa _{0}/2}
    \ind{\calI (\AA) = I}
  }\\
& \leq 
 \expec{\GGt , \hht , \aalpha }{
 \sum_{I \in \calI }  
   \calT ' \left(\AA  ,I, \aalpha ,
     \kappa , M \right)
    W \ind{\smin{\AAt _{I}} \geq \kappa _{0}/2}
  }\\
& \leq 
 \expec{\GGt , \hht , \aalpha }{
 \sum_{I \in \calI }  
   \calT ' \left(\AA  ,I, \aalpha ,
     \kappa , M \right)
  \Big|
    W \text{ and } \smin{\AAt _{I}} \geq \kappa _{0}/2
  }\\
& =
 \sum_{I \in \calI }  
 \expec{\GGt , \hht , \aalpha }{
   \calT ' \left(\AA  ,I, \aalpha ,
     \kappa , M \right)
  \Big|
   W \text{ and }
    \smin{\AAt _{I}} \geq \kappa _{0}/2
  }\\
& \leq 
 \sum_{I \in \calI }  
   (6+10^{-4}) \calD 
  \left(d, n, \frac{\tau_{1} }
         {(2 + 3\sqrt{d \ln n}\sigma ) (\sqrt{d} M^{2} / 4 \kappa M )} \right),
 & \text{by Lemma~\ref{lem:thoughtExp},}\\
& \leq 
 3 (6+10^{-4}) n d (\ln n)
\left(
     \calD 
  \left(d, n, \frac{4\tau_{1}  \kappa }
         {(2 + 3\sqrt{d \ln n}\sigma ) (\sqrt{d} M)} \right)\right).
\end{align*}

Thus, 
\begin{align*}
\eqref{eqn:lp'2} 
& \leq 
\expec{\calI, \AAt ,\yyt  }{V W 
 \sum_{\kappa \in \calK , M \in \calM}
 3 (6+10^{-4}) n d (\ln n)
     \calD 
  \left(d, n, \frac{4\tau_{1}  \kappa }
         {(2 + 3\sqrt{d \ln n}\sigma ) (\sqrt{d} M)} \right)} \\
& \leq 
\expec{\calI, \AAt ,\yyt  }{3 (6+10^{-4}) n d (\ln n)
(V\sizeof{\calM }) \sizeof{\calK }
 W    \calD 
  \left(d, n, \frac{4\tau_{1}  \min (\calK) }
         {(2 + 3\sqrt{d \ln n}\sigma ) (\sqrt{d} \max (\calM ))} \right)} \\
& \leq 
\expec{\calI, \AAt ,\yyt  }{
6 (6+10^{-4})  n d (\ln n) \sizeof{\calK }
 W     \calD 
  \left(d, n, \frac{4\tau_{1}  \min (\calK) }
         {(2 + 3\sqrt{d \ln n}\sigma ) (\sqrt{d} \max (\calM ))}\right)}\\
& \leq 
6 (6+10^{-4})  n d (\ln n) \sizeof{\calK }
 \calD 
  \left(d, n, \frac{2\tau_{1} \kappa_{0} }
         {\sqrt{d}(2 + 3\sqrt{d \ln n}\sigma )(1+6\sqrt{(d+1)\ln n}\sigma)}\right),
\end{align*}
where the last inequality follows from 
  $\max (\calM ) \leq 1 + 6 \sqrt{(d+1) \ln n} \sigma $ when $W$ is true.

To simplify, we first bound the third argument of the function $\calD
$ by:

\begin{align*}
&\frac{2\tau_{1} \kappa_{0} }
         {\sqrt{d}(2 + 3\sqrt{d \ln n}\sigma )(1+6\sqrt{(d+1)\ln
n}\sigma)}\\
& = \frac{1}{3d^{3}\sqrt{\ln n}}\frac{\kappa^{2}_{0} }
         {\sqrt{d}(2 + 3\sqrt{d \ln n}\sigma )(1+6\sqrt{(d+1)\ln
n}\sigma)} \\
& = \frac{1}{3d^{3.5}\sqrt{\ln n}} \left(\frac{1}{12d^{2}n^{7}\sqrt{\ln
n}} \right)^{2}\frac{\sigma^{2} (\min (1,\sigma ))^{2}}
         {(2 + 3\sqrt{d \ln n}\sigma )(1+6\sqrt{(d+1)\ln
n}\sigma)} \\
& \geq 
 \frac{1}{432d^{7.5}n^{14} (\ln n)^{1.5}} \frac{\min (1,\sigma^{4})}
         {(2 + 3\sqrt{d \ln n} )(1+6\sqrt{(d+1)\ln n})} \\
& \geq  \frac{1}{432d^{7.5}n^{14} (\ln n)^{1.5}} \frac{\min (1,\sigma^{4})}
         {30 d \ln n}\\
& =  \frac{\min (1,\sigma^{4})}{12,960 d^{8.5}n^{14}\ln^{2.5} n} 
\end{align*}
where the last inequality follows from the assumption that $n>d\geq 3$.

Applying Proposition \ref{pro:sizeK} to show
  $\sizeof{\calK } \leq 9\lg (dn/\min (1,\sigma ))$,
  we now obtain

%As the entries in $\calM$ and $\calK $ are powers of two, 
%  and the function $\calD $ is super-quadratic in the 
%  reciprocal of its third argument,
%\begin{multline}\label{eqn:lp'3}
%W \sum_{\kappa \in \calK , M \in \calM}
%  18 n d \log n
%  \left(
%     \calD 
%  \left(d, n, \frac{\tau_{1} \kappa }
%         {(2 + 3\sqrt{d \ln n}\sigma ) \sqrt{d} M} \right) + 2\right)\\
% \leq 
%W  36 n d \log n
%  \left(
%     \calD 
%  \left(d, n, \frac{\tau_{1} \min (\calK  ) }
%         {(2 + 3\sqrt{d \ln n}\sigma ) \sqrt{d} \max (\calM )} \right) + 2\right)
%\end{multline}

%So,
%\[
%  \frac{\min (\calK )}{\max (\calM )}
% \geq 
%  \frac{\kappa _{0}/2}{1 + 6 \sqrt{(d+1) \ln n} \sigma }
% \geq 
%  \frac{\min (1, \sigma ^{2})}
%       {168 d^{2} n^{15/2} \log n}.
%\]

\begin{align*}
\eqref{eqn:lp'2} 
& \leq 6 (6+10^{-4}) \sizeof{\calK }
 n d (\ln n) \calD 
  \left(d, n, \frac{\min (1,\sigma^{4})}{12,960 d^{8.5}n^{14}\ln^{2.5} n}\right)\\
& \leq 325 
 n d (\ln n) \lg (dn/\min (1,\sigma )) \calD 
  \left(d, n, \frac{\min (1,\sigma^{4})}{12,960 d^{8.5}n^{14}\ln^{2.5} n}\right).
\end{align*}
\end{proof}

\begin{lemma}[probability of small $\smin{\AAt _{\calI (\AA)}}$]\label{lem:sminAtilde}
For $\AA$, $\AAt $, and $\calI $ as defined in the proof of
  Lemma~\ref{lem:LP'},
\[
\prob{\calI , \AAt ,\GGt }{\smin{\AAt _{\calI (\AA)}}< \kappa _{0}/2}
< 
  0.42 \binom{n}{d}^{-1}.
\]
\end{lemma}

\begin{proof}
Let $I = \calI (\AA)$, we have
\begin{equation*}\label{eqn:sminAtilde}
\prob{}{\smin{\AAt _{I}}< \kappa _{0}/2}
\leq 
\frac{
    \prob{}{\smin{\AA _{I}}< \kappa _{0}}
  }{
    \prob{}{\smin{\AA _{I}}< \kappa _{0} \big| \smin{\AAt _{I}}< \kappa _{0}/2}
  }.
\end{equation*}

From Corollary~\ref{cor:MGC}, we have
\[
 \prob{}{\smin{\AA _{I}}< \kappa _{0}}
  \leq 
  0.417\binom{n}{d}^{-1},
\]
On the other hand,
  we have
\begin{align*}
\lefteqn{  \prob{}{\smin{\AA _{I}} \geq  \kappa _{0} \big| \smin{\AAt _{I}}< \kappa _{0}/2}}\\
 & \leq 
  \prob{}{\smin{\AA _{I}} \geq  \kappa _{0} \text{ and } \smin{\AAt _{I}}< \kappa _{0}/2}
  \\
 & \leq 
  \prob{}{\norm{\AA_{I} - \AAt_{I}} \geq \kappa_{0}/2},
  & \text{ by Proposition~\ref{pro:smin} (\ref{enu:matrixNormsSminDiff}),}
  \\
 & \leq 
  \prob{\AA}{\max_{i} \norm{\aa_{i} - \aat_{i}} \geq \kappa_{0}/2\sqrt{d}},
  & \text{by Proposition~\ref{pro:matrixNorms} (\ref{enu:matrixNormsMax}),}
 \\
& =
\prob{\AA}{\max_{i} \norm{\aa_{i} - \aat_{i}} 
          \geq 3 d^{5/2} \sqrt{\ln n} \tau_{1}}\\
& 
\leq n^{-2.9d + 1},
\end{align*}  
by Corollary~\ref{cor:chiSquare}.
Thus,
\begin{align*}\label{}
\eqref{eqn:sminAtilde} 
 \leq 
  \frac{
  0.417\binom{n}{d}^{-1}
  }{
    1 - n^{-2.9d + 1}
  }
 \leq 
  0.42 \binom{n}{d}^{-1},
\end{align*}
for $n > d \geq 3$.
\end{proof}

\begin{lemma}[From $\aat$]\label{lem:thoughtExp}
Let
  $I$ be a set in $\binom{[n]}{d}$ 
  and let $\vs{\aat}{1}{n}$ be points
  each of norm at most
  $1 + 3 \sqrt{(d+1) \ln n} \sigma $
  such that
\[
  \smin{\AAt _{I}} \geq \kappa _{0}/2.
\]
Then,
\begin{equation}\label{eqn:thoughtExp}
  \expec{\AA, \aalpha \in A_{1/d^{2}}}
        {\shadow{A_{I}\aalpha , \zz}{\vs{\aa}{1}{n}; \yy'}}
\leq 
    (6+10^{-4}) \calD 
  \left(d, n, \frac{\tau_{1}}
         {(2 + 3\sqrt{d \ln n}\sigma ) (\max _{i} y'_{i}/\min_{i} y'_{i})} \right).
\end{equation}
\end{lemma}
\begin{proof}
We apply Lemma~\ref{lem:comparisonPrime} to show
\begin{align*}
  \expec{\AA, \aalpha \in A_{1/d^{2}}}
        {\sizeof{\shadow{\AA_{I} \aalpha, \zz }
                   {\vs{\aa}{1}{n} ; \yyo}}}
& \leq 
 6 \expec{\AA, \aalphat \in A_{0}}
        {\sizeof{\shadow{\AAt_{I} \aalphat, \zz}
                   {\vs{\aa}{1}{n} ; \yyo}}}
 + 1\\
& \leq 
 6 \max _{\aalphat \in A_{0}} \expec{\AA }
        {\sizeof{\shadow{\AAt_{I} \aalphat, \zz}
                   {\vs{\aa}{1}{n} ; \yyo}}}
 + 1\\
& \leq 
    6 \calD 
  \left(d, n, \frac{\tau_{1}  }
         {(2 + 3\sqrt{d \ln n}\sigma ) (\max _{i} y'_{i}/\min_{i}
y'_{i})} \right) + 7\\
& \leq (6+10^{-4}) \calD 
  \left(d, n, \frac{\tau_{1}  }
         {(2 + 3\sqrt{d \ln n}\sigma ) (\max _{i} y'_{i}/\min_{i} y'_{i})} \right),
\end{align*}
by Corollary~\ref{cor:freeYi} and fact that
  $\calD (n,d,\sigma ) \geq 58,888,678$ for any positive $n,d,\sigma $.
\end{proof}

\begin{lemma}[Changing $\aalpha$ to $\aalphat$]\label{lem:comparisonPrime}
Let $I \in \binom{[n]}{d}$.
Let $\vs{\aa}{1}{n}$ be Gaussian random vectors
  in $\Reals{d}$ of standard deviation $\tau_{1}$,
  centered at points $\vs{\aat}{1}{n}$.
If $\smin{\AAt_{I}} \geq \kappa_{0}/ 2$, then
\[
  \expec{\AA, \aalpha \in A_{1/d^{2}}}
        {\sizeof{\shadow{\AA_{I} \aalpha, \zz }
                   {\vs{\aa}{1}{n} ; \yyo}}}
\leq 
 6 \expec{\AA, \aalphat \in A_{0}}
        {\sizeof{\shadow{\AAt_{I} \aalphat, \zz }
                   {\vs{\aa}{1}{n} ; \yyo}}}
 + 1.
\]
\end{lemma}
\begin{proof}
The key to our proof is Lemma~\ref{lem:simpSampling}.
To ready ourselves for the application of this lemma, we let
\[
  \calF_{\AA} (\tt) = 
      \sizeof{\shadow{\tt, \zz} {\vs{\aa}{1}{n}; \yyo} },
\]\index{FA@$\calF_{\AA} (\tt)$}%
and note that $\calF_{\AA} (\tt) = \calF_{\AA} (\tt / \norm{\tt})$.
If $\norm{\AAt - \AA} \leq 3 d \sqrt{\ln n} \tau_{1}$,
  then 
\[
  \norm{I - \AAt^{-1} \AA} 
\quad  \leq 
  \norm{\AAt^{-1}} \norm{\AAt - \AA}
\quad  \leq 
  \left(\frac{2}{\kappa_{0}} \right)
  3 d \sqrt{\ln n} \tau_{1}
\quad  \leq 
  \left(\frac{2}{\kappa_{0}} \right)
  \frac{3 d \sqrt{\ln n} \kappa_{0}}
       {12 d^{3} \sqrt{\ln n}}
\quad  \leq 
  \frac{1}
       {2 d^{2}}.
\]
By Proposition~\ref{pro:smin} (\ref{enu:matrixNormsSminDiff}),
\begin{align*}
\smin{\AA_{I}}
& \geq 
 \smin{\AAt_{I}} - \norm{\AAt  - \AA}\\
& \geq
  \kappa_{0}/2 - 3 d \sqrt{\ln n} \tau_{1}\\
& \geq
\frac{\kappa_{0}}{2}
\left(1 - \frac{1}{2 d^{2}} \right)\\
& \geq
\frac{\kappa_{0}}{2}
\left(\frac{17}{18} \right),
\end{align*}
for $d \geq 3$.
So, we can similarly bound
\begin{align*}
  \norm{I - \AA^{-1} \AAt } 
& \leq 
  \frac{9}{17 d^{2}}.  
\end{align*}
We can then apply Lemma~\ref{lem:simpSampling}
 to show
\[
  \expec{\aalpha \in A_{1/d^{2}}}
        {\sizeof{\shadow{\AA_{I} \aalpha, \zz }
         {\vs{\aa}{1}{n} ; \yyo}}}
\leq 
  6
  \expec{\aalphat \in A}
        {\sizeof{\shadow{\AAt_{I} \aalphat, \zz }
         {\vs{\aa}{1}{n} ; \yyo}}}.
\]
From Corollary~\ref{cor:chiSquare}
  and Proposition~\ref{pro:matrixNorms} (\ref{enu:matrixNormsMax}),
  we know that
  the probability that
  $\norm{\AAt - \AA} > 3 d \sqrt{\ln n} \tau_{1}$
  is at most $n^{-2.9d + 1}$.
As $\shadow{\AAt_{I} \aalphat, \zz}
         {\vs{\aa}{1}{n} ; \yyo } \leq \binom{n}{d}$, 
we can apply
Lemma~\ref{lem:probXYA} to show
\[
\expec{\AA}{  \expec{\aalpha \in A_{1/d^{2}}}
        {\sizeof{\shadow{\AA_{I} \aalpha, \zz }
         {\vs{\aa}{1}{n} ; \yyo}}}
}\leq 
  6
\expec{\AA}{  \expec{\aalphat \in A}
        {\sizeof{\shadow{\AAt_{I} \aalphat, \zz }
         {\vs{\aa}{1}{n}; \yyo}}}
  }
+ 1.
\]
\end{proof}

To compare the expected sizes of the shadows,
  we will show that the distribution 
  $\Span{\AA \aalpha , \zz}$
  is close to the distribution $\Span{\AAt  \aalphat , \zz}$.
To this end, we note that
 for a given $\aalphat \in A_{0} $
  the $\aalpha \in A$  for which
  $\AA \aalpha$  is a positive multiple of 
  $\AAt \aalphat$  
  is given by
\begin{equation}\label{eqn:rho}
  \aalpha = 
  \Psi (\aalphat)
  \defeq
  \frac{\AA^{-1} \AAt \aalphat}
       {\form{\AA^{-1} \AAt \aalphat}{\oone}}.
\end{equation} \index{Csi@$\Psi $}
To derive this equation, note that
  $\AAt \aalphat$ is the point in
  $\simp{\vs{\aat}{1}{d}}$ specified by
  $\aalphat$.
  $\AA^{-1} \AAt \aalphat$
  provides the coordinates of this point in the basis $\AA$.
  Dividing by 
  $\form{\AA^{-1} \AAt \aalphat}{\oone}$
  provides the $\aalpha \in A$ specifying
  the parallel point in $\aff{\vs{\aa}{1}{d}}$.
We can similarly derive
\begin{equation*}
  \Psi^{-1} (\aalpha) 
 = 
  \frac{\AAt^{-1} \AA \aalpha }
       {\form{\AAt^{-1} \AA \aalpha}{\oone}}.
\end{equation*}
Our analysis will follow from a bound on the Jacobian of
  $\Psi$.

\begin{lemma}[Approximation of $\aalpha$ by $\aalphat$]\label{lem:simpSampling}
Let $\calF (\xx)$ be a 
  non-negative
  function depending only on
  $\xx / \norm{\xx}$. \index{F@$\calF $}
If $\delta = 1/d^{2}$,
 $\norm{I - \AAt^{-1} \AA} \leq \epsilon $, and
 $\norm{I - \AA^{-1} \AAt} \leq \epsilon $, 
  where $\epsilon \leq 9/17d^{2}$, then
\[
 \expec{\aalpha \in A_{\delta }}{\calF (\AA \aalpha )}
\leq 
  6
 \expec{\aalphat \in A_{0}}{\calF (\AAt \aalphat )}
\]
\end{lemma}
\begin{proof}
Expressing the expectations as integrals, the lemma is equivalent to

\[
 \frac{1}{\vol{A_{\delta}}}
  \int_{\aalpha \in A_{\delta}} \calF (\AA \aalpha) \diff{\aalpha }
\leq 
 \frac{6}{\vol{A_{0}}}
  \int_{\aalphat \in A_{0}} \calF (\AAt \aalphat) \diff{\aalphat }.
\]

Applying Lemma~\ref{lem:inside} and setting
  $\aalpha = \Psi(\aalphat)$,
  we bound
\begin{align*}
 \frac{1}{\vol{A_{\delta}}}
  \int_{\aalpha \in A_{\delta}} \calF (\AA \aalpha) \diff{\aalpha }
& \leq
 \frac{1}{\vol{A_{\delta}}}
  \int_{\aalpha \in \Psi (A_{0})} \calF (\AA \aalpha) \diff{\aalpha }\\
& =
 \frac{1}{\vol{A_{\delta}}}
  \int_{\aalphat \in A_{0}} \calF (\AA \Psi (\aalphat)) 
  \abs{\frac{\partial \Psi (\aalphat)}{\partial \aalphat }}
  \diff{\aalphat }\\
& =
 \frac{1}{\vol{A_{\delta}}}
  \int_{\aalphat \in A_{0}} \calF (\AAt \aalphat) 
  \abs{\frac{\partial \Psi (\aalphat)}{\partial \aalphat }}
  \diff{\aalphat }\\
\intertext{(as $\AAt \aalphat$ is a positive multiple
  of $\AA \Psi (\aalphat)$ and $\calF (\xx)$ only depends on $\xx / \norm{\xx}$)}
& \leq 
  \max_{\aalphat \in A_{0}} 
 \left(\abs{\frac{\partial \Psi (\aalphat)}{\partial \aalphat }} \right)
 \frac{1}{\vol{A_{\delta}}}
  \int_{\aalphat \in A_{0}} \calF (\AAt \aalphat) 
  \diff{\aalphat }\\
& =
 \max_{\aalphat \in A_{0}} 
 \left(\abs{\frac{\partial \Psi (\aalphat)}{\partial \aalphat }} \right) 
\left( 
 \frac{\vol{A_{0} }}{\vol{A_{\delta}}}
\right)
 \frac{1}{\vol{A_{0}}}
  \int_{\aalphat \in A_{0}} \calF (\AAt \aalphat) 
  \diff{\aalphat }\\
& \leq 
  \frac{  (1 + \epsilon)^{d}}{(1- \epsilon \sqrt{d})^{d} (1-\epsilon )}
\left( \frac{1}{1 - d \delta } \right)^{d}
 \frac{1}{\vol{A_{0}}}
  \int_{\aalphat \in A_{0}} \calF (\AAt \aalphat) 
  \diff{\aalphat }\\
\intertext{(by Proposition~\ref{pro:Adelta} and
  Lemma~\ref{lem:jacobianRhoBound})}
& \leq 
  6
 \frac{1}{\vol{A_{0}}}
  \int_{\aalphat \in A_{0}} \calF (\AAt \aalphat) 
  \diff{\aalphat },
\end{align*}
for $\epsilon \leq 9 / 17d^{2}$, $\delta = 1/d^{2}$ and $d \geq 3$.
\end{proof}

\begin{proposition}[Volume dilation]\label{pro:Adelta}
\[
 \frac{\vol{A_{0} }}{\vol{A_{\delta}}}
=
 \left( \frac{1}{1 - d \delta } \right)^{d}.
\]
\end{proposition}
\begin{proof}
The set $A _{\delta }$  may be obtained by
  contracting the set $A _{0}$
  at the point $(1/d, 1/d, \ldots , 1/d)$
  by the factor $(1 - d \delta )$.
\end{proof}

\begin{lemma}[Proper subset]\label{lem:inside}
Under the conditions of Lemma~\ref{lem:simpSampling},
\[
   A_{\delta} \subset \Psi (A_{0}).
\]
\end{lemma}
\begin{proof}
We will prove
\[
   \Psi^{-1} (A_{\delta}) \subset A_{0}.
\]
Let $\aalpha \in A_{\delta}$,
  $\aalpha ' = \AAt^{-1}\AA \aalpha$
  and
  $\aalphat = \aalpha '/ \form{\aalpha '}{\oone}$. 
Using 
  Proposition~\ref{pro:vecNorms} to show
  $\norm{\aalpha} \leq \norm{\aalpha}_{1} = 1$
  and
  Proposition~\ref{pro:matrixNorms}~(\ref{enu:matrixNormsAx}),
  we bound
\begin{equation*}
 \alpha '_{i}
\quad \geq \quad 
 \alpha_{i} - \abs{\alpha_{i} - \alpha '_{i}}
\quad \geq \quad 
 \delta - \norm{\aalpha - \aalpha '}
\quad \geq \quad 
 \delta - \norm{I - \AAt^{-1} \AA} \norm{\aalpha }
\quad \geq \quad 
 \delta - \epsilon 
\quad > \quad  0.
\end{equation*}
So, all components of $\aalpha '$ are positive
  and therefore all components of
  $\aalphat = \aalpha ' / \form{\aalpha '}{\oone}$
  are positive.
\end{proof}

We will now begin a study of the Jacobian
 of $\Psi$.
This study will be simplified by
  decomposing $\Psi$ into the composition of two maps.
The second of these maps is given by:

\begin{definition}[$\Gamma_{\uu ,\vv}$] \index{Gam@$\Gamma_{\uu ,\vv}$}
Let $\uu$ and $\vv$ be vectors in $\Reals{d}$ and let
 $\Gamma_{\uu ,\vv } (\xx)$ be the map from
 $\setof{\xx : \form{\xx}{\uu} = 1}$
 to
 $\setof{\xx : \form{\xx}{\vv} = 1}$
by 
\[
  \Gamma_{\uu ,\vv } (\xx) = \frac{\xx}{\form{\xx}{\vv }}.
\]
\end{definition}

\begin{figure}[h]
 \noindent 
 \centering\epsfig{figure=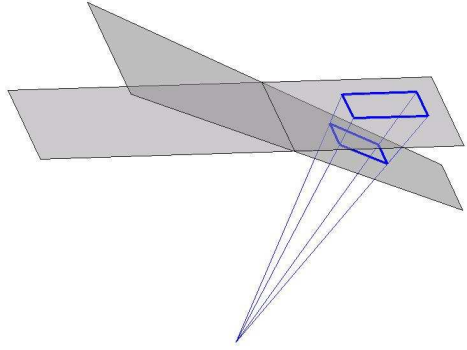,height=3in}
\caption{$\Gamma_{\uu , \vv}$ can be understood as the projection
  through the origin
  from one plane onto the other.}
\end{figure}

\begin{lemma}[Jacobian of $\Psi$]\label{lem:jacobianRhoDef}
\[
\abs{\frac{\partial \Psi (\aalphat)}{\partial \aalphat}}
=
  \det{\AA^{-1} \AAt}
  \frac{ \norm{\oone }}
       {\form{\AA^{-1} \AAt \aalphat }{\oone }^{d} 
        \norm{\left(\AAt^{-1} \AA \right)^{T} \oone }}.
\]
\end{lemma}
\begin{proof}
Let $\aalpha = \Psi (\aalphat)$ and
  let $\aalpha ' = \AA^{-1} \AAt \aalphat$.
As $\form{\aalphat}{\oone} = 1$,
  we have
\[
  \form{\aalpha '}{\left(\AAt^{-1} \AA \right)^{T} \oone } = 1.
\]
So, $\aalpha = \Gamma_{\uu , \vv } (\aalpha ')$,
  where 
  $\uu = \left(\AAt^{-1} \AA \right)^{T} \oone$
  and $\vv  = \oone$.
By Lemma~\ref{lem:twoPlanes},
\begin{align*}
\abs{\frac{\partial \aalpha}{\partial \aalphat}}
=
\abs{\frac{\partial \aalpha }{\partial \aalpha'}}
\abs{\frac{\partial \aalpha'}{\partial \aalphat}}
& =
\det{\frac{\partial \Gamma_{\uu , \vv}(\aalpha ')}{\partial \aalpha '}}
\det{\AA^{-1} \AAt}\\
& =
  \det{\AA^{-1} \AAt}
  \frac{ \norm{\oone }}
       {\form{\AA^{-1} \AAt \aalphat }{\oone }^{d} 
        \norm{\left(\AAt^{-1} \AA \right)^{T} \oone }}.
\end{align*}
\end{proof}

\begin{lemma}[Bound on Jacobian of $\Psi $]\label{lem:jacobianRhoBound}
Under the conditions of Lemma~\ref{lem:simpSampling},
\[
  \abs{\frac{\partial \Psi (\aalphat)}{\partial \aalphat}} 
\leq 
  \frac{  (1 + \epsilon)^{d}}{(1- \epsilon \sqrt{d})^{d} (1-\epsilon )}.
\]
for all $\aalphat \in A_{0}$.
\end{lemma}
\begin{proof}
The condition   
  $\norm{I - \AA^{-1} \AAt} \leq \epsilon $
  implies $\norm{\AA^{-1} \AAt} \leq 1 + \epsilon $,
so Proposition~\ref{pro:matrixNorms}~(\ref{enu:matrixNormsDet}) implies
\[
  \det{\AA^{-1} \AAt} \leq (1 + \epsilon)^{d}.
\]
Observing that $\norm{\oone} = \sqrt{d}$, and 
 $\norm{I - (\AAt^{-1} \AA)^{T}} = \norm{I - (\AAt^{-1} \AA)}$, 
 we compute
\[
  \norm{(\AAt^{-1} \AA)^{T} \oone }
\geq 
  \norm{\oone} - \norm{\oone - (\AAt^{-1} \AA)^{T} \oone }
\geq 
  \sqrt{d} - \norm{I - (\AAt^{-1} \AA)^{T}} \norm{\oone}
\geq 
  \sqrt{d} - \epsilon \sqrt{d}.
\]
So,
\[
  \frac{\norm{\oone}}{\norm{\left(\AAt^{-1} \AA \right)^{T} \oone}}
  \leq \frac{1}{1 - \epsilon} .
\]
Finally, as $\form{\aalphat}{\oone} = 1$
  and $\norm{\aalphat} \leq 1$, we have
\begin{align*}
  \form{\AA^{-1} \AAt \aalphat}{\oone }
& =
  \form{\aalphat}{\oone}
  + 
  \form{\AA^{-1} \AAt \aalphat - \aalphat }{\oone }\\
& =
  1
  +
  \form{(\AA^{-1} \AAt - I) \aalphat}{\oone }\\
& \geq 
  1
  -
  \norm{\AA^{-1} \AAt - I}
  \norm{\aalphat}
  \norm{\oone}\\
& \geq 
  1 - \epsilon \sqrt{d}.
\end{align*}
Applying Lemma~\ref{lem:jacobianRhoDef}, we have
\[
\abs{\frac{\partial \Psi (\aalphat)}{\partial \aalphat}}
=
  \det{\AA^{-1} \AAt}
  \frac{\norm{\oone }}
       {\form{\AA^{-1} \AAt \aalpha }{\oone }^{d} 
        \norm{\left(\AAt^{-1} \AA \right)^{T} \oone }}
\leq 
  \frac{  (1 + \epsilon)^{d}}{(1- \epsilon \sqrt{d})^{d} (1-\epsilon )}.
\]
\end{proof}

\begin{lemma}[Jacobian of $\Gamma_{\uu , \vv}$]\label{lem:twoPlanes}
\[
\abs{  \det{\frac{\partial \Gamma_{\uu , \vv}(\xx)}{\partial \xx}}}
 =
  \frac{\norm{\vv }}{\form{\xx}{\vv}^{d} \norm{\uu }}.
\]
\end{lemma}
\begin{proof}
Consider dividing $\Reals{d}$ into 
  $\Span{\uu , \vv}$ and the space orthogonal
  to $\Span{\uu , \vv}$.
In the $(d-2)$-dimensional orthogonal space, 
  $\Gamma_{\uu , \vv }$ acts as a multiplication
  by $1 / \form{\xx}{\vv }$.
On the other hand,
  the Jacobian of the restriction of
  $\Gamma_{\uu , \vv }$ to $\Span{\uu , \vv}$
  is computed by Lemma~\ref{lem:twoPlanesTwoDim}
  to be 
\[
  \frac{\norm{\vv }}{\form{\xx}{\vv}^{2} \norm{\uu }}.
\]
So,
\[
\abs{  \det{\frac{\partial \Gamma_{\uu , \vv}(\xx)}{\partial \xx}}}
 =
  \left(\frac{1}{\form{\xx}{\vv}} \right)^{d-2}
  \frac{\norm{\vv }}{\form{\xx}{\vv}^{2} \norm{\uu }}
=
  \frac{\norm{\vv }}{\form{\xx}{\vv}^{d} \norm{\uu }}.
\]
\end{proof}

\begin{lemma}[Jacobian of $\Gamma_{\uu , \vv}$ in 2D]\label{lem:twoPlanesTwoDim}
Let $\uu$ and $\vv$ be vectors in $\Reals{2}$ and let
 $\Gamma_{\uu , \vv}(\xx)$ be the map from
 $\setof{\xx : \form{\xx}{\uu} = 1}$
 to
 $\setof{\xx : \form{\xx}{\vv} = 1}$
by 
\[
  \Gamma_{\uu , \vv}(\xx) = \frac{\xx}{\form{\xx}{\vv }}.
\]
Then,
\[
\abs{  \det{\frac{\partial \Gamma_{\uu , \vv}(\xx)}{\partial \xx}}}
 =
  \frac{\norm{\vv }}{\form{\xx}{\vv}^{2} \norm{\uu }}.
\]
\end{lemma}
\begin{proof}
Let $\RR = \left(\begin{array}{rr}
                0 & -1\\
                1 & 0
                \end{array} \right)$,
the $90^{o}$ rotation counter-clockwise.
Let 
\[
  \uu ^{\bot} = \RR \uu  / \norm{\uu}
\qquad 
\text{ and }
\qquad 
  \vv ^{\bot} = \RR \vv  / \norm{\vv}.
\]
Express the $\xx$ such that $\form{\xx}{\uu} = 1$,
  as $\xx = \uu / \norm{\uu }^{2} + x \uu ^{\bot}$.
Similarly, parameterize the line
  $\setof{\xx : \form{\xx}{\vv } = 1}$
  by $\vv / \norm{\vv}^{2} + y \vv ^{\bot}$.
Then, we have
\[
  \Gamma _{\uu , \vv } 
  \left(u/ \norm{\uu }^{2} + x \uu ^{\bot} \right)
 = 
  \vv / \norm{\vv}^{2} + y \vv ^{\bot},
\]
where 
\[
  y = 
    \frac{\form{u/ \norm{\uu }^{2} + x \uu ^{\bot}}{\vv ^{\bot}}}
         {\form{u/ \norm{\uu }^{2} + x \uu ^{\bot}}{\vv}}
    =
    \frac{\form{u/ \norm{\uu }^{2} + x \uu ^{\bot}}{\vv ^{\bot}}}
         {\form{\xx}{\vv}}.
\]
So,
\begin{align*}
\abs{  \det{\frac{\partial \Gamma_{\uu , \vv } (\xx)}{\partial \xx}}}
& =
\abs{  \det{\frac{\partial y}{\partial x}}}\\
& = 
\abs{\frac{\Form{\uu ^{\bot}}{\vv ^{\bot}}
      \Form{\frac{\uu}{\norm{\uu}^2} + x \uu ^{\bot}}{\vv}
     -
      \Form{\uu ^{\bot}}{\vv}
      \Form{\frac{\uu}{\norm{\uu}^2} + x \uu ^{\bot}}{\vv ^{\bot}}
     }
     {\form{\xx}{\vv }^{2}}}\\
& = 
\abs{\frac{\Form{\uu ^{\bot}}{\vv ^{\bot}}
      \Form{\frac{\uu}{\norm{\uu}^2} }{\vv}
     -
      \Form{\uu ^{\bot}}{\vv}
      \Form{\frac{\uu}{\norm{\uu}^2} }{\vv ^{\bot}}
     }
     {\form{\xx}{\vv }^{2}}}\\
& = 
\abs{\frac{\norm{\vv} \left(\Form{\uu ^{\bot}}{\vv ^{\bot}}
      \Form{\frac{u}{\norm{\uu }}}{\frac{v}{\norm{\vv }}}
     -
      \Form{\uu ^{\bot}}{\frac{v}{\norm{\vv }}}
      \Form{\frac{u}{\norm{\uu }} }{\vv ^{\bot}} \right)
     }
     {\norm{\uu }
     \form{\xx}{\vv }^{2}}}\\
& = 
\abs{\frac{\norm{\vv} \left(
      \Form{\frac{u}{\norm{\uu }}}{\frac{v}{\norm{\vv }}}^{2}
     +
      \Form{\uu ^{\bot}}{\frac{v}{\norm{\vv }}}^{2}
      \right)
     }
     {\norm{\uu }
     \form{\xx}{\vv }^{2}}},
\text{ as $\RR$ is orthogonal and $\RR^{2} = -1$,}\\
& =
 \frac{\norm{\vv }}
      {\norm{\uu} \form{\xx}{\vv }^{2}},
\text{ as $\frac{\uu}{\norm{\uu}}, \uu^{\bot}$ is a basis.}
\end{align*}
\end{proof}

% Local Variables: ***
% TeX-master:"shadow.tex" ***
% End: ***

\subsection{Bounding the shadow of $LP^{+}$} \label{sec:lp+}

The main obstacle to proving a bound on the size of the shadow
  of $LP^{+}$ is
  that the vectors $\aap_{i} / \yp_{i} $ are not Gaussian random vectors.
To resolve this problem, we will
  show that, in almost every sufficiently small region, we can
  construct a family of Gaussian random
  vectors with distributions similar to the vectors $\aap _{i}/ \yp _{i}$.
We will then bound the expected size of the shadow of the vectors
  $\aap_{i}/ \yp_{i}$ by a small multiple of the expected size
  of the shadow of these Gaussian vectors.
These regions are defined by splitting the original perturbation
  into two, and letting the first perturbation define the region.

As in the analysis of $LP'$, a secondary obstacle is the
  correlation of $\kappa $ and $M$ with 
  $\AA$ and $\yy$.
We again overcome this obstacle by considering the sum of the
  expected sizes of the shadows when $\kappa $ and $M$
  are fixed to each of their likely values, and
  use the notation
\[
\calTp_{\zz} (\AA, \yy, \kappa , M )
   \defeq
\begin{cases}
  \sizeof{\shadow{(0,\zz),\zzp}{\vsDiv{\aap}{y^{+}}{1}{n} }},
& \text{if $\sqrt{d} M / 4 \kappa \geq 1$}\\
0 & \text{otherwise},
\end{cases}
\]\index{Tplus@$\calTp$}%
%\index{Tplus@$\calTp$}%
where
\begin{align*}
 \aap_{i} & = \left((y'_{i} - y_{i})/2, \aa_{i} \right),\\
 y^{+}_{i} & = 
         (y'_{i} + y_{i})/2,
 \text{ and}\\
  y_{i}' & =
   \begin{cases}
    M & \text{if $i \in I$}\\
    \sqrt{d}M^{2}/ 4 \kappa  & \text{otherwise.}
   \end{cases}
\end{align*}

By Lemma~\ref{pro:lp++} and Proposition~\ref{pro:kappaM}, we then have
\[
  \calSp_{\zz} (\AA, \yy , \calI  )
  = 
  \calTp_{\zz} \left(\AA ,\yy ,
      2^{\floor{\lg \smin{\AA_{\calI (\AA)}}}},
      2^{\ceiling{\lg (\max_{i} \norm{(y_{i},\aa_{i})})}+2} \right).
\]\index{Splus@$\calSp$}%

\begin{lemma}[LP+]\label{lem:LP+}
Let $d \geq 3$ and $n \geq d + 1$.
Let $\orig{\AA} = [\vs{\orig{\aa}}{1}{n}]\in \Reals{n\times d}$,
  $\orig{\yy} \in \Reals{n}$ and $\zz \in\Reals{d}$,
  satisfy
  $\max_{i} \norm{(\orig{y}_{i}, \aao_{i})} \in (1/2,1]$.
For any $\sigma > 0$, let
  $\AA$ be a Gaussian random matrix centered at $\orig{\AA}$
  of standard deviation $\sigma$,
  and let $\yy$ by a Gaussian random vector centered at
  $\orig{\yy}$ of standard deviation $\sigma $.
Let $\calI$ be a set of $3 n d \ln n$ randomly chosen $d$-subsets of
  $[n]$.
Then,
\[
\expec{\AA ,\yy ,\calI}{\calSp (\AA , \yy , \calI )}
\leq 
  49 \lg (n d / \min (\sigma ,1))
  \calD \left(d,n,
  \frac{\min (1,\sigma^{5}) }
       {2^{23} (d+1)^{11/2}n^{14} (\ln n)^{5/2}
       }
  \right)
 + n.
\]
where $\calD  (d,n, \sigma) $ is as given in Theorem \ref{thm:shadow}.
\end{lemma}
\begin{proof}
For $\rho _{0}$ and $\rho _{1}$ defined below, we let
  $\GG$ and $\GGt $ be Gaussian random matrices centered
  at the origin of standard deviations $\rho _{0}$ and $\rho _{1}$,
  respectively.
We then let $\AAt = \AAo  + \GG$ and $\AA = \AAt + \GGt $.  
  \index{Atilde@$\AAt$}
  \index{Gtilde@$\GGt$}
We similarly let $\hh$ and $\hht $ be Gaussian random vectors
  centered
  at the origin of standard deviations $\rho _{0}$ and $\rho _{1}$,
  respectively,
and let
  $\yyt = \yyo + \hh$ and $\yy = \yyt + \hht $.
  \index{htilde@$\hht$}
  \index{ytilde@$\yyt$}
If 
\[
  \sigma \leq \frac{3 \sqrt{1/4}}{\sqrt{2e} n (60 n (d+1)^{3/2} (\ln n)^{3/2})},
\]
we set $\rho_{1} =  \sigma$.
Otherwise, we set $\rho_{1}$ so that
\[
  \rho_{1} = 
  \frac{3 \sqrt{1/4 + d (\sigma^{2} - \rho_{1}^{2})}}
  {\sqrt{2e} n (60 n (d+1)^{3/2} (\ln n)^{3/2})},
\]\index{rho1@$\rho_{1}$}\index{rho0@$\rho_{0}$}%
and set $\rho_{0}^{2} = \sigma^{2} - \rho_{1}^{2}$.
We note that
\[
  \rho_{1} = \min \left(
  \sigma,
       \frac{3 \sqrt{1/4 + d \rho_{0}^{2}}}
  {\sqrt{2e} n (60 n (d+1)^{3/2} (\ln n)^{3/2})} \right).
\]

As in the proof of Lemma~\ref{lem:LP'}, we define the set
  of likely values for $M$:
\begin{multline*}
\calM
=
 \Bigg\{ 2^{\ceiling{\lg x}+2} :
  \left( \max _{i} \norm{(\yt_{i}, \aat _{i})} \right)
\left(1 - \frac{9 \sqrt{(d+1) \ln n}}{(60 n (d+1)^{3/2} (\ln n)^{3/2})} \right)
 \leq 
  x\\
 \leq 
  \left( \max _{i} \norm{(\yt_{i}, \aat _{i})} \right)
\left(1 + \frac{9 \sqrt{(d+1) \ln n}}{(60 n (d+1)^{3/2} (\ln n)^{3/2})} \right)
\Bigg\} .
\end{multline*}\index{M@$\calM$}%
%\[
%\calM
%=
% \setof{2^{\ceiling{x}+2} :
%  \left( \max _{i} \norm{(\yt_{i}, \aat _{i})} \right)
%\left(1 - \frac{9 \sqrt{(d+1) \ln n}}{(60 n (d+1)^{3/2} (\ln n)^{3/2})} \right)
% \leq 
%  x
% \leq 
%  \left( \max _{i} \norm{(\yt_{i}, \aat _{i})} \right)
%\left(1 + \frac{9 \sqrt{(d+1) \ln n}}{(60 n (d+1)^{3/2} (\ln n)^{3/2})} \right)
%}.
%\]
Observed that $\sizeof{\calM} \leq 2$.

As in the proof of Lemma~\ref{lem:LP'}, we define random variables:
\begin{align*}
W & =   \ind{\max _{i} \norm{(\yt _{i}, \aat _{i})} \leq 1 + 3 \sqrt{(d+1) \ln n} \rho_{0}},\\
X & =   \ind{\max _{i} \norm{(\yt _{i}, \aat _{i})} \geq 
   \frac{
   \sqrt{1/4 + d \rho_{0}^{2}}
   }{
   \sqrt{2 e} n
   }
},\\
Y & =    \ind{2^{\floor{\lg  \smin{\AA_{\calI (\AA)}}}} \in \calK  },
 \text{ and}\\
Z & =    \ind{2^{\ceiling{\lg \max_{i} \norm{(y_{i}, \aa_{i})}}+2} \in \calM }.
\end{align*}

In order to apply the shadow bound proved below in Lemma~\ref{lem:lp+2},
  we need
\[
  M \geq 3 \max _{i} \norm{(\yt _{i}, \aat _{i})},
\]
and
\[
  M \geq (60 n (d+1)^{3/2} (\ln n)^{3/2}) \rho_{1}.
\]
From the definition of $\calM $ and
  the inequality
  $1 - 9 \sqrt{(d+1) \ln n}/(60 n (d+1)^{3/2} (\ln n)^{3/2}) \geq 3/4$,
 the first of these inequalities
  holds if $Z$ is true.
Given that $Z$ is true, the second inequality holds if $X$
  is also true.

From Corollary~\ref{cor:K}, we know
\begin{equation}\label{eqn:lp+Y}
  \prob{\AA ,\calI}{\mathrm{not} (Y)}
\leq 
  \prob{\AA ,\calI } 
       {2^{\floor{\lg  \smin{\AA_{\calI (\AA)}}}} \not \in \calK}
\leq 
 0.42 \binom{n}{d}^{-1}
\leq 
 0.42 n \binom{n}{d+1}^{-1}
 .
\end{equation}
From Corollary~\ref{cor:chiSquare} we have
\begin{equation}\label{eqn:lp+W}
  \prob{\AAt, \yyt }{\mathrm{not} (W)} 
  \leq  
  n^{-2.9 (d+1) + 1}
  \leq 
  0.0015 \binom{n}{d+1}^{-1}.
\end{equation}
From Proposition~\ref{pro:noncentralgaussianNear},
  we know
\begin{equation}\label{eqn:lp+X}
\prob{}{\mathrm{not} (X)}
= 
\prob{}{
\max_{i} \norm{(\yt_{i},\aat_{i})} <
  \frac{
   \sqrt{1/4 + d \rho_{0}^{2}}
   }{
   \sqrt{2 e} n
   }
}
<
 n^{- (d+1)}
\leq \frac{1}{24}\binom{n}{d+1}^{-1}.
\end{equation}

To bound the probability that $Z$ fails,
  we note that
\[
\max_{i} \norm{(\yt_{i},\aat_{i})} \geq 
  \frac{
   \sqrt{1/4 + d \rho_{0}^{2}}
   }{
   \sqrt{2 e} n
   }
\]
and
\[
\max_{i} \norm{(y_{i} - \yt_{i},\aa_{i} - \aat_{i})} \leq 
  \rho_{1} 3 \sqrt{(d+1) \ln n},  
\]
imply $Z$ is true.
Hence, by Corollary~\ref{cor:chiSquare} and \eqref{eqn:lp+X},
\begin{equation}\label{eqn:lp+Z}
  \prob{}{\mathrm{not (Z)}}
  \leq   n^{-2.9 (d+1) + 1} + n^{- (d+1)} 
  \leq .044 \binom{n}{d+1}^{-1}.
\end{equation}

As in the proof of Lemma~\ref{lem:LP'},
  we now expand
\begin{equation}\label{eqn:lp+twoTerms}
\expec{\calI , A, \yy }
      {\calSp (\AA ,\yy ,\calI )}
 = 
\expec{\calI , A, \yy }
      {\calSp (\AA ,\yy ,\calI ) W X Y Z
      }
+
\expec{\calI , A, \yy  }
      {\calSp (\AA ,\yy ,\calI ) (1 - W X Y Z)
      }.
\end{equation}
To bound the second term by $n$, we apply
  \eqref{eqn:lp+W} , 
  \eqref{eqn:lp+X} , 
  \eqref{eqn:lp+Y}  and
  \eqref{eqn:lp+Z} to
  show
\[
  \prob{\AA ,\calI }{
  \mathrm{not} (W) \text{ or }
  \mathrm{not} (X) \text{ or }
  \mathrm{not} (Y) \text{ or }
  \mathrm{not} (Z)
}
 \leq 
  n \binom{n}{d+1}^{-1},
\]
and then combine this inequality with
    Proposition~\ref{pro:trivial}.

To bound the first term of \eqref{eqn:lp+twoTerms},
  we note
\begin{align}
\lefteqn{\expec{\calI , A, \yy }
      {\calSp (\AA ,\yy ,\calI ) W X Y Z
      }}  \nonumber \\
&  \leq 
\expec{\calI, \AAt ,\yyt }{ W X
 \sum_{\kappa \in \calK , M \in \calM}
 \expec{\GGt , \hht  }{
  \calTp \left(\AA ,\yy ,
     \kappa , M \right)
  X Z}
}
\nonumber \\
&  \leq 
\expec{\calI, \AAt ,\yyt }{ W X
 \sum_{\kappa \in \calK , M \in \calM}
 \expec{\GGt , \hht  }{
  \calTp \left(\AA ,\yy ,
     \kappa , M \right)
  \Big|   X Z}
}\nonumber \\
&  \leq 
\expec{\calI, \AAt ,\yyt }{ W X
 \sum_{\kappa \in \calK , M \in \calM}
  e \calD \left(d,n,\frac{\rho _{1} \min_{i} y'_{i}}
  {3 (\max_{i} y'_{i})^{2} } \right) + 1,
  }
  \text{ by Lemma~\ref{lem:lp+2} }\\
&  \leq 
\expec{\calI, \AAt ,\yyt }{ W X
 \sum_{\kappa \in \calK , M \in \calM}
 e  \calD \left(d,n,\frac{\sigma  M}
  {3 (M^{2}/4 \kappa )^{2} } \right) + 1
  }\nonumber \\
&  \leq 
\expec{\calI, \AAt ,\yyt }{ W X
 \sizeof{\calK} \sizeof{M}
 e  \calD \left(d,n,
  \frac{16 \sigma  \min (\calK)^{2}}
  {3 \max (\calM)^{3}} \right) + 1
  }.\label{eqn:lp+a}
\end{align}
As $\min (\calK) \geq \kappa_{0}/2$
  and
 $W$ implies
  $\max (\calM ) \leq 9 \left(1 + 3\sqrt{(d+1) \ln n}\sigma  \right)$,
\begin{align*}
  \frac{16 \sigma \min (\calK)^{2}}
       {3 \max (\calM)^{3}}
& \geq 
  \frac{16 \sigma ^{3}\min (1,\sigma)^{2} }
       {3 \cdot 4 \left(9 (1 + 3\sqrt{(d+1) \ln n} \sigma  \right)^{3}
     \left(12 d^{2} n^{7} \sqrt{\ln n} \right)^{2}
  }\\
& \geq 
  \frac{16 \min (1,\sigma^{5}) }
       {3 \cdot 4 \left(9 (1 + 3\sqrt{(d+1) \ln n} \right)^{3}
     \left(12 d^{2} n^{7} \sqrt{\ln n} \right)^{2}
  }\\
& \geq 
  \frac{\min (1,\sigma^{5}) }
       {2^{23} (d+1)^{11/2}n^{14} (\ln n)^{5/2}
       }.
\end{align*}
Applying this inequality, Proposition~\ref{pro:sizeK}, and
  the fact that $X$ implies $\sizeof{\calM} \leq 2$,
  we obtain
\[
  \eqref{eqn:lp+a} \leq 
  49 \lg (n d / \min (\sigma ,1))
  \calD \left(d,n,
  \frac{\min (1,\sigma^{5}) }
       {2^{23} (d+1)^{11/2}n^{14} (\ln n)^{5/2}
       }
  \right).
\]
\end{proof}

\begin{lemma}[$LP^{+}$ Shadow, part 2]\label{lem:lp+2}
Let $d \geq 3$ and $n \geq d+1$.
Let $\yy$ be a Gaussian random vector of standard deviation
  $\rho_{1}$ centered at a point $\yyt$,
  and let $\vs{\aa}{1}{n}$ be Gaussian random vectors
  in $\Reals{d}$
  of standard deviation $\rho_{1}$
  centered at $\vs{\aat}{1}{n}$ respectively.
Under the conditions
\begin{align}
y_{i}' & > 3 (\norm{\yt_{i},\aat_{i}}), \forall i,  \text{ and}
\label{eqn:lp+2max}
\\
y_{i}' & > 60 n (d+1)^{3/2} (\ln n)^{3/2} \rho_{1}, \forall i.
\label{eqn:lp+2rho}
\end{align}
Let 
\begin{align*}
\aap_{i} & = \left((y'_{i} - y_{i})/2, \aa_{i} \right), \text{and} \\
\yp_{i} & =  (y'_{i} + y_{i})/2.
\end{align*}
Then,
\[
  \expec{(y_{1}, \aa _{1}), \ldots ,(y_{n}, \aa _{n}) }
   {\sizeof{\shadow{(0,\zz), \zzp }
               {\vsDiv{\aap }{y^{+}}{1}{n}}}}
\leq 
 e  \calD \left(d,n,\frac{\rho _{1}  \min_{i} y'_{i} }
  {3 (\max_{i} y'_{i})^{2}} \right) + 1.
\]
\end{lemma}

\begin{proof}
We use the notation
\[
(p_{i,0} (\htil_{i}), \pp _{i} (\htil_{i}, \ggt_{i}))  
= 
  \aap_{i}/ \yp_{i} 
=
\left(  \frac{y_{i}' - \yt_{i} - \htil_{i}}{y_{i}' + \yt_{i} + \htil_{i}},
  \frac{2 (\aat_{i} + \ggt_{i})}
      {y_{i}'+ \yt_{i} + \htil_{i}} \right),
\]\index{pi0@$p_{i,0}$}\index{pi@$\pp_{i}$}%
where $\vs{\ggt}{1}{n}$ are the columns of $\GGt$
  and $(\vs{\htil}{1}{n}) = \hht $ \index{gtilde@$\ggt_{i}$}\index{htilde@$\htil_{i}$}
  as defined in the proof of Lemma~\ref{lem:LP+}.

The Gaussian random vectors that we will use to approximate
  these will come from their first-order approximations:
\[
  ( \ph _{i,0} (\htil_{i}),  \pph (\htil_{i}, \ggt_{i}))
 =
\left(  \frac{y_{i}' - \yt_{i} - \htil_{i} ( 2 y_{i}'/ ( y_{i}' + \yt_{i}))}
  {y_{i}' + \yt_{i}},
  \frac{2\aat_{i} + 2\ggt_{i} - \htil_{i} ( 2 \aat_{i}/(y_{i}' + \yt_{i}))}
     {y_{i}' + \yt_{i}}
 \right)
\]\index{phati0@$\ph_{i,0}$}\index{phati@$\pph_{i}$}%

Let $\nuh _{i} (\ph _{i,0}, \pph _{i})$
  be the induced density on $(\ph _{i,0}, \pph _{i})$.
In Lemma~\ref{lem:almostGaussian}, we prove that there exists
  a set $B$ \index{B@$B$}
  of $\left((p_{1,0}, \pp _{1}),\ldots,(p_{n,0}, \pp _{n})\right)$
 such that
\[
\prob{\prod_{i=1}^{n}\nu_{i} (p_{i,0}, \pp _{i})}
 {\left((p_{1,0}, \pp _{1}),\ldots,(p_{n,0}, \pp_{n})\right) \in B}
   \geq 1 - 0.0015 \binom{n}{d+1}^{-1},
\]
and
 for $((p_{1,0}, \pp _{1}),\ldots, (p_{n,0}, \pp _{n})) \in B$,
\[
   \prod_{i=1}^{n}\nu_{i} (p_{i,0}, \pp _{i})
  \leq e\prod_{i=1}^{n} \mu '_{i} (p_{i,0}, \pp _{i}).
\]
Consequently, 
 Lemma \ref{lem:approximate} allows us to prove
\begin{multline*}
  \expec{\prod_{i=1}^{n}\nu_{i} (p_{i,0}, \pp _{i})}
   {\sizeof{\shadow{(0,\zz), \zzp }
               {(p_{1,0}, \pp _{1}),\ldots, (p_{n,0}, \pp _{n}) }}}\\
\leq 
e  \expec{\prod_{i=1}^{n}\nuh_{i} (p_{i,0}, \pp _{i})}
   {\sizeof{\shadow{(0,\zz), \zzp }
               {(p_{1,0}, \pp _{1}),\ldots, (p_{n,0}, \pp _{n}) }}} + 1.
\end{multline*}
By Lemma~\ref{lem:almostGaussianNuhat}, the densities
  $\nuh_{i}$ represent Gaussian distributions centered
  at points of norm at most
\[
\norm{\left(  \frac{y_{i}' - \yt_{i}}
  {y_{i}' + \yt_{i}},
  \frac{2\aat_{i}}{y_{i}' + \yt_{i}}
 \right)}
\leq 
  \sqrt{5},
\quad \text{(by condition \eqref{eqn:lp+2max})}
\]
 whose covariance matrices have eigenvalues
  at most
\[
\left(9 \rho _{1} / 2 y_{i}' \right)^{2}
\leq 
\left(9  / 2 (60 n (d+1)^{3/2} (\ln n)^{3/2}) \right)^{2}
\leq 
 1  / 9 d \ln n, 
\quad \text{(by condition \eqref{eqn:lp+2rho})}
\]
and at least
\[
  \left(9 \rho _{1} / 8 y_{i}' \right)^{2}.
\]
Thus, we can apply Corollary~\ref{cor:freeYi} to bound
\begin{align*}
\lefteqn{ \expec{\prod_{i=1}^{n}\nuh_{i} (p_{i,0}, \pp _{i})}
   {\sizeof{\shadow{(0,\zz), \zzp }
               {(p_{1,0}, \pp _{1}),\ldots, (p_{n,0}, \pp _{n}) }}}}\\
& \leq 
 e  \calD \left(d,n,\frac{9 \rho _{1} / 8 \max _{i}y'_{i} }
  {(1 + \sqrt{5}) (\max_{i} y'_{i} / \min_{i} y'_{i}) } \right) + 1,\\
& \leq 
 e  \calD \left(d,n,\frac{\rho _{1}  \min_{i} y'_{i} }
  {3 (\max_{i} y'_{i})^{2}} \right) + 1,
\end{align*}
thereby proving the Lemma.
\end{proof}

\begin{lemma}[$\nuh $]\label{lem:almostGaussianNuhat}
Under the conditions of Lemma~\ref{lem:lp+2},
  the vector $  ( \ph _{i,0} (\htil_{i}),  \pph (\htil_{i}, \ggt_{i}))$ is a Gaussian
  random  vector centered at
\[
\left(  \frac{y_{i}' - \yt_{i}}
  {y_{i}' + \yt_{i}},
  \frac{2\aat_{i}}{y_{i}' + \yt_{i}}
 \right), 
\]
and has a covariance matrix with eigenvalues between
 $\left(9 \rho _{1} / 8 y_{i}' \right)^{2}$ and 
 $\left(9 \rho _{1} / 2 y_{i}' \right)^{2}$.
\end{lemma}

\begin{proof}\label{}
Because $(\ph _{i,0} (\htil_{i}), \pph (\htil_{i},
  \ggt_{i}))$ is linear in $(\htil_{i},\ggt_{i})$
  and $(\htil_{i},\ggt_{i})$ is
  a Gaussian random vector,
  $(\ph _{i,0} (\htil_{i}),  \pph (\htil_{i}, \ggt_{i}))$ is a Gaussian vector.
The statement about the center of the distributions follows
  immediately from the fact that $(\htil_{i}, \ggt_{i})$ is centered
  at the origin.
To construct the covariance matrix, we 
  note that the matrix corresponding to the transformation
  from $(\htil_{i}, \ggt_{i})$ to 
  $(\ph _{i,0} (\htil_{i}),  \pph (\htil_{i}, \ggt_{i}))$ 
  is
\[\Large
  C_{i} \defeq 
\left(
  \begin{array}{c | c}
  \frac{-2 y'_{i}}{(y'_{i} +\yt_{i})^{2}},
  & 0,  \hdots, 0\\
  \hline
  \begin{array}{l}
\frac{-2 \tilde{a}_{i,1}}{(y'_{i} + \yt_{i})^{2}}\\
\frac{-2 \tilde{a}_{i,2}}{(y'_{i} + \yt_{i})^{2}}\\
\vdots\\
\frac{-2  \tilde{a}_{i,d}}{(y'_{i} + \yt_{i})^{2}}
  \end{array}
 &
 \frac{2}{y'_{i} + \yt_{i}}  I
  \end{array}
 \right)
\]\index{Ci@$C_{i}$}%
Thus, the covariance matrix of $(\ph _{i,0} (\htil_{i}),  \pph (\htil_{i}, \ggt_{i}))$
  is  given by $\rho_{1}^{2}C_{i}^{T}C_{i}$.

We now note that
\[\Large
  \frac{y'_{i}+ \yt_{i}}{2} C_{i} -
\left(
  \begin{array}{c | c}
  -1
  & 0,  \hdots, 0\\
  \hline
  \begin{array}{l}
  0\\
  0\\
\vdots\\
  0
  \end{array}
 &
  I
  \end{array}
 \right)
 = 
\left(
  \begin{array}{c | c}
  \frac{\yt_{i}}{y'_{i} +\yt_{i}},
  & 0,  \hdots, 0\\
  \hline
  \begin{array}{l}
- \frac{\tilde{a}_{i,1}}{y'_{i} + \yt_{i}}\\
- \frac{\tilde{a}_{i,2}}{y'_{i} + \yt_{i}}\\
\vdots\\
- \frac{\tilde{a}_{i,d}}{y'_{i} + \yt_{i}}
  \end{array}
 &
   0
  \end{array}
 \right)
\]
As all the singular values of the middle matrix are 1,
  and the norm of the right-hand matrix is
  $\norm{(\yt _{i}, \aat _{i})} / (y'_{i} + \yt_{i})$,
  all the singular values of $C_{i}$ lie between
\[
\frac{2 }{y'_{i}+\yt_{i}}
\left(1-\frac{\norm{(\yt _{i}, \aat _{i})}}
             {y'_{i}+\yt_{i}} \right)  \ \mbox{and}
\ 
\frac{2}{y'_{i}+\yt_{i}}
\left(1+\frac{\norm{(\yt _{i}, \aat _{i})}}
             {y'_{i}+\yt_{i}} \right) 
\]
The stated bounds now follow from inequality \eqref{eqn:lp+2max}.
\end{proof}

\begin{lemma}[Almost Gaussian]\label{lem:almostGaussian}
Under the conditions of Lemma~\ref{lem:lp+2}, 
  let $\nu_{i} (p_{i,0}, \pp _{i})$ be the induced density on
  $(p_{i,0}, \pp _{i})$,
  and let $\nuh _{i} (\ph _{i,0}, \pph _{i})$
  be the induced density on $(\ph _{i,0}, \pph _{i})$.
Then, there exists a set $B$ of 
   $\left((p_{1,0}, \pp _{1}),\ldots,(p_{n,0}, \pp _{n})\right)$
  such that
\begin{enumerate}
\item [$(a)$]$\prob{}{\left((p_{1,0}, \pp _{1}),\ldots,(p_{n,0}, \pp
_{n})\right) \in B} \geq 1- 0.0015 \binom{n}{d+1}^{-1}$; and 
\item [$(b)$]  for all $\left((p_{1,0}, \pp _{1}),\ldots,(p_{n,0}, \pp _{n})
  \right) \in B$, 
\[
   \prod_{i=1}^{n}\nu_{i} (p_{i,0}, \pp _{i})
  \leq e\prod_{i=1}^{n} \nuh _{i} (p_{i,0}, \pp _{i}).
\]
\end{enumerate}
\end{lemma}

\begin{proof}
Let 
\[
B = \setof{
  \begin{array}{l}
  ( 
  (p_{1,0} (\htil_{1}), \pp_{1} (\htil_{1}, \ggt_{1})), \ldots ,
  (p_{n,0} (\htil_{n}), \pp_{n} (\htil_{n}, \ggt_{n}))
  )\\
\qquad  \qquad  \text{such that}
  \norm{(\htil_{i}, \ggt_{i})} \leq 3 \sqrt{(d+1) \ln n} \rho _{1},
  \text{ for $1 \leq i \leq n$}
\end{array}
}.
\]\index{B@B}%
From inequalities \eqref{eqn:lp+2max} and \eqref{eqn:lp+2rho},
  and the assumption $\abs{\htil_{i}} \leq 3 \sqrt{(d+1) \ln n} \rho_{1}$,
  we can show  $y' _{i} + \yt_{i} + \htil_{i} > 0$, and so
  the map from
$( \htil_{1},\ggt_{1}), \ldots ,( \htil_{n},\ggt_{n})$ to 
 $(p_{1,0}, \pp _{1}), \ldots , (p_{n,0}, \pp _{n})$
  is invertible for  
 $(p_{1,0}, \pp _{1}), \ldots , (p_{n,0}, \pp _{n}) \in B$. 
 Thus, we may apply Corollary~\ref{cor:chiSquare} to 
  establish part $(a)$.

Part $(b)$ of follows directly Lemma~\ref{lem:almostGaussianSingle}.
\end{proof}

\begin{lemma}[Almost Gaussian, single variable]\label{lem:almostGaussianSingle}
Under the conditions of Lemma~\ref{lem:lp+2}, 
  for all $\htil_{i}$ and $\ggt_{i}$ such that
  $\norm{(\htil_{i},\ggt_{i})} \leq  3 \sqrt{(d+1) \ln n} \rho_{1}$,
\[
  \nu_{i} (p_{i,0} (\htil_{i}), \pp _{i} (\htil_{i}, \ggt_{i}))
  \leq e^{1/n} \nuh _{i} (p_{i,0} (\htil_{i}), \pp _{i} (\htil_{i}, \ggt_{i})).
\]
\end{lemma}
\begin{proof}
Let $\mu (\htil_{i}, \ggt_{i})$ be the density on $(\htil_{i}, \ggt_{i})$.
As observed in the proof of Lemma~\ref{lem:almostGaussian},
  the map from $(\htil_{i}, \ggt_{i})$ to
  $(p_{i,0} (\htil_{i}), \pp_{i} (\htil_{i}, \ggt_{i}))$ is injective for
  $\norm{(\htil_{i},\ggt_{i})} \leq  3 \sqrt{(d+1) \ln n} \rho_{1}$;
 so, by Proposition~\ref{pro:cov}, the induced density 
  on $\nu_{i}$ is
\[
  \nu_{i} (p_{i,0}, \pp_{i})
  =
\frac{1}{ \Jacob{p_{i,0}, \pp_{i}}{\htil_{i}, \ggt_{i} }}
  \mu (\htil_{i}, \ggt_{i}),
\text{where $(p_{i,0}, \pp_{i}) = (p_{i,0} (\htil_{i}), \pp_{i} (\htil_{i}, \ggt_{i}))$.}
\]
Similarly, 
\[
  \nuh _{i} (\ph _{i,0}, \pph _{i})
  =
\frac{1}{ \Jacob{\ph _{i,0}, \pph _{i}}{\hhat_{i}, \ggh _{i} }}
  \mu (\hhat_{i}, \ggh _{i}),
\text{where $(\ph _{i,0}, \pph _{i}) = (\ph _{i,0} (\hhat_{i}), \pph _{i} (\hhat_{i} , \ggh _{i}))$.}
\]
The proof now follows from 
  Lemma~\ref{lem:almostGaussianPointwise}, which tells us that
\[
  \frac{\mu (\htil_{i},\ggt_{i})}{\mu (\hhat_{i} , \ggh _{i})} \leq e^{0.81 / n},
\]
and
  Lemma~\ref{lem:almostGaussianJac}, which tells us that
\[
  \frac{ \Jacob{\ph _{i,0}, \pph _{i}}{\hhat_{i}, \ggh _{i} }}{\Jacob{p_{i,0}, \pp_{i}}{\htil_{i}, \ggt_{i} }} \leq e^{1/10n}.
\]
\end{proof}

\begin{lemma}[Almost Gaussian, pointwise]\label{lem:almostGaussianPointwise}
Under the conditions of Lemma~\ref{lem:almostGaussianSingle},
If $p_{i,0} (\htil_{i}) = \ph _{0} (\hhat_{i})$,
  $\pp_{i}(\htil_{i}, \ggt_{i} ) = \pph _{i} (\hhat_{i}, \ggh _{i} )$,
  and
  $\norm{\htil_{i}, \ggt_{i}} \leq 3 \sqrt{(d+1) \ln n} \rho_{1} $,
  then
\[
  \frac{\mu (\htil_{i}, \ggt_{i})}{\mu (\hhat_{i}, \ggh _{i})} \leq e^{0.81/n}.
\]
\end{lemma}
\begin{proof}
We first observe that the conditions of the lemma imply
\[
  \hhat_{i} = \frac{\htil_{i} (y_{i}' + \yt_{i})}{y_{i}' + \yt_{i} + \htil_{i}},
\qquad \mbox{ and} \qquad
  \ggh _{i} = \frac{\ggt_{i}  (y_{i}' + \yt_{i})}{y_{i}' + \yt_{i} + \htil_{i}}.
\]
We then compute
\begin{align}\label{eqn:almostGaussianPointwise}
  \frac{\mu (\htil_{i}, \ggt_{i})}{\mu (\hhat_{i}, \ggh _{i})}
& =
 \mathrm{exp} \left(
  \frac{-1}{2 \rho_{1}^{2}}
   \norm{(\htil_{i}, \ggt_{i})}^{2}
   \left(\frac{2 \htil_{i} (y_{i}' + \yt_{i}) + \htil_{i}^{2}}
              {(y_{i}' + \yt_{i} + \htil_{i})^{2}} \right)
 \right).
\end{align}
Assuming   $\norm{(\htil_{i}, \ggt_{i})} \leq 3 \sqrt{(d+1) \ln n} \rho_{1} $,
  the absolute value of the exponent in \eqref{eqn:almostGaussianPointwise}
  is at most
\[
\frac{9 (d+1) \ln n}{2}
  \left(\frac{2 \htil_{i} (y_{i}' + \yt_{i}) + \htil_{i}^{2}}
             {(y_{i}' + \yt_{i} + \htil_{i})^{2}} 
  \right).
\]
From inequalities \eqref{eqn:lp+2max} and \eqref{eqn:lp+2rho},
  we find
\[
  \frac{y_{i}' + \yt_{i}}{(y_{i}' + \yt_{i} + \htil_{i})^{2}}
  \leq 
  \frac{40}{(37)^{2} n (d+1)^{3/2} (\ln n)^{3/2} \rho _{1}}.
\]
Observing that $\htil_{i} \leq (1/40) (y_{i}' + \yt_{i})$,
  we can now lower bound the exponent in \eqref{eqn:almostGaussianPointwise}
  by
\begin{align*}
\frac{9 (d+1) \ln n}{2}
  \left(\frac{2 \htil_{i} (81/80) 40}{(37)^{2} n (d+1)^{3/2} (\ln n)^{3/2} \rho _{i}}
  \right)
\leq
0.81 / n.
\end{align*}
\end{proof}

\begin{lemma}[Almost Gaussian, Jacobians]\label{lem:almostGaussianJac}
Under the conditions of Lemma~\ref{lem:almostGaussianSingle},
\[
\frac{ \Jacob{\ph _{0}, \pph _{i}}{\hhat_{i}, \ggh _{i} }}
     {\Jacob{p_{i,0}, \pp_{i}}{\htil_{i}, \ggt_{i} }} 
  \leq e^{.0094 / n}
\]
\end{lemma}
\begin{proof}
We first note that
\[
 \Jacob{\ph _{0}, \pph _{i}}{\hhat_{i}, \ggh _{i} } 
  = \abs{\det{C_{i}}} 
  = \frac{2^{d+1} y_{i}'}{(y_{i}' +\yt_{i})^{d+2}}.
\]
To compute $ \Jacob{p_{i,0}, \pp_{i}}{\htil_{i}, \ggt_{i} } $,
   we note that
\begin{align*}
  \jacobn{p_{i,0}}{\htil_{i}} & = \frac{-2 y_{i}'}{(y_{i}' + \yt_{i} + \htil_{i})^{2}}, \text{and}\\
  \jacobn{\pp_{i,j} (\htil_{i}, g_{i,k})}{g_{i,k}} & = 
   \begin{cases}
    0 & \text{if $j \not = k$}\\
  \frac{2}{y_{i}' + \yt_{i} + \htil_{i}} & \text{otherwise}.
   \end{cases}
\end{align*}
Thus, the matrix of partial derivatives
   is lower-triangular, and its determinant has absolute value
\[
 \Jacob{p_{i,0}, \pp_{i}}{\htil_{i}, \ggt_{i} }
=
 \frac{2^{d+1} y_{i}'}{(y_{i}' +\yt_{i} + \htil_{i})^{d+2}}.
\]
Thus,
\begin{align*}
\frac{ \Jacob{\ph _{0}, \pph _{i}}{\hhat_{i}, \ggh _{i} }}{\Jacob{p_{i,0}, \pp_{i}}{\htil_{i}, \ggt_{i} }}
& = 
\left(\frac{y_{i}' + \yt_{i} + \htil_{i}}{y_{i}' + \yt_{i}} \right)^{d+2}\\
& = 
\left(1 + \frac{\htil_{i}}{y_{i}' + \yt_{i}} \right)^{d+2}\\
& \leq 
\left(1 + \frac{3 \htil_{i}}{2y_{i}'} \right)^{d+2},
  & \text{by \eqref{eqn:lp+2max} }\\
& \leq 
e^{\frac{3(d+2) \htil_{i}}{2 y_{i}' }}\\
& \leq 
e^{0.094 / n}, 
  & \text{by $d \geq 3$ and \eqref{eqn:lp+2rho}.}
\end{align*}
\end{proof}

% Local Variables: ***
% TeX-master:"shadow.tex" ***
% End: ***

\section{Discussion and Open Questions}\label{sec:conclusions}

The results proved in this paper support the assertion that
  the shadow-vertex simplex algorithm usually runs in polynomial
  time.
However, our understanding of the performance of the simplex algorithm
  is far from complete.
In this section, we discuss problems in the analysis of the simplex
  algorithm and in the smoothed analysis of algorithms that deserve
  further study.

\subsection{Practicality of the analysis}\label{ssec:concPractical}
While we have demonstrated that the smoothed complexity of the
  shadow-vertex algorithm is polynomial, the polynomial
  we obtain is quite large.
Yet, we believe that the present analysis provides
  some intuition for why the shadow-vertex simplex algorithm should
  run quickly.
It is clear that the proofs in this paper are very loose and
  make many worst-case assumptions that are unlikely to be
  simultaneously valid.
We did not make any attempt to optimize
  the coefficients or exponents of the polynomial we obtained.
We have not attempted such optimization for two reasons:
  they would increase the length of the paper and probably make it
  more difficult to read; and, we believe that it should be possible
  to improve the bounds in this paper by \textit{simplifying}
  the analysis rather than making it more complicated.
Finally, we point out that most of our intuition comes from
  the shadow size bound, which is not so bad as the bound
  for the two-phase algorithm.

\subsection{Further analysis of the simplex
  algorithm}\label{ssec:concSimplex}

\begin{itemize}
\item While we have analyzed the shadow-vertex pivot rule,
  there are many other pivot rules that are more commonly used
  in practice.
 Knowing that one pivot rule usually takes polynomial time
  makes it seem reasonable that others should as well.
 We consider the maximum-increase and steepest-increase rules,
   as well as randomized pivot rules,
  to be good candidates for smoothed analysis.
 However, the reader should note that there is a reason that
  the shadow-vertex pivot rule was the first to be analyzed:
  there is a simple geometric description of the vertices  encountered
  by the algorithm.
 For other pivot rules, the only obvious characterization of the
  vertices encountered is by iterative application of the pivot rule.
 This iterative characterization introduces dependencies that make
  probabilistic analysis difficult.

\item Even if we cannot perform a smoothed analysis of other pivot
  rules, we might be able to 
   measure the diameter of a polytope under smoothed
  analysis.  
  We conjecture that it is expected polynomial
  in $m$, $d$, and $1/\sigma $.

\item 
  Given that the shadow-vertex simplex algorithm can solve
  the perturbations of linear programs efficiently,
  it seems natural to ask if we can follow the solutions as
  we \emph{unperturb} the linear programs.
For example, having solved an instance of type $(\ref{eqn:lpEnumerated2})$,
  it makes sense to follow the solution as we let $\sigma $
  approach zero.
Such an approach is often called a \emph{homotopy} or 
  \emph{path-following} method.
So far, we know of no reason that there should exist
  an $\AA$ for which  one cannot follow
  these solutions in expected polynomial time, where the expectation
  is taken over the choice of $\GG $.
Of course, if one could follow these solutions in expected polynomial
  time for every $\AA$, then one would have a randomized
  strongly-polynomial time algorithm for linear programming!

\end{itemize}

\subsection{Degeneracy}\label{ssec:concDegeneracy}
One criticism of our model is that it does not allow for
  degenerate linear programs.
It is an interesting problem to find a model of local perturbations
  that will preserve meaningful degeneracies.
It seems that one might be able to expand upon the ideas
  of Todd~\cite{ToddModels} to construct such a model.
Until such a model presents itself and is analyzed, we make the following
  observations about types of degeneracies.

\begin{itemize}
\item In primal degeneracy, a single feasible vertex may
  correspond to multiple bases, $I$.
  In the polar formulation, this corresponds to
  an unexpectedly large number of the $\aa_{i}$s lying
  in a $(d-1)$-dimensional affine subspace.
In this case, a simplex method may cycle---spending many steps
  switching among bases for this vertex, failing to make progress
  toward the objective function.
Unlike many simplex methods, the shadow-vertex method may still be
  seen to be making progress in this situation: each successive basis
  corresponds to a simplex that maps to an edge further along
  the shadow.
It just happens that these edges are co-linear.

A more severe version of this phenomenon 
  occurs when the set of feasible points of a linear program
  lies in an affine subspace of fewer then $d$ dimensions.
By considering perturbations to the constraints under the condition
  that they do not alter the affine span of the set of feasible points,
  the results on the sizes of shadows obtained in Section~\ref{sec:shadow}
  carry over unchanged.  However, how such a restriction would
  affect the results in Section~\ref{sec:phaseI} is presently unclear.

\item In dual degeneracy, the optimal solution of the linear
  program is a face of the polyhedron rather than a vertex.
  This does not appear to be a very strong condition,
  and we expect that one could extend our analysis
  to a model that preserves such degeneracies.

\end{itemize}

\subsection{Smoothed Analysis}\label{ssec:concSmoothed}

We believe that many algorithms will be better understood
  through smoothed analysis.
Scientists and engineers routinely use algorithms with poor worst-case 
  performance.
Often, they solve problems that appear intractable from the
  worst-case perspective.
While we do not expect smoothed analysis to explain every such instance,
  we hope that it can explain away a significant fragment of the
  discrepancy between the algorithmic intuitions 
  of engineers and theorists.
To make it easier to apply smoothed analyses, we briefly discuss
  some alternative definitions of smoothed analysis.

\textbf{Zero-preserving perturbations: }
One criticism of \textit{smoothed complexity} as defined in
  Section~\ref{ssec:smooth} 
  is that the additive Gaussian perturbations destroy any zero-structure
  that the problem has, as it will replace the zeros with small
  values.
One can refine the model to fix this problem by studying  
  \textit{zero-preserving perturbations}.
In this model, one applies Gaussian perturbations only 
  to non-zero entries.
Zero entries remain zero.

\textbf{Relative perturbations: }
A further refinement is the model of
  \textit{relative perturbations}.
Under a relative perturbation, an input is mapped to a constant
  multiple of itself.
For example, a reasonable definition would be to map each variable by
\[
   x \mapsto x ( 1 + \sigma g),
\]
where $g$ is a Gaussian random variable of mean zero and variance 1.
Thus, each number is usually mapped to one of similar magnitude,
  and zero is always mapped to zero.
When we measure smoothed complexity under relative perturbations,
  we call it \textit{relative smoothed complexity}.
Smooth complexity as defined in Section~\ref{ssec:smooth} above can be called 
  \textit{absolute smoothed complexity} if clarification is necessary.
It would be very interesting to know if the simplex method
  has polynomial relative smoothed complexity.

%It is possible to generalize the definition of perturbation 
%  and smoothed analysis.
%We briefly outline an extreme generalization of these definitions.

\textbf{$\epsilon$-smoothed-complexity:}
Even if we cannot bound the expectation 
  of the running time of an algorithm 
  under perturbations,
  we can still obtain computationally meaningful results for an
  algorithm by proving that it has
  $\epsilon$-smoothed-complexity $f (n,\sigma, \epsilon )$, 
  by which we mean that
  the probability that it takes time more than $f (n,\sigma, \epsilon  )$
  is at most $\epsilon $:n
 \index{epss@$\epsilon$-smoothed-complexity}%
\[
    \forall _{x \in X_{n}}
       \prob{g}{C (A,x+\sigma \max (x) g) \leq f (n,\sigma )}
     \geq 1 - \epsilon.
\]

\section{Acknowledgments}\label{sec:ack}
We thank Chloe for support in writing this paper,
  and Donna, Julia and Diana for their patience while
  we wrote, wrote, and re-wrote.
We thank Marcos Kiwi for his comments on the 
  earliest draft of this paper, and
  Mohammad Madhian for his detailed comments
  on the first draft.
We thank Alan Edelman, Charles Leiserson, and Michael Sipser
  for helpful conversations.
We thank Alan Edelman for proposing the
  name ``smoothed analysis''.
Finally, we thank the referees for extrodinary efforts
  and many helpful suggestions.

% Local Variables: ***
% TeX-master:"shadow.tex" ***
% End: ***

\bibliographystyle{alpha}
\bibliography{shadow}

%\printindex[not]
\printindex

\end{document}